\documentclass[
superscriptaddress,
amsmath,
amssymb,
prx,
twocolumn,
aps]{revtex4-2}
\usepackage{bbm}
\usepackage{graphicx}
\usepackage{dcolumn}
\usepackage{bm}
\usepackage{xcolor}
\definecolor{mylinkcolor}{RGB}{68, 114, 196}  
\usepackage{hyperref}
\hypersetup{
    colorlinks=true,
    linkcolor=mylinkcolor,
    urlcolor=mylinkcolor,
    citecolor=mylinkcolor
}
\usepackage[normalem]{ulem}

\begin{document}
\title{Decomposing Electronic Structures in Twisted Multilayers: Bridging Spectra and Incommensurate Wave Functions}

\author{Citian Wang}
\affiliation{School of Physics, Peking University, Beijing 100871, China}

\author{Huaqing Huang}
\email[Corresponding author: ]{huaqing.huang@pku.edu.cn}
\affiliation{School of Physics, Peking University, Beijing 100871, China}
\affiliation{Collaborative Innovation Center of Quantum Matter, Beijing 100871, China}
\affiliation{Center for High Energy Physics, Peking University, Beijing 100871, China}

\date{\today}

\begin{abstract}
Twisted multilayer systems, encompassing materials like twisted bilayer graphene (TBG), twisted trilayer graphene, and twisted bilayer transition metal dichalcogenides, have garnered significant attention in condensed matter physics.
Despite this interest, a comprehensive description of their electronic structure, especially in dealing with incommensurate wave functions, poses a persistent challenge.
Here, we introduce a unified theoretical framework for efficiently describing the electronic structure covering both spectrum and wave functions in twisted multilayer systems, accommodating both commensurate and incommensurate scenarios.
Our analysis reveals that physical observables in these systems can be systematically decomposed into contributions from individual layers and their respective $\bm{k}$ points, even in the presence of intricate interlayer coupling.
This decomposition facilitates the computation of wave function-related quantities relevant to response characteristics beyond spectra.
We propose a local low-rank approximated Hamiltonian in the truncated Hilbert space that can achieve any desired accuracy, offering numerical efficiency compared to the previous large-scale calculations.
The validity of our theory is confirmed through computations of spectra and optical conductivity for moir\'e TBG, and demonstrating its applicability in incommensurate quasicrystalline TBG.
Our findings provide a generic approach to studying the electronic structure of twisted multilayer systems and pave the way for future research.
\end{abstract}

\maketitle


\section{Introduction}
Twisted multilayer systems, formed by stacking two-dimensional (2D) periodic layers with a twist, encompassing materials like twisted bilayer graphene (TBG)~\cite{cao2018unconventional,cao2018correlated,yankowitz2019tuning,sharpe2019emergent,lu2019superconductors,kerelsky2019maximized,jiang2019charge,xie2019spectroscopic,choi2019electronic,serlin2020intrinsic,stepanov2020untying,zondiner2020cascade,wong2020cascade, PhysRevLett.124.076801, park2021tunable},
twisted trilayer graphene~\cite{chen2019evidence,chen2019signatures, PhysRevLett.122.016401,hao2021electric, PhysRevX.13.041007,guerci2023chern},
and twisted bilayer transition metal dichalcogenides~\cite{liu2014evolution, PhysRevLett.121.266401,jin2019observation,wang2020correlated,zhang2020flat,huang2021correlated}, have garnered significant attention in condensed matter physics. These systems provide a highly tunable platform for exploring exotic quantum phenomena,
including unconventional superconductivity~\cite{PhysRevLett.121.087001, PhysRevLett.121.257001, PhysRevLett.122.026801, PhysRevX.8.041041,balents2020superconductivity,torma2022superconductivity,tian2023evidence},
correlated insulating phases~\cite{PhysRevLett.117.116804, PhysRevX.8.031089,regan2020mott, PhysRevLett.124.097601, PhysRevB.103.035427,PhysRevLett.128.247402},
and topological phases of matters~\cite{PhysRevLett.121.037702, PhysRevB.99.075127, PhysRevB.99.195455, PhysRevB.99.155415, PhysRevX.9.021013, PhysRevLett.122.086402, PhysRevLett.123.036401, PhysRevLett.123.216803}.
The recent observation of fractional anomalous Hall states in twisted bilayer $\mathrm{MoTe_2}$~\cite{Cai2023Signatures, Zeng2023Therm, PhysRevX.13.031037} has reignited interest among both theoretical and experimental physicists in these twisted multilayer systems. However, providing a comprehensive description of their electronic structure remains challenging, primarily due to intricate interlayer coupling, especially in incommensurate cases without a moir\'e superlattice.

Previous methods for calculating electronic structure in twisted multilayer systems generally fall into two categories. One approach is the moir\'e effective method, which approximates usually incommensurate noncrystalline systems as translational symmetric systems using a low-energy continuum model~\cite{PhysRevLett.99.256802, PhysRevB.81.161405,bistritzer2011moire, PhysRevB.86.155449, Koshino_2015,PhysRevB.93.235153, PhysRevB.93.035452, PhysRevB.97.035306, PhysRevX.8.031087, PhysRevX.8.031088, PhysRevB.99.205134, PhysRevB.100.035101, PhysRevResearch.1.013001, PhysRevB.103.205411, PhysRevB.107.125112}. This includes the well-known Bistritzer-MacDonald (BM) model for TBG with small twist angles~\cite{bistritzer2011moire}.
The other method relies on large-scale calculation techniques, such as direct supercell calculations and density-functional calculations with generalized unfolding techniques~\cite{PhysRevB.82.121407,trambly2010localization, PhysRevB.85.195458, PhysRevB.87.205404, PhysRevB.90.155406, PhysRevB.90.155451, PhysRevB.95.085420, PhysRevB.95.075420, PhysRevMaterials.2.010801, PhysRevB.99.195419, yu2019dodecagonal, PhysRevB.105.125127, PhysRevX.12.021055}. Both methods have their advantages and disadvantages. The moir\'e effective approach is computationally efficient but becomes less valid for large twist angles or larger energy scales due to the absence of microscopic details. On the other hand, the large-scale simulation approach offers high accuracy but involves extremely high computational costs. Worse still, it cannot handle incommensurate cases without a single moir\'e superlattice, which are generally encountered.

Experimental observations of incommensurate twisted multilayer systems, including quasicrystalline 30$^\circ$  TBG~\cite{science.aar8412,pnas.1720865115,Pezzini2020CVD,Deng2020intelayer,pezzini202030}, graphene on top of the $\mathrm{BN}$ layer~\cite{PhysRevLett.116.126101}, and trilayer moir\'e quasicrystal~\cite{Uri2023moireQuasicrystal}, highlights the limitations of existing methods. To address this, Moon \textit{et al.} introduced a $\bm{k}$-space tight-binding model~\cite{PhysRevB.99.165430}, successfully calculating the density of states and offering some insights into interlayer coupling. However, this model lacks a quasi-band structure for direct comparison with angle-resolved photoemission spectroscopy (ARPES) data and the ability to extract wave function information essential for computing physical observables beyond spectra. In summary, substantial gaps exist between existing models and experimental observations, and a unified, accurate, and computationally efficient theory for electronic structure in both commensurate and incommensurate twisted multilayer systems is still absent, significantly hindering further research.

In this paper, we introduce a unified theoretical framework for comprehensively describing the electronic structure in multilayer systems, accommodating both commensurate and incommensurate scenarios, bridging gaps in previous methods. The workflow of our theory, depicted in Fig.~\ref{fig:workflow}, is simple and straightforward. Based on the Hamiltonian under the \textit{composite} Bloch basis, we demonstrate that physical observables in trace form in twisted multilayer systems can be decomposed into contributions from individual layers and their respective $\bm{k}$ points, even with intricate interlayer coupling. This decomposition, along with our proposed local low-rank approximated Hamiltonian in the truncated Hilbert space, enables the computation of wave function-related quantities relevant to response characteristics beyond spectra in the desired accuracy. We validate our theory by computing the density of states, quasi-band structure, and optical conductivity for commensurate moir\'e TBG and demonstrate its application in incommensurate quasicrystalline TBG. Our theory offers a versatile and computationally efficient approach to investigating the electronic structure of twisted multilayer systems.

\section{Hamiltonian in composite Bloch basis}
\begin{figure}
\includegraphics[width=1.0\columnwidth, keepaspectratio]{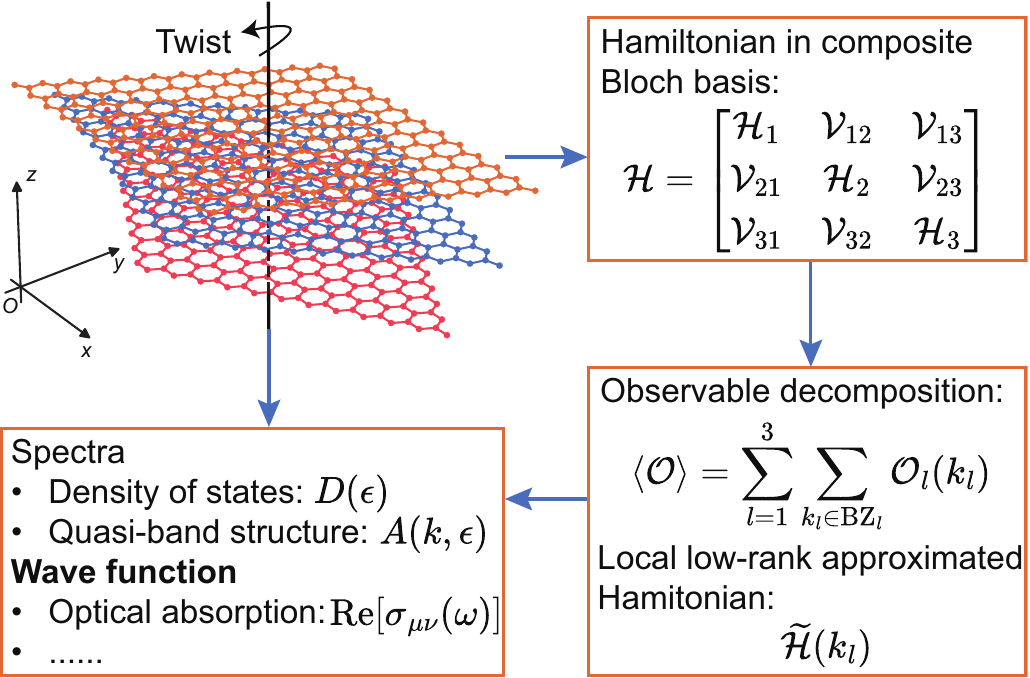}
\caption{\label{fig:workflow} Schematic illustration of the workflow in our unified theory for electronic structure in twisted multilayer systems (using twisted trilayer graphene as an example). Given the twisted multilayer systems, no matter commensurate or incommensurate, its Hamiltonian $\mathcal{H}$ is composed of single layer parts $\mathcal{H}_l$ and interlayer coupling $\mathcal{V}_{l,l^\prime}$ under the composite Bloch basis. Observable $\langle \mathcal{O}\rangle$ can be decomposed into contributions from individual layers and their respective $\bm{k}$ points $\mathcal{O}_{l}(\bm{k}_l)$. By combining our local low-rank approximated Hamiltonian $\widetilde{\mathcal{H}}(\bm{k}_l)$ in the truncated Hilbert space, the spectra and wave function-related quantities can be obtained in desired accuracy.}
\end{figure}

In this section, we demonstrate the Hamiltonian of a general twisted multilayer system under a composite Bloch basis. We also emphasize the numerical challenges resulting from the intricate interlayer coupling.

We first elucidate the basic structure in our model. Consider a general twisted multilayer system, where each 2D periodic layer is denoted by subscript $l$ and has a height vector $h_{l}\bm{e}_z$ along the stacking $z$ axis. The lattice vectors $\bm{R}_l$ of layer-$l$ are defined as:
\begin{equation}
\label{eq_rlat}
\bm{R}_{l} = n^1_{l} \bm{a}^1_{l} + n^2_{l} \bm{a}^2_{l},
\end{equation}
where $n^1_{l},n^2_{l} \in \mathbb{Z}$, $\bm{a}^1_{l}$ and $\bm{a}^2_{l}$ are the 2D primitive real lattice vectors for layer-$l$ perpendicular to the stacking $z$ axis, as shown in Fig.~\ref{fig:workflow}. The reciprocal lattice vectors of layer-$l$ are defined as:
\begin{equation}
\label{eq_klat}
\bm{G}_{l}=m^1_{l} \bm{b}^1_{l} + m^2_{l} \bm{b}^2_{l},
\end{equation}
where $m^1_{l},m^2_{l} \in \mathbb{Z}$, $\bm{b}^1_{l}$ and $\bm{b}^2_{l}$ are the 2D primitive reciprocal lattice vectors for layer-$l$ that satisfy $\bm{a}^i_{l} \cdot \bm{b}^j_{l} = 2\pi \delta_{i,j}$. This relation ensures $e^{i \bm{R}_{l} \cdot \bm{G}_{l}} = 1$ for any pair of real and reciprocal lattice vectors in the same layer.

Then we introduce a single layer local atomic basis in real space, denoted as $\{|\bm{R}_{l} \bm{t}_{l} \alpha \rangle\}$, which satisfies orthonormality conditions:
\begin{equation}
\label{eq_real_orthonormal}
\langle {\bm{R}}^{\prime}_{l^{\prime}} {\bm{t}}^{\prime}_{l^\prime} \beta | \bm{R}_{l} \bm{t}_{l}\alpha \rangle = \delta_{l,l^{\prime}} \delta_{ \bm{R}_l, \bm{R}^\prime_{l^\prime}} \delta_{ \bm{t}_l,\bm{t}^\prime_{l^\prime}} \delta_{\alpha, \beta}.
\end{equation}
Here, $\bm{t}_{l}$ represents the positions of different degrees of freedom within each unit cell of layer-$l$, such as atom sites or sublattice positions. And the internal orbital or spin degrees of freedom are uniquely labeled by $\alpha$. This also straightforwardly leads to position matrix elements:
\begin{equation}
\label{eq_real_position_matrix}
\langle {\bm{R}}^{\prime}_{l^{\prime}} {\bm{t}}^{\prime}_{l^\prime} \beta | \bm{r} | \bm{R}_{l} \bm{t}_{l}\alpha \rangle = ( \bm{R}_{l} \bm+\bm{t}_{l}+h_l\bm{e}_z) \delta_{l,l^{\prime}} \delta_{ \bm{R}_l, \bm{R}^\prime_{l^\prime}} \delta_{ \bm{t}_l,\bm{t}^\prime_{l^\prime}} \delta_{\alpha, \beta}.
\end{equation}

Since individual layers still possess translational symmetry, the Bloch theorem remains valid in single-layer 2D reciprocal space (as shown in Eq.~(\ref{eq_kplusG})). Therefore, we are allowed to construct a single layer Bloch basis $\{|\bm{k}_{l} \bm{t}_{l}\alpha\rangle\}$ for layer-$l$ by defining
\begin{equation}
\label{eq_kbasis}
|\bm{k}_{l} \bm{t}_{l} \alpha \rangle = \frac{1}{\sqrt{N_{l}}} \sum_{\bm{R}_{l}} e^{i\bm{k}_{l}\cdot(\bm{R}_l+\bm{t}_{l})} |\bm{R}_{l} \bm{t}_{l} \alpha \rangle,
\end{equation}
where $\bm{k}_{l}$ resides in the first Brillouin zone (BZ) of layer-$l$ (i.e., $\bm{k}_{l} \in \mathrm{BZ}_l$), and we consider layer-$l$ contains $N_l$ cells with Born-von Karman boundary condition. Furthermore, the single layer Bloch basis $\{|\bm{k}_{l} \bm{t}_{l}\alpha\rangle\}$ satisfies orthonormality conditions:
\begin{equation}
\label{eq_reciprocal_orthonormal}
\langle \bm{k}^\prime_{l} \bm{t}^\prime_{l} \beta |\bm{k}_{l} \bm{t}_{l}\alpha\rangle = \delta_{l,l^\prime} \delta_{\bm{k},\bm{k}^\prime} \delta_{\bm{t},\bm{t}^\prime} \delta_{\alpha, \beta}.
\end{equation}
Note that we adopt a convention incorporating the phase factor $e^{i \bm{k}_{l} \cdot \bm{t}_{l}}$. Different conventions for this phase factor are discussed in Appendix~\ref{app:convention}. Also, $|\bm{k}_{l}+\bm{G}_{l},\bm{t}_l \alpha\rangle$ is equivalent to $|\bm{k}_{l}\bm{t}_l\alpha\rangle $ with a phase factor $e^{i\bm{G}_l\cdot\bm{t}_l}$, as shown below:
\begin{eqnarray}
\label{eq_kplusG}
|\bm{k}_{l}+\bm{G}_{l},\bm{t}_l \alpha\rangle
&=& \frac{1}{\sqrt{N_{l}}}\sum_{\bm{R}_{l}} e^{i (\bm{k}_{l}+\bm{G}_{l}) \cdot (\bm{R}_{l}+\bm{t}_{l})} |\bm{R}_{l} \bm{t}_{l} \alpha \rangle, \nonumber \\
&=& \frac{1}{\sqrt{N_{l}}}\sum_{\bm{R}_{l}} e^{i\bm{k}_{l}\cdot (\bm{R}_{l}+\bm{t}_{l})+i\bm{G}_l \cdot \bm{t}_{l}} |\bm{R}_{l} \bm{t}_{l}\alpha \rangle, \nonumber \\
&=& e^{i\bm{G}_l\cdot\bm{t}_{l}} |\bm{k}_{l}\bm{t}_l \alpha\rangle,
\end{eqnarray}
This equivalence results from the identity $e^{\bm{G}_l \cdot \bm{R}_{l}}=1$. Hence, the single layer Bloch wavevectors $\bm{k}_l$ within $\mathrm{BZ}_l$ provide a complete description for layer-$l$. This indicates that the composite Bloch basis $\coprod_{l}\{|\bm{k}_{l} \bm{t}_{l}\alpha\rangle\}$ also provides a complete description for the entire twisted multilayer system.

Under the composite Bloch basis, the Hamiltonian $\mathcal{H}$ of a general twisted multilayer system can be expressed concisely as follows:
\begin{equation}
\label{eq_full_H}
\mathcal{H}=\sum_{l} \mathcal{H}_{l} + \sum_{l,l^{\prime}} \mathcal{V}_{l,l^{\prime}}.
\end{equation}
Here, $\mathcal{H}_{l}$ is the single-layer Hamiltonian of layer-$l$, while $\mathcal{V}_{l,l^{\prime}}$ denotes the interlayer coupling between layer-$l$ and layer-$l^{\prime}$. After some algebra (see Appendix~\ref{app:intralayer} for details), we can show that $\mathcal{H}_{l}$ exhibits a block-diagonal structure:
\begin{eqnarray}
\label{eq_intralayer}
\langle\bm{k}^\prime_{l}\bm{t}^\prime_{l}\beta|\mathcal{H}_{l}|\bm{k}_{l}\bm{t}_{l}\alpha\rangle
&=&  \delta_{\bm{k}_l, \bm{k}^\prime_l} \sum_{\bm{R}_l} e^{i \bm{k}_l \cdot ( \bm{R}_l +\bm{t}_l-\bm{t}^\prime_l )} \nonumber \\
& & \times \mathcal{H}_{\beta,\alpha}(\bm{R}_{l}+\bm{t}_l-\bm{t}^\prime_l),
\end{eqnarray}
where $\mathcal{H}_{\beta,\alpha}(\bm{r})$ donates the hopping integral between $\alpha$ and $\beta$ with a displacement $\bm{r}$.

However, when it comes to interlayer coupling matrix elements, they are significantly more complex than the intralayer terms. A direct calculation of these elements provides insight into their intricate nature:
\begin{widetext}
\begin{equation}
\label{eq_interlayer_direct}
\langle\bm{k}^\prime_{l^\prime}\bm{t}^\prime_{l^\prime}\beta|\mathcal{V}_{l,l^\prime}|\bm{k}_{l}\bm{t}_{l}\alpha\rangle= \frac{1}{\sqrt{N_l {N}_{l^\prime}}} \sum_{\bm{R}_l,\bm{R}^\prime_{l^\prime}} e^{-i\bm{k}^\prime_{l^\prime} \cdot ( \bm{R}^\prime_{l^\prime} + \bm{t}^\prime_{l^\prime} )} e^{i\bm{k}_l \cdot (\bm{R}_l+\bm{t}_l)} \mathcal{H}_{\beta,\alpha}(\bm{R}_{l}-\bm{R}^\prime_{l^\prime}+\bm{t}_{l}-\bm{t}^\prime_{l^\prime}+\Delta h_{l,l^\prime} \bm{e}_z),
\end{equation}
\end{widetext}
where $l \neq l^\prime$. Notably, when $\bm{R}_l$ and $\bm{R}^\prime_{l^\prime}$ are incommensurate lattice vectors (not forming a moir\'e superlattice), the quantity $\bm{R}_{l}-\bm{R}^\prime_{l^\prime}$ is no longer periodic. Thus, the change of variables technique, successfully applied to intralayer terms, doesn't work for interlayer coupling terms (see Appendix~\ref{app:intralayer}). This failure occurs because we cannot apply the Poisson summation formula [see Eq.~(\ref{eq_appen_Poisson_delta_final}) in Appendix~\ref{app:Poisson}] to decouple the states $|\bm{k}^\prime_{l^\prime} \rangle$ and $| \bm{k}_{l} \rangle$. Essentially, a single state $| \bm{k}_{l} \rangle$ in layer-$l$ becomes entangled with all $|\bm{k}^\prime_{l^\prime} \rangle$ states over the entire BZ of another layer-$l^\prime$. As a result, direct diagonalization of the Hamiltonian $\mathcal{H}$ becomes extremely challenging due to intricate interlayer coupling.

By employing a Fourier transformation of the hopping integral, a concise and elegant description of interlayer coupling in twisted bilayer systems through generalized Umklapp processes~\cite{Koshino_2015} in reciprocal space is achieved. More specifically, the interlayer coupling between layer-$l$ and layer-$l^{\prime}$ can be expressed in our notation as (see Appendix~\ref{app:interlayer} for details):
\begin{eqnarray}
\label{eq_interlayer}
\langle\bm{k}^\prime_{l^\prime}\bm{t}^\prime_{l^\prime}\beta|\mathcal{V}_{l,l^\prime}|\bm{k}_{l}\bm{t}_{l}\alpha\rangle
&=& \sum_{\bm{G}_l,\bm{G}^\prime_{l^\prime}}
\mathcal{V}^{l^\prime,l}_{\beta,\alpha}(\bm{k}_l + \bm{G}_l)
e^{-i\bm{G}_l \cdot \bm{t}_{l} +i \bm{G}^\prime_{l^\prime} \cdot \bm{t}^\prime_{l^\prime}} \nonumber \\
& & \times \delta_{\bm{k}_l+\bm{G}_l,\bm{k}^\prime_{l^\prime}+\bm{G}^\prime_{l^\prime}}.
\end{eqnarray}
It reveals that coupling between a state $|\bm{k}_{l}\rangle$ in layer-$l$ and a state $|\bm{k}^\prime_{l^\prime}\rangle$ in layer-$l^\prime$ is nonzero only when a pair of reciprocal lattice vectors $\bm{G}_l$ and ${\bm{G}}^\prime_{l^\prime}$ satisfying the following condition:
\begin{equation}
\label{eq_Umklapp}
\bm{k}_{l} + \bm{G}_l=\bm{k}^\prime_{l^\prime} + \bm{G}^\prime_{l^\prime}.
\end{equation}

\section{Hilbert space truncation}
Before proceeding, we have a few remarks on Eq.~(\ref{eq_interlayer}). It shows that for a state at $\bm{k}_{l}$ in layer-$l$, all states in layer-$l^\prime$ coupled to it are located at $\bm{k}^\prime_{l^\prime}=\bm{k}_{l}+\bm{G}_l-\bm{G}^\prime_{l^\prime}, \forall \bm{G}_l,\bm{G}^\prime_{l^\prime}$, and the coupling can be interpreted as virtual scattering processes involving $| \bm{k}_{l} + \bm{G}_l \rangle$ and $|\bm{k}^\prime_{l^\prime} + \bm{G}^\prime_{l^\prime} \rangle$. However, all $\bm{G}_l-\bm{G}^\prime_{l^\prime}$ also form a dense point set in reciprocal space due to the inherent incommensurability from $\bm{R}_l-\bm{R}^\prime_{l^\prime}$. Thus, Eq.~(\ref{eq_interlayer}) is merely an identity transformation in mathematics to Eq.~(\ref{eq_interlayer_direct}) and doesn't alter the physical situation: We still face the dilemma of $|\bm{k}_{l}\rangle$ remains entangled with all $| \bm{k}^\prime_{l^\prime} \rangle$ states within $\mathrm{BZ}_{l^\prime}$, making the diagonalization of the Hamiltonian $\mathcal{H}$ impossible. 

However, our reinterpretation of Eq.~(\ref{eq_interlayer}) reveals an intriguing insight:
the intensity of each virtual scattering process, denoted as $\mathcal{V}_{l^\prime,l}^{\beta,\alpha}(\bm{k}_l + \bm{G}_l)$, is solely determined by $|\bm{k}_l + \bm{G}_l|$ and decreases monotonically with $r$ when $\mathcal{H}_{\beta,\alpha}(\bm{r}+\Delta h_{l,l^\prime} \bm{e}_z)$ is centrosymmetric. This suggests a possible theoretical route to introduce a uniform cutoff $G_{\mathrm{cut}}$ for large $|\bm{k}_l + \bm{G}_l|$, allowing the truncation of the working Hilbert space with desired accuracy. Such approximation might simplify the task of diagonalizing a matrix with size $\mathcal{O}(N_l)$ ($N_l \to \infty$) to a series of diagonalizations of $N_{\mathrm{cut}} \times N_{\mathrm{cut}}$ matrices, where the truncated Hilbert space dimension $N_{\mathrm{cut}}$ is finite with $N_{\mathrm{cut}} \ll N_l$.

On this theoretical route, several works have been conducted for both incommensurate and commensurate cases. In the incommensurate case, in addition to the previously mentioned $\bm{k}$-space tight-binding model~\cite{PhysRevB.99.165430}, Koshino's scheme for computing the density of states $D(\epsilon)$ in incommensurate TBG~\cite{Koshino_2015} is a notable success. It provides the quasi-band structure and the density of states of an incommensurate TBG with a large rotation angle, which cannot be treated as a long-range moir\'e superlattice. However, similar to the $\bm{k}$-space tight-binding model, it can't extract the required wave function information for calculating physical observables beyond spectra. Besides, there are lingering theoretical ambiguities that require further clarification with more rigorous formalism (see Appendix~\ref{app:Koshino} for details). In the commensurate case, the recently introduced truncated atomic plane wave method~\cite{PhysRevB.107.125112} is efficient and accurate for small twist angles with moir\'e periodicity, as a generalization of the BM model. However, it's not applicable in incommensurate cases. All these analyses underscore the evident limitations in existing methods and reveal an urgent need for a unified theory with a comprehensive description that encompasses both the spectrum and the wave functions, accommodating both commensurate and incommensurate scenarios.

Our theory, building upon the core idea of truncating the working Hilbert space with an interlayer coupling cutoff, takes a significant step forward in this technical route while satisfying all the mentioned requirements. Two key components in our unified theory, as illustrated in Fig.~\ref{fig:workflow}, are:
\begin{itemize}
    \item Decomposition of physical observables.
    \item Construction of a local low-rank approximation Hamiltonian.
\end{itemize}
The first component provides a rigorous decomposed expression for physical observables, allowing us to maximize the use of single-layer periodicity. Additionally, it offers a weight factor [see Eq.~(\ref{eq_expectation_divide_Okw})] that transforms the global approximation problem into a local one and guides the construction of the local low-rank approximation Hamiltonian. In turn, the resulting local low-rank approximation Hamiltonian efficiently computes spectra and wave function-related quantities with desired accuracy, as demonstrated below.

\section{Decomposition of physical observables}
\subsection{General formalism}
In this section, we demonstrate how to decompose the physical observables in general twisted multilayer systems. The motivation for this decomposition is clear: we aim to utilize single-layer periodicity to mitigate the challenges posed by the complex multilayer Hamiltonian (Eq.~(\ref{eq_full_H})), characterized by its immense size and intricate interlayer coupling. 
And the result exceeds initial expectations, offering theoretical insight into observable decomposition and guiding numerical approximations, as shown below.

We start by introducing the eigenstate basis $\{ |\psi_n\rangle\ \}$ of the twisted multilayer system. Given the immense size and intricate interlayer coupling of the full Hamiltonian [Eq.~(\ref{eq_full_H})], it is impractical to represent the Hamiltonian matrix explicitly, let alone diagonalize it. Nonetheless, we posit the existence of eigenstates $|\psi_n\rangle$ with energy eigenvalues $\epsilon_n$, satisfying:
\begin{equation}
\mathcal{H} |\psi_n\rangle = \epsilon_n |\psi_n\rangle.
\end{equation}
Note that since there is generally no translational symmetry, the conventional band structure indices $\bm{k} n$ are reduced to a single label $n$ for distinct eigenstates.

We state that if a physical observable $\langle \mathcal{O} \rangle$ of an arbitrary operator $\mathcal{O}$ in the twisted multilayer system can be expressed as a trace, it can be decomposed into contributions from individual layers, and further into contributions from their respective $\bm{k}$ points as:
\begin{eqnarray}
\label{eq_expectation_divide_O}
\langle \mathcal{O} \rangle &=& \sum_{l} \sum_{\bm{k}_l\in \mathrm{BZ}_{l}} \mathcal{O}_{l}(\bm{k}_l),\\
\label{eq_expectation_divide_Ok}
\mathcal{O}_{l}(\bm{k}_l) &=&\sum_{n} \sum_{m} w_{nm}(\bm{k}_l) \mathcal{O}_{nm}, \\
\label{eq_expectation_divide_Okw}
w_{nm}(\bm{k}_l) &=& \sum_{\bm{t}_l \alpha} \langle \bm{k}_l \bm{t}_l \alpha| \psi_n \rangle  \langle \psi_m |\bm{k}_l \bm{t}_l \alpha \rangle,
\end{eqnarray}
where $\mathcal{O}_{nm}= \langle \psi_n | \mathcal{O}| \psi_m\rangle$ is the matrix element of operator $\mathcal{O}$ in the eigenstates basis $\{| \psi_n \rangle\}$, $w_{nm}(\bm{k}_l)$ is the weight factor at $\bm{k}_l$ of layer-$l$.

Now, let's prove the above conclusion for an arbitrary operator $\mathcal{O}$ with $\langle \mathcal{O} \rangle$ can be expressed as a trace:
\begin{equation}
\label{eq_O_eigen}
\langle \mathcal{O} \rangle = \mathrm{Tr}(\mathcal{O})=\sum_{n} \langle \psi_n | \mathcal{O}| \psi_n \rangle.
\end{equation}
Here, $\mathrm{Tr}(\cdot)$ denotes the trace operation, which is independent of the choice of orthonormal basis due to its cyclic property. 
For example, if we have a matrix $\mathcal{A}$ represented in one basis $\{ |\bm{e}_j\rangle \}$, it can be equivalently expressed as $\mathcal{S}^\dagger \mathcal{A} \mathcal{S}$ in another orthonormal basis $\{ |\widetilde{\bm{e}}_k\rangle \}$, where $\mathcal{S}$ is a unitary matrix connecting the two basis with $\mathcal{S} \mathcal{S}^\dagger = \mathbbm{1}$. As a result, the trace of $\mathcal{A}$ remains invariant under basis transformations: $\mathrm{Tr}(\mathcal{S} ^\dagger \mathcal{A} \mathcal{S} ) = \mathrm{Tr}( \mathcal{A} \mathcal{S} \mathcal{S}^\dagger) = \mathrm{Tr}(\mathcal{A})$. This invariance ensures that the trace operation consistently yields the same physical results, regardless of the chosen basis. Therefore, instead of using the eigenstates basis $\{| \psi_n \rangle\}$, Eq.~(\ref{eq_O_eigen}) can be alternatively expressed using the composite Bloch basis $\coprod_{l}\{|\bm{k}_{l} \bm{t}_{l}\alpha\rangle\}$:
\begin{equation}
\label{eq_O_k}
\langle \mathcal{O} \rangle= \mathrm{Tr}(\mathcal{O}) = \sum_{l} \sum_{\bm{k}_l\in \mathrm{BZ}_{l}} \sum_{\bm{t}_{l}\alpha} \langle \bm{k}_{l}\bm{t}_{l}\alpha |\mathcal{O}| \bm{k}_{l}\bm{t}_{l}\alpha\rangle.
\end{equation}
Comparing this expression with Eq.~(\ref{eq_expectation_divide_O}), we can readily express the contribution from $\bm{k}_l$ of layer-$l$ to $\langle \mathcal{O} \rangle$ as
\begin{eqnarray}
\label{eq_O_k_tr}
\mathcal{O}_{l}(\bm{k}_{l})
&=& \sum_{\bm{t}_l \alpha} \langle \bm{k}_l\bm{t}_l \alpha |\mathcal{O}| \bm{k}_l\bm{t}_l \alpha \rangle, \nonumber \\
&=& \sum_{\bm{t}_{l}\alpha} \sum_{n} \sum_{m} \langle \bm{k}_l\bm{t}_l \alpha | \psi_n \rangle \langle \psi_n | \mathcal{O}| \psi_m\rangle  \langle \psi_m | \bm{k}_l\bm{t}_l \alpha \rangle, \nonumber \\
&=& \sum_{n} \sum_{m} \left( \sum_{\bm{t}_{l}\alpha} \langle \bm{k}_l\bm{t}_l \alpha | \psi_n \rangle  \langle \psi_m | \bm{k}_l\bm{t}_l \alpha \rangle \right) \mathcal{O}_{nm}, \nonumber \\
&=& \sum_{n} \sum_{m} w_{nm}(\bm{k}_l) \mathcal{O}_{nm},
\end{eqnarray}
where we use the closure relation $\sum_{n}| \psi_n \rangle \langle \psi_n | = \sum_{m} | \psi_m\rangle \langle \psi_m | = \mathbbm{1}$. Note that Eq.~(\ref{eq_O_k_tr}) is equivalent to Eq.~(\ref{eq_expectation_divide_Ok}) and Eq.~(\ref{eq_expectation_divide_Okw}), which conclude the proof. 

Without specific expressions, Eq.~(\ref{eq_expectation_divide_O})-(\ref{eq_expectation_divide_Okw}) already offer valuable insights into the electronic structure of twisted multilayer systems. They enable the direct decomposition of an operator's expectation value into contributions from individual layers and further into contributions from their respective $\bm{k}$ point. This decomposition remains valid even when $\mathcal{H} \neq \bigoplus_{l} \mathcal{H}_l$, where tight coupling exists between different layers and when all $|\bm{k}^\prime_{l^\prime} \rangle$ and $| \bm{k}_{l} \rangle$ states are mixed, all without the need for any additional approximations. Our derivation also introduces the weight factor $w_{nm}(\bm{k}_l)$, which naturally emerges during expectation value calculations, signifying the contribution of the overlap of two specific eigenstates for $\mathcal{O}$ within an individual $\bm{k}$ point. Importantly, our approach doesn't rely on any ensemble-like assumption, and the weight factor's specific form is precisely defined by Eq.~(\ref{eq_expectation_divide_Okw}) without ambiguity (see Appendix~\ref{app:Koshino}).

Specially, when the matrix representation of operator $\mathcal{O}$ in the eigenstate basis $\{| \psi_n \rangle\}$ is diagonal, i.e.,
\begin{equation}
\mathcal{O}_{nm}= \langle \psi_n | \mathcal{O}| \psi_m\rangle =\mathcal{O}_{n} \delta_{n,m},
\end{equation}
substituting this into Eq.~(\ref{eq_expectation_divide_Ok}) yields:
\begin{equation}
\label{eq_expectation_diag}
\mathcal{O}_l(\bm{k}_l) = \sum_{n} \sum_{m} w_{nm}(\bm{k}_l) \mathcal{O}_{n} \delta_{n,m}= \sum_{n} w_{n}(\bm{k}_l) \mathcal{O}_{n}.
\end{equation}
Here, $w_{n}(\bm{k}_l)$ is the reduced weight factor indicating the contribution of a specific eigenstate $| \psi_n \rangle$ for $\mathcal{O}$ within an individual $\bm{k}_l$ point, as the overlap of different eigenstates for $\mathcal{O}$ is zero:
\begin{equation}
\label{eq_reduce_w}
w_{n}(\bm{k}_l) = \sum_{\bm{t}_l \alpha} |\langle \bm{k}_l \bm{t}_l \alpha| \psi_n \rangle  |^2.
\end{equation}

\subsection{Decomposition of spectral quantities}
Now, we demonstrate the application of the decomposition formula for spectral quantities. As an example, we consider the density of states defined as
\begin{equation}
D(\epsilon)=\sum_{n} \delta(\epsilon-\epsilon_n).
\end{equation}
It is straightforward to express this by a trace on the density of states operator $\mathcal{D}(\epsilon)=\delta(\epsilon-\mathcal{H})$, which is diagonal in the eigenstate basis as $\mathcal{D}(\epsilon)_{nm}=\delta(\epsilon-\epsilon_n) \delta_{n,m}$. Applying this to Eq.~(\ref{eq_expectation_diag}), we have
\begin{eqnarray}
& & D(\epsilon) = \langle \mathcal{D}(\epsilon) \rangle = \sum_{l} \sum_{\bm{k}_l\in \mathrm{BZ}_{l}} \mathcal{D}_{l}(\bm{k}_l,\epsilon),\\
\label{eq_Dl}
& & \mathcal{D}_{l}(\bm{k}_l,\epsilon) = \sum_{n} w_{n}(\bm{k}_l) \delta(\epsilon-\epsilon_n),
\end{eqnarray}
where the reduced weight factor $w_{n}(\bm{k}_l)$ is given by Eq.~(\ref{eq_reduce_w}), and $\mathcal{D}_{l}(\bm{k}_l,\epsilon)$ is nothing but the single-layer spectral function of layer-$l$.

Therefore, the summation of all layers at the same $\bm{k}$ point yields the total spectral function $A(\bm{k},\epsilon)$ presenting the quasi-band structure of the twisted multilayer systems, which can be directly compared with ARPES results:
\begin{eqnarray}
\label{eq_A}
A(\bm{k},\epsilon)
&=& \sum_{l} \mathcal{D}_{l}(\bm{k}_l=\bm{k},\epsilon), \nonumber \\
&=& \sum_{l} \sum_{n} \sum_{\bm{t}_l \alpha} |\langle \bm{k}_l=\bm{k}, \bm{t}_l \alpha| \psi_n \rangle  |^2 \delta(\epsilon-\epsilon_n).
\end{eqnarray}
When taking $l=2$ for TBG, the above definition corresponds to the exact form of Eq.~(\ref{eq_Koshino_A}) from Koshino's scheme without any approximation (see Appendix~\ref{app:Koshino}).

Although we lack the specific expression for $w_{n}(\bm{k}_l)$ due to the unknown nature of the eigenstates $| \psi_n \rangle$, Eq.~(\ref{eq_Dl}) already offers valuable insights into the quasi-band structure in twisted multilayer systems. It reveals that at any point $\bm{k}$ in incommensurate case, the single layer spectral function of layer-$l$ actually contains the spectra of the entire multilayer system, modulated by the reduced weight factors $w_{n}(\bm{k}_l)$.
This spectral property, inherent in incommensurate systems and arising from multifractality, as observed in other systems such as 1D Fibonacci quasicrystals ~\cite{RevModPhys.93.045001}, demonstrates the comprehensive capabilities of our theory in describing the spectra of incommensurate twisted multilayer systems.

\subsection{Decomposition of wave function-related quantities}
Next, we apply our decomposition formula to wave function-related quantities. As mentioned before, previous research primarily focused on the spectral characteristics of the twisted multilayer system, offering limited insights into the corresponding eigenstates. However, a comprehensive description requires capturing both the spectra (energy eigenvalues) and the wave functions (eigenstates), with a particular emphasis on the latter. Because eigenstates are essential for computing physical observables and are closely linked to the system's response properties and topology~\cite{vanderbilt_2018}. Our decomposition formula for twisted multilayer systems enables the computation of wave function-related quantities associated with response characteristics beyond spectra, as demonstrated below.

For instance, we consider the frequency-dependent optical conductivity, denoted as $\sigma_{\mu \nu}(\omega)$, which can be calculated using the Kubo-Greenwood formula~\cite{LOUIE20069}
\begin{equation}
\label{eq_optical}
\sigma_{\mu \nu} (\omega) = \frac{e^2 \hbar}{iS} \sum_{n,m} \frac{f(\epsilon_n)-f(\epsilon_m )}{\epsilon_n -\epsilon_m} \frac{\langle  \psi_n|v_\mu|  \psi_m\rangle \langle  \psi_m|v_\nu|  \psi_n\rangle}{\hbar \omega+\epsilon_n-\epsilon_m +i\eta}.
\end{equation}
Here the velocity operator $\bm{v}$ is defined as $\bm{v}=(i/\hbar)[\mathcal{H},\bm{r}]$, $f(\epsilon_n)$ is the Fermi-Dirac distribution function representing the occupation number of the eigenstate $| \psi_n\rangle$ labeled by index $n$ with corresponding energy $\epsilon_n$. The parameter $\eta$ means $0^{+}$, and $S$ is the total area of the sample. Usually, our focus is on the real part of the optical conductivity, denoted as $\mathrm{Re}[\sigma_{\mu \nu}(\omega)]$, which is related to the optical absorption intensity at photon energy $\hbar \omega$~\cite{LOUIE20069}.

To perform the decomposition, we first rewrite $\mathrm{Re}[\sigma_{\mu \nu}(\omega)]$ as a trace. This might seem unexpected initially, as the optical conductivity is derived from linear response theory and not directly linked to a single operator. However, it is possible to establish a connection between $\sigma_{\mu \nu}(\omega)$ and a trace by defining an optical conductivity operator $\mathcal{C}_{\mu \nu}$. We will demonstrate this procedure at zero temperature, where the Fermi-Dirac distribution function reduces to a Heaviside step function as $f(\epsilon_n)=\Theta(\epsilon_{\mathrm{F}}-\epsilon_n)$, and extending it to finite temperature is straightforward. First, we simplify the expression by utilizing Sokhotsky's formula~\cite{1976MoIzNQV}
\begin{equation}
\lim_{\eta \to 0^{+}} \frac{1}{x+i\eta} = \mathrm{p.v.} \frac{1}{x} - i\pi \delta(x),
\end{equation}
where $\mathrm{p.v.}$ donates the Cauchy principal value. Then we obtain
\begin{widetext}
\begin{eqnarray}
\sigma_{\mu \nu}(\omega)
&=& \frac{e^2 \hbar}{iS}
\sum_{n,m} \frac{f(\epsilon_n)-f(\epsilon_m )}{\epsilon_n -\epsilon_m} \langle  \psi_n|v_\mu|  \psi_m\rangle \langle  \psi_m|v_\nu|  \psi_n\rangle
\left[\mathrm{p.v.} \frac{1}{\hbar \omega+\epsilon_n-\epsilon_m } - i\pi \delta(\hbar \omega+\epsilon_n-\epsilon_m)\right],  \nonumber \\
&=& -\frac{\pi e^2 \hbar}{S}
\sum_{n,m} \frac{\Theta(\epsilon_{\mathrm{F}}-\epsilon_n)-\Theta(\epsilon_{\mathrm{F}}-\epsilon_m)}{\epsilon_n -\epsilon_m} \langle  \psi_n|v_\mu|  \psi_m\rangle \langle  \psi_m|v_\nu|  \psi_n\rangle
\delta(\hbar \omega+\epsilon_n-\epsilon_m)  \nonumber \\
& & - i \frac{e^2 \hbar}{S}
\mathrm{p.v.} \sum_{n,m} \frac{\Theta(\epsilon_{\mathrm{F}}-\epsilon_n)-\Theta(\epsilon_{\mathrm{F}}-\epsilon_m)}{\epsilon_n -\epsilon_m}
\frac{\langle  \psi_n|v_\mu|  \psi_m\rangle \langle  \psi_m|v_\nu|  \psi_n\rangle}{\hbar \omega+\epsilon_n-\epsilon_m }.
\end{eqnarray}
Under the constrain of the delta function $\delta(\hbar \omega+\epsilon_n-\epsilon_m)$ for given photon energy $\hbar \omega$, we have $\epsilon_n-\epsilon_m=-\hbar \omega < 0~\to~\epsilon_n<\epsilon_m$.
Therefore, the only nonzero contribution of $\Theta(\epsilon_{\mathrm{F}}-\epsilon_n)-\Theta(\epsilon_{\mathrm{F}}-\epsilon_m)$ is given by $\epsilon_n<\epsilon_{\mathrm{F}}<\epsilon_m$, resulting in $\Theta(\epsilon_{\mathrm{F}}-\epsilon_n)=1$ and $\Theta(\epsilon_{\mathrm{F}}-\epsilon_m)=0$. This indicates that the summation for $n$ is over all occupied states, while for $m$ is over all unoccupied states, simplifying $\mathrm{Re}[\sigma_{\mu \nu}(\omega)]$ as follows:
\begin{eqnarray}
\mathrm{Re}[\sigma_{\mu \nu}(\omega)]
&=& \frac{\pi e^2 }{\omega S} \sum_{n\in \mathrm{occ}} \sum_{ m \in \mathrm{unocc}} \langle  \psi_n|v_\mu|  \psi_m\rangle \langle  \psi_m|v_\nu|  \psi_n\rangle \delta(\hbar \omega+\epsilon_n-\epsilon_m ),\nonumber \\
&=& \frac{\pi e^2 }{\omega S} \sum_{n\in \mathrm{occ}} \langle  \psi_n|v_\mu \left(\sum_{ m \in \mathrm{unocc}} |\psi_m\rangle \langle \psi_m| \right)
\delta(\hbar \omega+\epsilon_n-\mathcal{H} ) v_\nu|  \psi_n\rangle ,\nonumber \\
&=& \frac{\pi e^2 }{\omega S} \sum_{n} \langle  \psi_n| \mathcal{P_{\mathrm{occ}}} v_\mu \mathcal{P_{\mathrm{unocc}}} \delta(\hbar \omega+\epsilon_n-\mathcal{H} ) v_\nu|  \psi_n\rangle ,\nonumber \\
&=& \frac{\pi e^2 }{\omega S} \mathrm{Tr}(\mathcal{C}_{\mu \nu}), \\
\mathcal{C}_{\mu \nu} &=&\sum_{n} |\psi_n\rangle\langle \psi_n | \langle \psi_n |\mathcal{P}_{\mathrm{occ}} v_\mu \mathcal{P}_{\mathrm{unocc}} \delta(\hbar \omega+\epsilon_n- \mathcal{H}) v_\nu|\psi_n\rangle.
\end{eqnarray}
Here, $\mathcal{P}_{\mathrm{occ}}$ and $\mathcal{P}_{\mathrm{unocc}}$ are the projection operators for the occupied and unoccupied subspaces, respectively:
\begin{equation}
\mathcal{P}_{\mathrm{occ}} = \sum_{n\in \mathrm{occ}}|\psi_n\rangle\langle \psi_n |~~\text{and}~~\mathcal{P}_{\mathrm{unocc}} = \sum_{ m \in \mathrm{unocc}} |\psi_m\rangle\langle \psi_m|.
\end{equation}
Notably, the matrix representation of the operator $\mathcal{C}_{\mu \nu}$ in the eigenstate basis $\{| \psi_n \rangle\}$ is diagonal:
\begin{equation}
\mathcal{C}_{\mu \nu}^{nm}=\delta_{n,m}\langle \psi_n |\mathcal{P}_{\mathrm{occ}} v_\mu \mathcal{P}_{\mathrm{unocc}} \delta(\hbar \omega+\epsilon_n- \mathcal{H}) v_\nu|\psi_n\rangle,
\end{equation}
substituting it to Eq.~(\ref{eq_expectation_diag}) leads to the precise decomposition for the optical absorption $\mathrm{Re}[\sigma_{\mu \nu}(\omega)]$ and the $\bm{k}$-resolved optical absorption of layer-$l$ $\mathrm{Re}[\sigma^l_{\mu \nu}(\bm{k}_l,\omega)]$:
\begin{eqnarray}
\label{eq_Re_total}
\mathrm{Re}[\sigma_{\mu \nu}(\omega)]
&=& \sum_{l} \sum_{\bm{k}_l \in \mathrm{BZ}_l} \mathrm{Re}[\sigma^l_{\mu \nu}(\bm{k}_l,\omega)], \\
\label{eq_ReS}
\mathrm{Re}[\sigma^l_{\mu \nu}(\bm{k}_l,\omega)]
&=& \frac{\pi e^2}{\omega S} \sum_{n\in \mathrm{occ}} w_{n}(\bm{k}_l) \sum_{ m \in \mathrm{unocc}} \langle  \psi_n|v_\mu|  \psi_m\rangle
\langle  \psi_m|v_\nu|  \psi_n\rangle
\delta(\hbar \omega+\epsilon_n-\epsilon_m ),
\end{eqnarray}
where $\langle \psi_n |v_\mu| \psi_m\rangle$ can be directly expanded by the newly derived velocity operator matrix elements under the composite Bloch basis $\coprod_{l}\{|\bm{k}_{l} \bm{t}_{l}\alpha\rangle\}$ for twisted multilayer systems (see Appendix~\ref{app:velocity} for details):
\begin{eqnarray}
\label{eq_app_v_k}
\langle\bm{k}^\prime_{l}\bm{t}^\prime_{l}\beta|\bm{v}|\bm{k}_{l}\bm{t}_{l}\alpha\rangle
&=&  \frac{1}{\hbar}
\nabla_{\bm{k}_l} \left( \langle\bm{k}_{l}\bm{t}^\prime_{l}\beta|\mathcal{H}|\bm{k}_{l}\bm{t}_{l}\alpha\rangle \right)
\delta_{\bm{k}_l,\bm{k}^\prime_{l}},\\
\label{eq_app_v_kp}
\langle\bm{k}^\prime_{l^\prime}\bm{t}^\prime_{l^\prime}\beta|\bm{v}|\bm{k}_{l}\bm{t}_{l}\alpha\rangle
&=& \sum_{\bm{G}_l,\bm{G}^\prime_{l^\prime}}\frac{1}{\hbar}
\left[\nabla_{\bm{q}}\mathcal{V}^{l^\prime,l}_{\beta,\alpha}(\bm{q})\right]_{\bm{q}=\bm{k}_l + \bm{G}_l}  e^{-i\bm{G}_l \cdot \bm{t}_{l} +i \bm{G}^\prime_{l^\prime} \cdot \bm{t}^\prime_{l^\prime}}
\delta_{\bm{k}_l+\bm{G}_l,\bm{k}^\prime_{l^\prime}+\bm{G}^\prime_{l^\prime}}.
\end{eqnarray}
\end{widetext}

\section{Local low-rank approximation of Hamiltonian}
\begin{figure*}
\includegraphics[width=2\columnwidth, keepaspectratio]{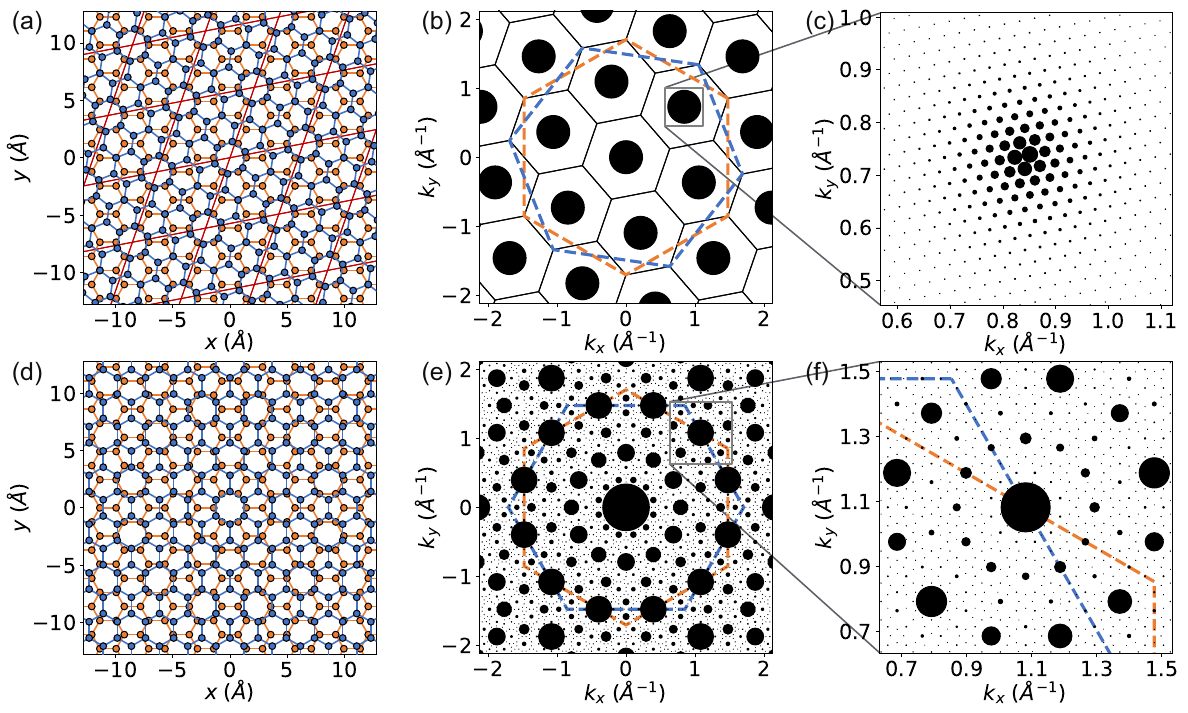}
\caption{\label{fig:GmG_both} Structure of interlayer coupling kernel $\{\bm{G}_1-\bm{G}^\prime_2\}$ in TBG. At a commensurate twisted angle of $\theta(1,1)\approx 21.79^\circ$, (a) a moir\'e superlattice and (b) a moir\'e reciprocal lattice is formed, with the moir\'e BZ outlined by black solid lines. The regions marked by blue and orange dotted lines represent $\mathrm{BZ}_1$ and $\mathrm{BZ}_2$, respectively. (c) A zoomed-in view of the gray area as a single point in (b) shows moir\'e lattice disruption with slight deviation $\Delta \theta\approx0.21^\circ$. At an incommensurate twisted angle of 30$^\circ$, (d) a dodecagonal quasicrystal is formed without moir\'e superlattice, and (e) the kernel $\{\bm{G}_1-\bm{G}^\prime_2\}$ densely populates whole reciprocal plane, but dominant one clearly stands out from others. (f) A zoomed-in view of the gray area in (d) shows a self-similar quasicrystalline structure. The size of the black dots indicates the relative amplitude of $|G_1|$.}
\end{figure*}

\subsection{General considerations}
Until now, our derivations have remained rigorous, without approximations or assumptions, using the eigenstates basis $\{| \psi_n \rangle\}$. However, obtaining these eigenstates directly is impractical due to the large matrix size and intricate interlayer coupling in the full Hamiltonian $\mathcal{H}$.
Therefore, we need an alternative set of states $\{| \tilde{\psi}_n \rangle\}$ and energies $\{\tilde{\epsilon}_n\}$ that effectively approximates $\{| \psi_n \rangle\}$ and $\{\epsilon_n\}$ while remaining numerically feasible. The weight factor $w_{nm}(\bm{k}_l)$ in our formalism provides a natural guide for this approximation. It implies that we can select distinct sets of states $\{| \tilde{\psi}_n(\bm{k}_l) \rangle\}$ and corresponding energies $\{\tilde{\epsilon}_n(\bm{k}_l)\}$ for different $\bm{k}_l$ points in different layers, accurately representing $\{|\psi_n\rangle\}$ and $\{\epsilon_n\}$ with significant weight in $|\bm{k}_l\rangle$. Consequently, we can exclude $\{| \psi_n \rangle\}$ states with negligible contributions at a specific $\bm{k}_l$ point, greatly simplifying numerical computations.
Such a problem is mathematically known as a generalized weighted low-rank approximation problem~\cite{srebro2003weighted}. To gain some insights, we'll first revisit the basic low-rank approximation problem and then explore the generalized weighted version, facilitated by the weight factor $w_{nm}(\bm{k}_l)$ in our formalism.

We revisit the basic low-rank approximation problem, aiming to minimize a cost function between matrix $\mathcal{H}$ and its low-rank approximation $\widetilde{\mathcal{H}}$ with a rank reduction constraint as $\min_{\mathrm{rank}(\widetilde{\mathcal{H}})\leq r}\|\mathcal{H}-\widetilde{\mathcal{H}}\|$. It can be analytically solved by employing singular value decomposition (SVD) on $\mathcal{H}$, which is widely recognized as the Eckart-Young-Mirsky theorem~\cite{Eckart1936, MIRSKY1960}. Suppose $\mathrm{rank}(\mathcal{H})=N$ and consider the SVD of $\mathcal{H}$ as $\mathcal{H}=\mathcal{U} \Sigma\mathcal{V}^{\top} \in \mathbb{C}^{N \times N}$, where $\Sigma= \operatorname{diag}\left(\sigma_1, \ldots, \sigma_{N}\right)$ is the diagonal matrix with singular values $\sigma_1 \geq \ldots \geq \sigma_{N}$. For a given rank $r \in\{1, \ldots, N-1\}$, we partition $\mathcal{U}$, $\Sigma$, and $\mathcal{V}$ as $\mathcal{U} =[\mathcal{U}_1,\mathcal{U}_2]$, $\Sigma=\mathrm{diag}[\Sigma_1 ,\Sigma_2]$, and $\mathcal{V}=[\mathcal{V}_1,\mathcal{V}_2]$, where $\mathcal{U}_1\in \mathbb{C}^{N \times r}$, $\Sigma_1\in \mathbb{R}^{r \times r}$, and $\mathcal{V}_1\in \mathbb{C}^{r \times N}$. The Eckart-Young-Mirsky theorem conclude that the rank-$r$ matrix $\widetilde{\mathcal{H}}_{\mathrm{EYK}}$, obtained through the truncated SVD as
\begin{equation}
\widetilde{\mathcal{H}}_{\mathrm{EYK}}=\mathcal{U}_1 \Sigma_1 \mathcal{V}_1^{\top},
\end{equation}
satisfies $\|\mathcal{H}-\widetilde{\mathcal{H}}_{\mathrm{EYK}}\|=\min _{\operatorname{rank}(\widetilde{\mathcal{H}}) \leq r}\|\mathcal{H}-\widetilde{\mathcal{H}}\|=\sqrt{\sigma_{r+1}^2+\cdots+\sigma_{N}^2}$.
Thus, $\widetilde{\mathcal{H}}_{\mathrm{EYK}}$ is the solution of the basic low-rank approximation problem.

The Eckart-Young-Mirsky theorem enables a global low-rank approximation $\widetilde{\mathcal{H}}_{\mathrm{EYK}}$ for matrix $\mathcal{H}$. However, its computational demands, particularly the need for a truncated SVD of the entire matrix, pose substantial challenges for large matrices like the Hamiltonian Eq.~(\ref{eq_full_H}). Our theoretical framework addresses this challenge by introducing a key weight factor $w_{nm}(\bm{k}_l)$, which transforms the global approximation problem into a local one, greatly simplifying numerical computations. Specifically, we aim for a local low-rank approximation, donated as $\widetilde{\mathcal{H}}(\bm{k}_l)$, for specific $\bm{k}_l$ points. This is a generalized weighted low-rank approximation problem in mathematics, for which there is no analytical solution via SVD. In order to solve it, we draw inspiration from CUR decomposition~\cite{CUR_pnas, HAMM20201088}, which approximates a matrix $\mathcal{H}$ by selecting a small number of its columns and rows based on their contribution to the matrix's overall structure:
\begin{equation}
\widetilde{\mathcal{H}}_{\mathrm{CUR}} = \mathcal{C} \mathcal{U}^{-1} \mathcal{R},
\end{equation}
with column submatrix $\mathcal{C}$, row submatrix $\mathcal{R}$ and their intersection $\mathcal{U}$. This technique is particularly valuable for large, sparse matrices, especially when SVD, even truncated SVD, is impractical due to computational demands. The essence of CUR decomposition lies in selecting columns and rows that preserve the essential information. While typically it's a global approximation, we can adapt it to a local low-rank approximation around specific $\bm{k}_l$ points without relying on SVD.

Now, we start to construct a local low-rank approximation $\widetilde{\mathcal{H}}(\bm{k}_l)$ around $\bm{k}_l$ following the essence of CUR decomposition. The initial step involves selecting relevant columns and rows that contain essential information about $|\bm{k}_l\rangle$ from the original Hamiltonian $\mathcal{H}$.  These selected elements serve as building blocks to construct $\widetilde{\mathcal{H}}(\bm{k}_l)$. To guide our selection, a criterion is necessary. In mathematics, identifying informative columns often relies on metrics like cosine similarity to quantify the correlation between each matrix column and the target vector as
\begin{equation}
C(\mathcal{H}_{:, i},\bm{k}_l) = \frac{\langle \mathcal{H}_{:, i} |\bm{k}_l\rangle}{ \langle \mathcal{H}_{:, i}| \mathcal{H}_{:, i}\rangle \langle \bm{k}_l| \bm{k}_l\rangle},
\end{equation}
where $|\mathcal{H}_{:, i}\rangle$ is the $i$-th column of $\mathcal{H}$. With $\langle \bm{k}_l| \bm{k}_l\rangle=1$ and nearly uniform $\langle \mathcal{H}_{:, i}| \mathcal{H}_{:, i}\rangle$ across columns, we select columns dominated by $\langle \mathcal{H}_{:, i} |\bm{k}_l\rangle$ to capture essential information about $|\bm{k}_l\rangle$. The selection of relevant rows is similar. Specifically,
for selecting columns responding to layer-$l^\prime$ within the context of Eq.~(\ref{eq_full_H}), we need to compare amplitudes of $\mathcal{V}_{\bm{k}_l,\bm{k}^\prime_{l^\prime}}$ for all $\bm{k}^\prime_{l^\prime} \in \mathrm{BZ}_{{l^\prime}}$ to identify the relevant $\bm{k}^\prime_{l^\prime}$. Thus, a throughout investigation of the interlayer coupling $\mathcal{V}_{l,l^\prime} $ is needed to finish such principal component analysis.

Recall the interlayer coupling matrix elements described by Eq.~(\ref{eq_interlayer}), for a given $\bm{k}_l$ in layer-$l$, all $|\bm{k}^\prime_{l^\prime}\rangle$ coupled to $|\bm{k}_l\rangle$ form a set $\{|\bm{k}^\prime_{l^\prime}\rangle|\bm{k}^\prime_{l^\prime}=\bm{k}_l+\bm{G}_l-\bm{G}^\prime_{l^\prime}\}$ in layer-$l^\prime$. This set's structure is solely determined by the kernel $\{\bm{G}_l-\bm{G}^\prime_{l^\prime}\}$ [see Fig.~(\ref{fig:GmG_both}) for intuition]. In the commensurate case, $\{\bm{G}_l-\bm{G}^\prime_{l^\prime}\}$ forms a discrete set of points in reciprocal space, corresponding to the moir\'e reciprocal superlattice. The mapping from $\{\bm{G}_l-\bm{G}^\prime_{l^\prime}\}$ to $\{|\bm{k}^\prime_{l^\prime}\rangle|\bm{k}^\prime_{l^\prime}=\bm{k}_l+\bm{G}_l-\bm{G}^\prime_{l^\prime}\}$ is many-to-one. However, in the incommensurate scenario, $\{\bm{G}_l-\bm{G}^\prime_{l^\prime}\}$ densely populates reciprocal space. Notably, each $\bm{k}^\prime_{l^\prime}$ point uniquely corresponds to specific combinations of $\bm{G}_l$ and $\bm{G}^\prime_{l^\prime}$, establishing a one-to-one mapping from $\{\bm{G}_l-\bm{G}^\prime_{l^\prime}\}$ to $\{|\bm{k}^\prime_{l^\prime}\rangle|\bm{k}^\prime_{l^\prime}=\bm{k}_l+\bm{G}_l-\bm{G}^\prime_{l^\prime}\}$. These distinct mappings, corresponding to different cutoff schemes, are discussed below in our unified theory.

\subsection{Incommensurate case}

We discuss the incommensurate case first, as it is more common. In contrast, the set of commensurate angles has a measure of zero, and even slight deviations can disrupt the commensurate structure, causing coupling between $|\bm{k}_l\rangle$ and $|\bm{k}^\prime_{l^\prime}\rangle$ across the entire $\mathrm{BZ}_{l^\prime}$, as depicted in Fig.~\ref{fig:GmG_both}(c).

\begin{figure}
\includegraphics[width=1\columnwidth, keepaspectratio]{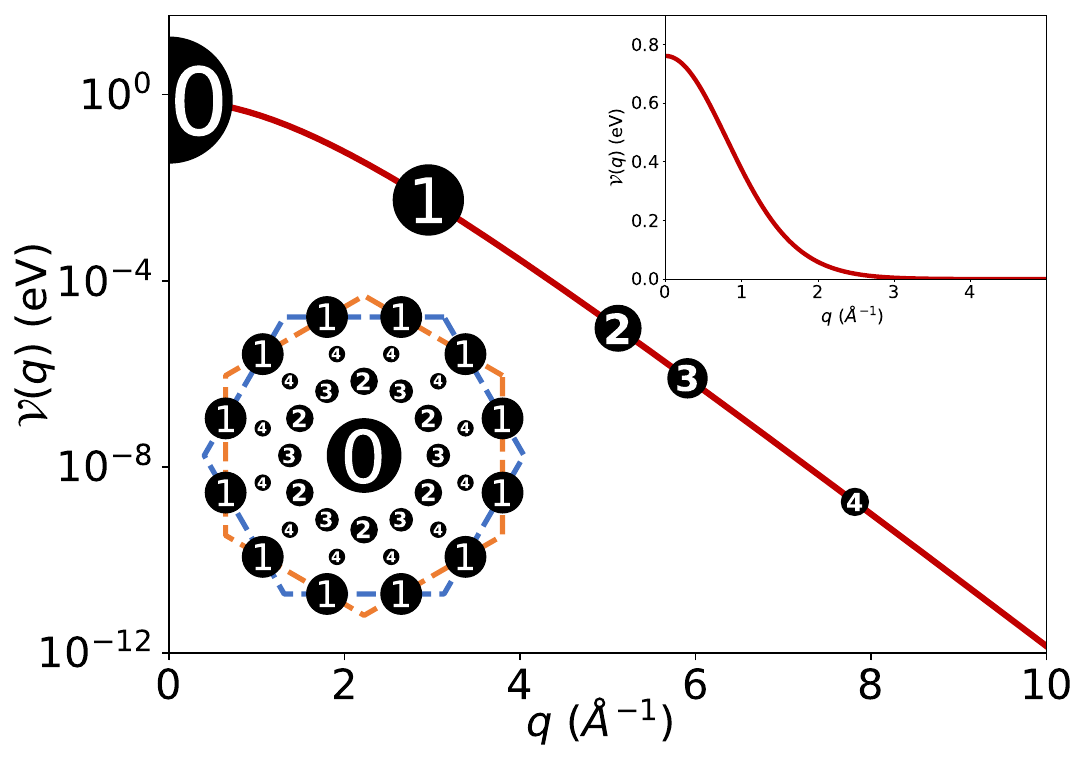}
\caption{\label{fig:30_decay} The interlayer coupling strength $\mathcal{V}(q)$ in the quasicrystalline 30$^\circ$ TBG as a function of $q=|\bm{G}_1|$ for $\bm{k}_1=\bm{0}$ presented on a logarithmic scale, with additional inserts depicting the data in different formats.}
\end{figure}

In the incommensurate scenario, although the kernel $\{\bm{G}_l-\bm{G}^\prime_{l^\prime}\}$ densely populates the entire reciprocal plane, dominant ones clearly stand out from others [see in Fig.~\ref{fig:GmG_both}(e)]. This highlights that for a given $\bm{k}_l$ in layer-$l$, although coupled $\bm{k}^\prime_{l^\prime}$ spans the entire $\mathrm{BZ}_{l^\prime}$, only a select few dominate. Further analysis reveals an exponential decay pattern in coupling strength. For example, in the case of quasicrystalline 30$^\circ$ TBG, dominant ones exhibit a dodecagonal pattern, as shown in Fig.~\ref{fig:30_decay}. Thus, we can introduce a cutoff based on coupling strength to select relevant $\mathcal{V}_{\bm{k}_l,\bm{k}^\prime_{l^\prime}}$, guiding our selection of a subset of columns and rows to effectively capture the essential information about $|\bm{k}_l\rangle$.

Specifically, for a wave vector $\bm{k}_l$ in layer-$l$, once the cutoff of coupling strength $G_{\mathrm{cut}}$ is given, we obtain $N_{\mathrm{cut}}$ dominant interlayer coupling elements, $\mathcal{V}_{\bm{k}_l,\bm{k}^j_{l^\prime}}~(j = 1, 2, \dots, N_{\mathrm{cut}})$, corresponding to layer-$l^\prime$, satisfying $|\bm{k}_l+\bm{G}_l|<G_{\mathrm{cut}}$. The resulting dominant relevant $\bm{k}^\prime_{l^\prime}$ in layer-$l^\prime$ coupled to $\bm{k}_l$ are donated as $\{ \bm{k}^\prime_{l^\prime}\}_{\bm{k}_l}$, sorting in descending order according to amplitudes of $\mathcal{V}_{\bm{k}_l,\bm{k}^\prime_{l^\prime}}$. In other words, the cutoff truncates the dense infinite kernel $\{\bm{G}_l-\bm{G}^\prime_{l^\prime}\}$ into a low-rank sparse finite set. And this selection results in $N_{\mathrm{cut}}$ relevant columns and rows for $\bm{k}_l$ corresponding to layer-$l^\prime$. Repeating this process for all layers provides all dominant columns and rows for $\bm{k}_l$ within the cutoff. Note that the choice of the number of columns and rows to truncate depends on the trade-off between dimensionality reduction and information preservation. Selecting more columns and rows increases accuracy but also incurs higher numerical costs.

\begin{figure*}
\includegraphics[width=2\columnwidth, keepaspectratio]{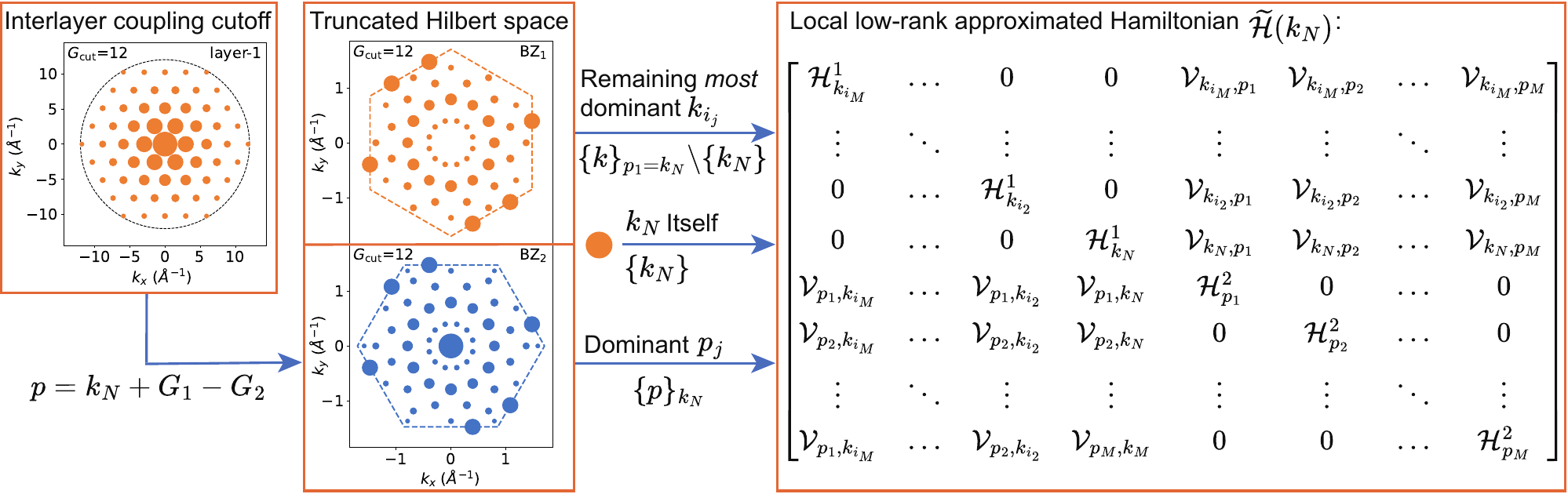}
\caption{\label{fig:30_matrix} Schematic illustration of constructing a local low-rank approximated Hamiltonian $\mathcal{H}(\bm{k}_N)$ for $\bm{k}_N=\bm{0}$ in layer-1 of quasicrystalline 30$^\circ$ TBG (see Appendix~\ref{app:TBG_matrix} for details). At $G_{\mathrm{cut}}=12$~\AA$^{-1}$, all $\bm{G}_1$ within the cutoff (black dotted circle) determine the dominant $\bm{p}_j=\bm{k}_N+\bm{G}_1-\bm{G}^\prime_2 \in \mathrm{BZ}_2$ (blue dotted hexagon) in layer-2. Together with the remaining \textit{most} dominant $\bm{k}_{i_j} \in \{ \bm{k}\}_{\bm{p}_1=\bm{k}_N} \backslash \{\bm{k}_N\}$ inside $\mathrm{BZ}_1$ (orange dotted hexagon) and $\{\bm{k}_N\}$ itself, it forms the truncated Hilbert space spanned by $\{ \bm{p}\}_{\bm{k}_N}\coprod \{ \bm{k}\}_{\bm{p}_1=\bm{k}_N}$, housing the low-rank approximated Hamiltonian $\mathcal{H}(\bm{k}_N)$.}
\end{figure*}

Having the selected columns and rows, the remaining challenge is constructing the approximated Hamiltonian $\widetilde{H}(\bm{k}_l)$. One direct approach is to take their intersection, denoted as $\widetilde{\mathcal{H}}_0(\bm{k}_l)$. For a specific $\bm{k}_l=\bm{k}^1_l$, the selection results in a truncated Hilbert space spanned by $\{\bm{k}^1_l\}\coprod_{l^\prime \neq l} \{ \bm{k}^\prime_{l^\prime}\}_{\bm{k}^1_l}$. However, such a method may be considered overly simplistic (see Appendix~\ref{app:TBG_matrix}).
In this work, we propose a method to construct the local approximated Hamiltonian $\widetilde{\mathcal{H}}(\bm{k}_l)$ beyond $\widetilde{H}_0(\bm{k}_l)$, by including the remaining \textit{most} dominant $\bm{k}_l( \neq \bm{k}^1_l)$ coupled to all $\bm{k}^\prime_{l^\prime} \in \coprod_{l^\prime \neq l} \{ \bm{k}^\prime_{l^\prime}\}_{\bm{k}^1_l}$, which is usually $\bm{k}_l=\bm{k}^\prime_{l^\prime}$. This consideration is based on the fact that the leading order coupling is generally given by $\bm{G}_l=\bm{G}^\prime_{l^\prime}=\bm{0}$, indicating the most dominate $|\bm{k}^\prime_{l^\prime}\rangle$ coupled to $|\bm{k}_l\rangle$ is located at $\bm{k}^\prime_{l^\prime}=\bm{k}_l$. Thus, the most dominant $\bm{k}_l$ associated with $\bm{k}^\prime_{l^\prime}$ is usually  $\bm{k}_l=\bm{k}^\prime_{l^\prime}$ itself. Thus, we will work on an enlarging truncated Hilbert space spanned by
\begin{equation}
\label{eq_truncated_Hilbert}
\{\bm{k}^1_l\}
\coprod_{l^\prime \neq l}
\left(
\{ \bm{k}^\prime_{l^\prime}\}_{\bm{k}^1_l}
\coprod
\{ \bm{k}_{l}\}_{\bm{k}^\prime_{l^\prime}=\bm{k}^1_l}
\backslash
\{\bm{k}^1_l\}
\right).
\end{equation}
Mathematically, our local approximated Hamiltonian corresponds to the Rayleigh-Ritz methods for eigenvalue problems:
\begin{equation}
\widetilde{\mathcal{H}}(\bm{k}_l)= \mathcal{S}^\dagger(\bm{k}_l)\mathcal{H}\mathcal{S}(\bm{k}_l),
\end{equation}
where the transform matrix $\mathcal{S}(\bm{k}_l)$ is chosen as the projection matrix from the full Hilbert space to the truncated subspace spanned by Eq.~(\ref{eq_truncated_Hilbert})
as discussed in Appendix~\ref{app:Ritz}.

In summary, our approach for constructing the local low-rank approximated Hamiltonian involves four key steps:
\begin{itemize}
    \item Investigating the interlayer coupling environment $\mathcal{V}_{l,l^\prime}$ for a given $\bm{k}_l=\bm{k}^1_l$, applying a cutoff $G_{\mathrm{cut}}$ to distinguish the $N_{\mathrm{cut}}$ dominant relevant coupling terms $\mathcal{V}_{\bm{k}^1_l,\bm{k}^j_{l^\prime}}$ ($\bm{k}^j_{l^\prime} \in \{\bm{k}^\prime_{l^\prime}\}_{\bm{k}^1_l}$) from the irrelevant ones.
    \item Repeating this process for all layer-$l^\prime~(\neq l)$, and selecting all relevant columns and rows, which offer the building blocks for further construction.
    \item Further examining the inverse interlayer coupling environment $\mathcal{V}_{l^\prime,l}$ for each $\bm{k}^j_{l^\prime} \in \{\bm{k}^\prime_{l^\prime}\}_{\bm{k}_l}$,
    choosing the \textit{most} dominant $\bm{k}_l( \neq \bm{k}^1_l)$ (usually $\bm{k}_l=\bm{k}^j_{l^\prime}$) and neglecting other less relevant ones.
    \item The final local approximated Hamiltonian $\widetilde{\mathcal{H}}(\bm{k}_l)$ is given by the submatrix around $\bm{k}_l$, i.e., $\widetilde{\mathcal{H}}(\bm{k}_l)=\mathcal{S}^\dagger(\bm{k}_l)\mathcal{H}\mathcal{S}(\bm{k}_l)$ with  $\mathcal{S}(\bm{k}_l)$ the projection matrix of the truncated Hilbert space.
\end{itemize}
The abstract discussion above is clarified in an incommensurate bilayer example (see Fig.~\ref{fig:30_matrix} and Appendix~\ref{app:TBG_matrix}), with straightforward generalization to the multilayer case.

\subsection{Commensurate case}

\begin{figure}
\includegraphics[width=1\columnwidth, keepaspectratio]{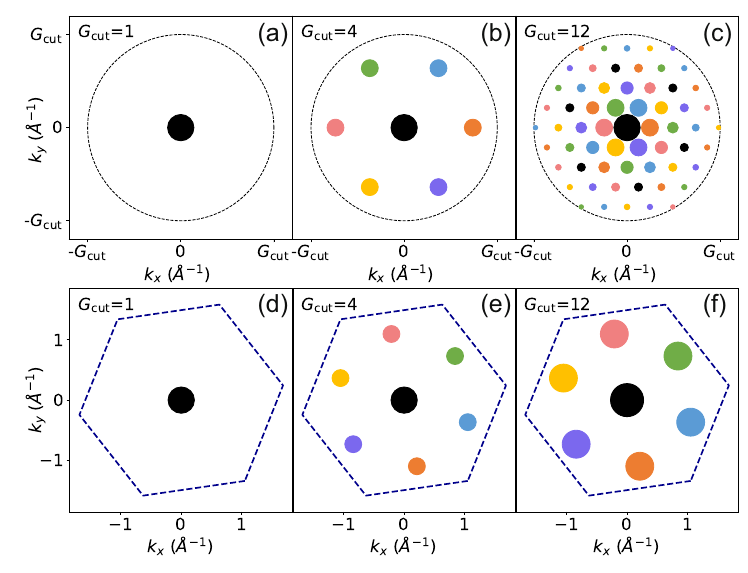}
\caption{\label{fig:21.79_converge} Schematic illustration of convergence of approximated Hamiltonian $\mathcal{H}(\bm{k}_N)$ for $\bm{k}_N=\bm{0}$ in layer-1 of 21.79$^\circ$ TBG with moir\'e lattice. All $\bm{G}_1$ within the cutoff (black dotted circle) for (a) $G_{\mathrm{cut}}=1$~\AA$^{-1}$, (b) $G_{\mathrm{cut}}=4$~\AA$^{-1}$, and (c) $G_{\mathrm{cut}}=12$~\AA$^{-1}$ are represented color-coded dots, corresponding to the folded $\{\bm{p}\}_{\bm{k}=\bm{0}}$ in $\mathrm{BZ}_2$ (blue dotted hexagon) with (d) $N_{\mathrm{cut}}=1$, (e) $N_{\mathrm{cut}}=7$, and (f) $N_{\mathrm{cut}}=61$, respectively. The size of the dots indicates the relative amplitude, and we omit the phase factor $\exp(-i\bm{G}_1 \cdot \bm{t}_1 +i \bm{G}^\prime_2 \cdot \bm{t}^\prime_2) $ in Eq.~(\ref{eq_interlayer}) for better illustration.
}
\end{figure}

Turning to the commensurate case, the first step in constructing the local low-rank approximated Hamiltonian $\widetilde{\mathcal{H}}(\bm{k}_l)$ also involves examining the interlayer coupling environment $\mathcal{V}_{l,l^\prime}$ around $\bm{k}_l$. Unlike the incommensurate case, the mapping from $\{\bm{G}_l-\bm{G}^\prime_{l^\prime}\}$ to $\{|\bm{k}^\prime_{l^\prime}\rangle|\bm{k}^\prime_{l^\prime}=\bm{k}_l+\bm{G}_l-\bm{G}^\prime_{l^\prime}\}$ is many-to-one in the commensurate case due to the moir\'e periodicity. Consequently, only a finite number of $\bm{k}^\prime_{l^\prime}$ points in $\mathrm{BZ}_{l^\prime}$ of layer-${l^\prime}$ will couple to $\bm{k}_l$ in layer-$l$. The rank of $\{\bm{k}^\prime_{l^\prime}\}_{\bm{k}_l}$ is simply $N_{l^\prime}/N_{\mathrm{sc}}$, indicating the number of unit cells from layer-$l^\prime$ the moir\'e supercell contains. Nevertheless, Eq.~(\ref{eq_interlayer}) is valid in both commensurate and incommensurate scenarios. Thus, we can determine the interlayer coupling environment in the commensurate case through straightforward calculation, similar to the incommensurate case.

\begin{figure*}[ht]
\includegraphics[width=2\columnwidth, keepaspectratio]{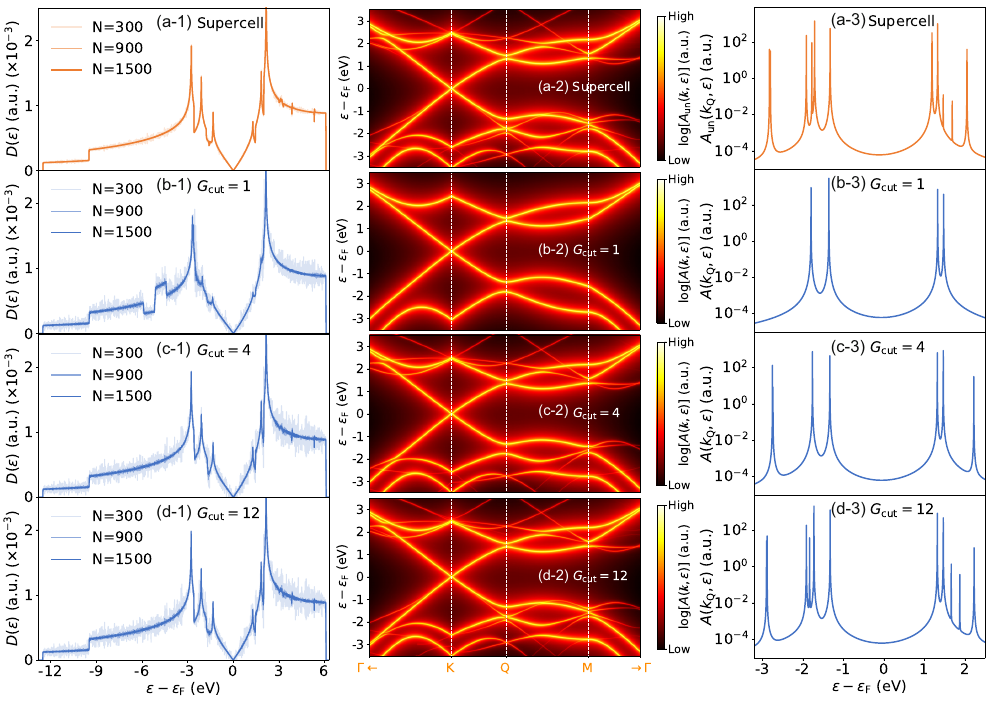}
\caption{\label{fig:21.79_DOS} Comparing spectral results for 21.79$^\circ$ TBG with moir\'e superlattice obtained by our theory and supercell unfolding calculations. (1) The normalized density of states $D(\epsilon)$, (2) the quasi-band structure, and (3) the spectra at the $\mathrm{M}$ point from (a) supercell unfolding calculations and from our theory with cutoffs (b) $G_{\mathrm{cut}}=1$~\AA$^{-1}$, (c) $G_{\mathrm{cut}}=4$~\AA$^{-1}$, and (d) $G_{\mathrm{cut}}=12$~\AA$^{-1}$. $D(\epsilon)$ is depicted with color-coded lines, corresponding to different N$\times$N $\bm{k}$-mesh. The spectra are presented on a logarithmic scale, and the $\bm{k}$-path in $\mathrm{BZ}_1$ used here is consistent with that in Fig.~\ref{fig:21.79_unfold} at Appendix~\ref{app:numerical_TBG}.}
\end{figure*}

After determining the interlayer coupling environment, the next step is to apply a cutoff.
In the incommensurate case, where $\{\bm{G}_l-\bm{G}^\prime_{l^\prime}\}$ densely populates reciprocal space, the rationale for the cutoff is clear. We aim to approximate the infinite set $\{ \bm{k}^\prime_{l^\prime} = \bm{k}_l + \bm{G}_l - \bm{G}^\prime_{l^\prime} \}$ with a finite set $\{ \bm{k}^\prime_{l^\prime}\}_{\bm{k}_l}$ of rank $N_{\mathrm{cut}}$ while maintaining high accuracy. Therefore, we restrict consideration to $\bm{k}^\prime_{l^\prime}$ points satisfying $|\bm{k}_l+\bm{G}_l(\bm{k}^\prime_{l^\prime})| \leq G_{\mathrm{cut}}$, where the dominant coupling strength $\mathcal{V}(\bm{k}_l+\bm{G}_l(\bm{k}^\prime_{l^\prime}))$ guides the truncation. Here, $\bm{G}_l(\bm{k}^\prime_{l^\prime})$ is uniquely determined by Eq.~(\ref{eq_Umklapp}).
In contrast, in the commensurate case, where $|\{\bm{k}^\prime_{l^\prime}=\bm{k}_l+\bm{G}_l-\bm{G}^\prime_{l^\prime}\} | = N_{l^\prime}/N_{\mathrm{sc}}$ (note that $\bm{k}^\prime_{l^\prime} \in \mathrm{BZ}_{l^\prime}$), a finite set already exists. Then the question arises: what does applying a cutoff mean in this context, what should be cut off, and how to define $N_{\mathrm{cut}}$? To address this, we examine Eq.~(\ref{eq_interlayer}), which becomes an infinite series summation in the commensurate case.

Consider a specific $\bm{k}_l$, all $\bm{k}_l+\bm{G}_l$ unfolding into $\bm{k}^\prime_{l^\prime}=\bm{k}_l+\bm{G}_l-\bm{G}^\prime_{l^\prime}$ naturally defines the infinite set of $ \bm{G}_l$. The cutoff is applied to this infinite set, approximating the infinite series summation with partial sums. The truncation is determined by $|\bm{k}_l+\bm{G}_l| \leq G_{\mathrm{cut}}$ for all possible $\bm{k}^\prime_{l^\prime}$. In other words, in the commensurate case, $N_{\mathrm{cut}}$ represents the total number of $\bm{G}_l$ used for the partial sums. Note that increasing $G_{\mathrm{cut}}$ results in a power-law growth of $N_{\mathrm{cut}}$, while the interlayer coupling strength decays exponentially with $G_{\mathrm{cut}}$ (see Fig.~\ref{fig:30_decay}). Consequently, a small $G_{\mathrm{cut}}$ is sufficient for a well-converged approximated Hamiltonian, yielding a manageable number of $N_{\mathrm{cut}}$, as shown in Fig.~\ref{fig:21.79_converge}. Then the local low-rank Hamiltonian is constructed based on the resulting truncated Hilbert space, following the standard procedure.
Therefore, our theory offers a unified description of the Hilbert space truncation in both commensurate and incommensurate scenarios, necessitating distinct interpretations as discussed above:
\begin{widetext}
\begin{eqnarray}
\langle\bm{k}^\prime_{l^\prime}\bm{t}^\prime_{l^\prime}\beta|\mathcal{V}_{l,l^\prime}|\bm{k}_{l}\bm{t}_{l}\alpha\rangle
= \sum_{\bm{G}_l,\bm{G}^\prime_{l^\prime}}
\mathcal{V}^{l^\prime,l}_{\beta,\alpha}(\bm{k}_l + \bm{G}_l)
e^{-i\bm{G}_l \cdot \bm{t}_{l} +i \bm{G}^\prime_{l^\prime} \cdot \bm{t}^\prime_{l^\prime}}
\delta_{\bm{k}_l+\bm{G}_l,\bm{k}^\prime_{l^\prime}+\bm{G}^\prime_{l^\prime}} \Theta( G_{\mathrm{cut}}-|\bm{k}_l+\bm{G}_l|).
\end{eqnarray}
\end{widetext}

\begin{figure*}
\includegraphics[width=2\columnwidth, keepaspectratio]{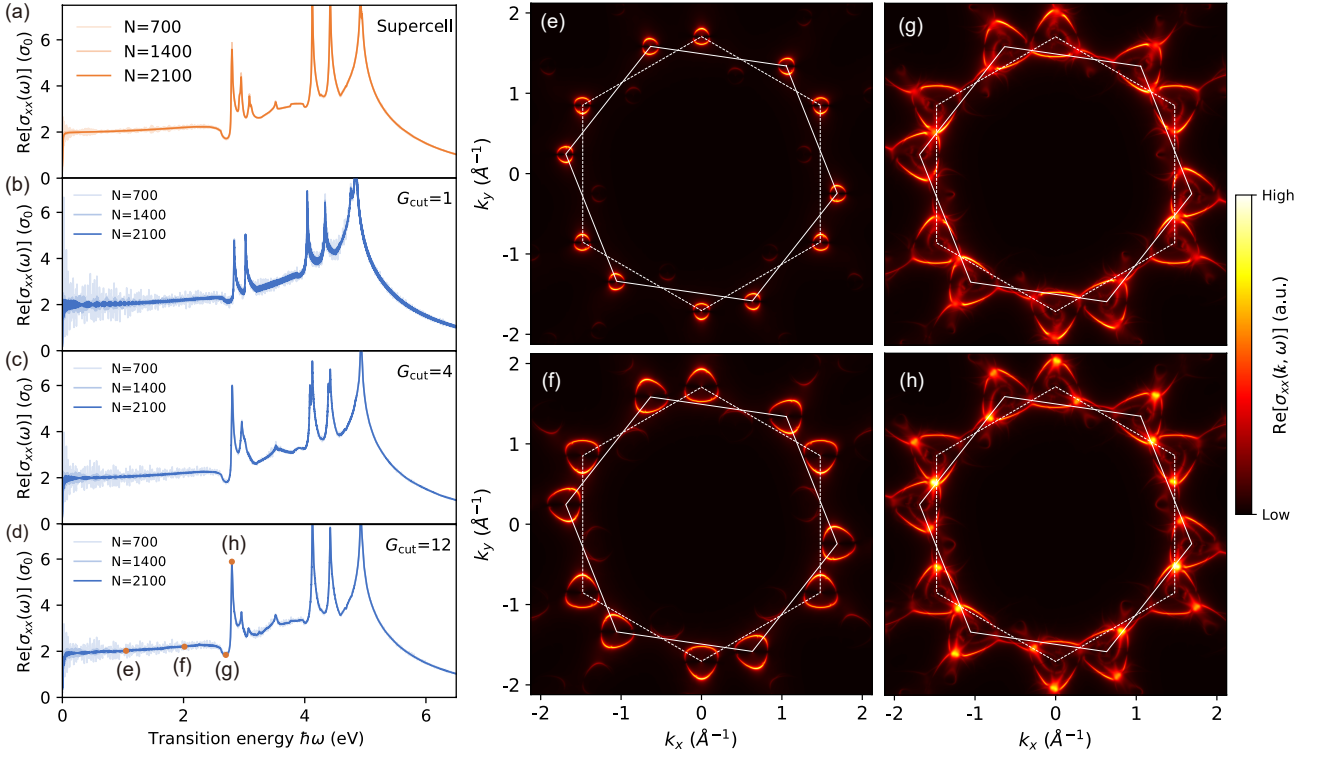}
\caption{\label{fig:21.79_ReS} Optical conductivity for 21.79$^\circ$ TBG with moir\'e superlattice obtained from (a) supercell unfolding calculations and from our theory with cutoffs (b) $G_{\mathrm{cut}}=1$~\AA$^{-1}$, (c) $G_{\mathrm{cut}}=4$~\AA$^{-1}$, and (d) $G_{\mathrm{cut}}=12$~\AA$^{-1}$. $\mathrm{Re}[\sigma_{xx}(\omega)]$ is depicted with color-coded lines, corresponding to different N$\times$N $\bm{k}$-mesh. And profiles of $\mathrm{Re}[\sigma_{xx}(\omega)]$ at various transition energies [highlighted by orange dots in (d)] for 21.79$^\circ$ TBG obtained from our theory with cutoffs $G_{\mathrm{cut}}=12$~\AA$^{-1}$: (e) $\hbar \omega=1.0$ eV, (f) $\hbar \omega=2.0$ eV, (g) $\hbar \omega=2.7$ eV, and (h) $\hbar \omega=2.8$ eV. The white dotted and solid lines in (e) represent boundaries of $\mathrm{BZ}_1$ and $\mathrm{BZ}_2$, respectively.}
\end{figure*}

\section{Numerical studies}
\subsection{Commensurate case}

In this section, we validate our theory in the commensurate case by comparing it to exact supercell calculations, using a 21.79$^\circ$ TBG with a moir\'e superlattice as an example (see Appendix~\ref{app:numerical_TBG} for details).

We first demonstrate our theory's ability to extract spectral information.  We start by examining the density of states $D(\epsilon)$. In Fig.~\ref{fig:21.79_DOS}(a-1), the density of states from supercell calculations is well-converged and exhibits 6 Van Hove singularity (VHS) peaks~\cite{PhysRev.89.1189}. Our local approximated theory, even when considering only leading-order contributions at $G_{\mathrm{cut}}=1$~\AA$^{-1}$ [see Fig.~\ref{fig:21.79_DOS}(b-1)], captures the overall trend of $D(\epsilon)$ effectively. When we increase the cutoff $G_{\mathrm{cut}}$ to $4$~\AA$^{-1}$, all VHS peaks are accurately reproduced, as depicted in Fig.~\ref{fig:21.79_DOS}(c-1). Furthermore, at $G_{\mathrm{cut}}=12$~\AA$^{-1}$ [see Fig.~\ref{fig:21.79_DOS}(d-1)], more detailed features emerge, highlighting the high accuracy of our method with a reasonable increase in computational cost.

We proceed to analyze the quasi-band structure to understand the impact of interlayer coupling on the electronic structure of 21.79$^\circ$ TBG. For better visualization, the spectral functions are presented on a logarithmic scale, which enhances the identification and comparison of fine details arising from the interlayer coupling. As shown in Fig.~\ref{fig:21.79_DOS}(a-2), the band structure retains a Dirac cone at the $\mathrm{K}$ point, with numerous small gaps introduced due to the interlayer coupling. Turning to our theory, the leading-order approximation provides a minimal four-band model at $G_{\mathrm{cut}}=1$~\AA$^{-1}$, capturing the primary gap at the $\mathrm{Q}$ point [see Fig.~\ref{fig:21.79_DOS}(b-2)]. Increasing the cutoff to $G_{\mathrm{cut}}=4$~\AA$^{-1}$ reveals detailed subbands in Fig.~\ref{fig:21.79_DOS}(c-2). Similarly, more subband gaps open up at $G_{\mathrm{cut}}=12$~\AA$^{-1}$ [see Fig.~\ref{fig:21.79_DOS}(d-2)], which closely resembles the benchmark results, underscoring the high accuracy of our theory.

To further illustrate the increasing accuracy with the cutoff, we examine the evolution of the single $\bm{k}$ point spectrum. We focus on the $\mathrm{Q}$ point, located at the intersection of $\mathrm{BZ}_1$ and $\mathrm{BZ}_2$, where strong interlayer hybridization occurs, as depicted in Fig.~\ref{fig:21.79_DOS}(a-3). As mentioned above, the four-band model at $G_{\mathrm{cut}}=1$~\AA$^{-1}$ in Fig.~\ref{fig:21.79_DOS}(b-3) considers only $\bm{k}_{\mathrm{Q}}$ and $\bm{p}=\bm{k}_{\mathrm{Q}}$ two wave vectors, representing the leading-order approximation. When we increase the cutoff to $G_{\mathrm{cut}}=4$~\AA$^{-1}$, additional 12 wave vectors are introduced to provide corrections and achieve higher accuracy, as shown in Fig.~\ref{fig:21.79_DOS}(c-3). Finally, with the cutoff $G_{\mathrm{cut}}$ set to $12$~\AA$^{-1}$, our theory exhibits very high accuracy and nearly captures all the details of the benchmark results, as demonstrated in Fig.~\ref{fig:21.79_DOS}(d-3). All these results confirm the validity of our theory in describing spectral properties in the case of commensurate 21.79$^\circ$ TBG with a moir\'e superlattice.

Next, we illustrate our theory's capability to extract information from eigenvectors beyond energy eigenvalues, enabling us to compute physical observables beyond spectral quantities. As an example, we compute the optical conductivity, specifically $\sigma_{xx}(\omega)$, where its real part directly correlates with optical absorption. Our formula for $\mathrm{Re}[\sigma_{xx}(\omega)]$ is directly given by Eq.~(\ref{eq_ReS}), where the velocity matrix elements in the twisted multilayer system are given by Eq.~(\ref{eq_app_v_k}) and Eq.~(\ref{eq_app_v_kp}) (see Appendix~\ref{app:velocity} for detailed derivations).
$\mathrm{Re}[\sigma_{xx}(\omega)]$ as a bulk property is determined by the wave functions and can be utilized to assess our theory's capability to compute physical observables beyond spectra. In Fig.~\ref{fig:21.79_ReS}(a), the optical absorption obtained from supercell calculation exhibits a plateau at $2\sigma_0$, with $\sigma_0=e^2/4\hbar$ representing the optical conductivity of a single-layer graphene under the linear band regime~\cite{PhysRevB.87.205404}. Additionally, there are several absorption peaks around 3 eV and in the 4-5 eV range in the optical conductivity. Our local approximated theory, considering leading-order contributions at $G_{\mathrm{cut}}=1$~\AA$^{-1}$ as shown in Fig.~\ref{fig:21.79_ReS}(b), effectively captures the general behavior of $\mathrm{Re}[\sigma_{xx}(\omega)]$. With higher cutoff values, our theory consistently provides highly accurate results [see Fig.~\ref{fig:21.79_ReS}(c)(d)], reaffirming its applicability for calculating response quantities involving eigenstates for twisted multilayer systems.

Importantly, our theory provides a comprehensive description of the wave function, enabling multiscale-resolved calculations for wave function-related physical observables. For instance, we compute the $\bm{k}$-resolved $\mathrm{Re}[\sigma_{xx}(\omega)]$ to gain some insights into the optical absorption in the 21.79$^\circ$ TBG. In Fig.~\ref{fig:21.79_ReS}, we present several $\mathrm{Re}[\sigma_{xx}(\bm{k},\omega)]$ profiles at different transition energies (marked in Fig.~\ref{fig:21.79_ReS}(d) as orange dots), highlighting the $\bm{k}$ points' contribution to optical transitions (see Fig.~\ref{fig:more_ReS_cuts} in Appendix~\ref{app:numerical_30} for more $\mathrm{Re}[\sigma_{xx}(\omega)]$ profiles at different transition energies). At $\hbar \omega=1$ eV and $\hbar \omega=2$ eV, Fig.~\ref{fig:21.79_ReS}(e)(f) clearly shows that the optical absorption plateau at $2\sigma_0$ originates from the linear band regime around the Dirac cones at $\mathrm{K}_{1,2}$ and $\mathrm{K}^\prime_{1,2}$ of layer-1 and layer-2, which is exactly as expected. Additionally, the increase in the optical absorption between $\hbar \omega=2.7$ eV and $\hbar \omega=2.8$ eV is explained by the $\mathrm{Re}[\sigma_{xx}(\bm{k},\omega)]$ profiles in Fig.~\ref{fig:21.79_ReS}(g)(h) (see Supplemental Material (SM) for more profiles at different transition energies~\footnote{\label{fn}See Supplemental Material file at http://link.aps.org/supplemental/xxx for animations illustrating the $\mathrm{Re}[\sigma_{xx}(\bm{k},\omega)]$ evolution in commensurate moir\'e 21.79$^\circ$ TBG and the density of states and Fermi surfaces evolution in incommensurate quasicrystalline 30$^\circ$ TBG.}). This rise is mainly attributed to bright spots around the $\mathrm{Q}$ point, corresponding to the transition from the second lowest valence quasi-band to the highest conduction quasi-band [see Fig.~\ref{fig:21.79_DOS}(d-2)]. As mentioned before, the physics around the $\mathrm{Q}$ point is dominated by the interlayer coupling. Therefore, our theory attributes this increase to interlayer coupling-induced optical absorption. This physical origin is not apparent in supercell calculations, where similar peaks are attributed to spots near $\mathrm{M}_{\mathrm{sc}}$ of the supercell BZ~\cite{PhysRevB.87.205404}. This example illustrates that our theory enables a more in-depth analysis of wave function-related physical observables, providing additional insights into the electronic structure of twisted multilayer systems.

In summary, the calculations for spectral and wave function-related quantities validate our theory in the commensurate case. Given the general nature of our theory, its validity in incommensurate situations beyond the moir\'e superlattice is demonstrated below.

\begin{figure*}
\includegraphics[width=2\columnwidth, keepaspectratio]{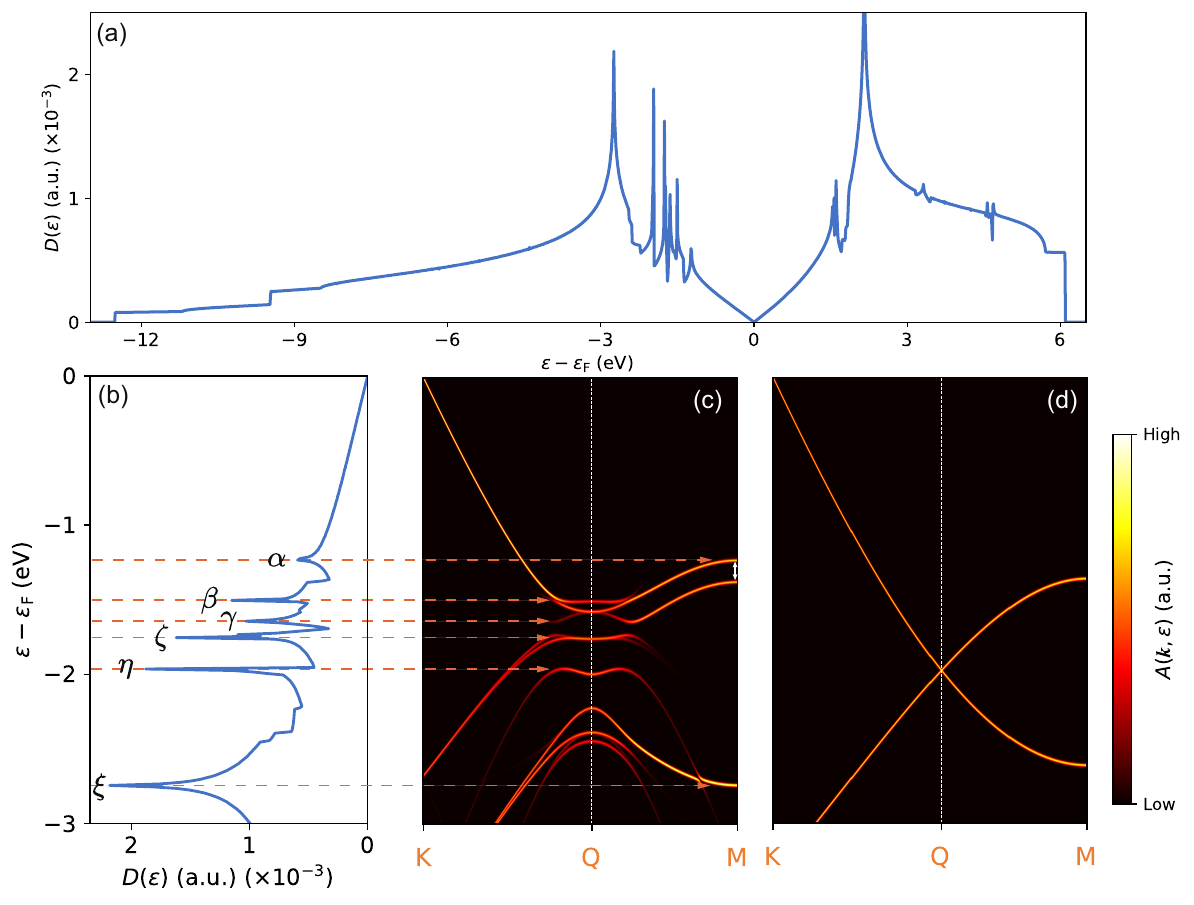}
\caption{\label{fig:30_DOS_band} Density of states $D(\epsilon)$ of incommensurate quasicrystalline 30$^\circ$ TBG in (a) complete energy range and in (b) 0 to 3 eV below the Fermi energy. In (b), VHS peaks are labeled by the corresponding Greek alphabet. Quasi-band structure around the Q point (c) with and (d) without interlayer coupling. The cutoff parameter utilized is $G_{\mathrm{cut}}=12$ \AA$^{-1}$. Dotted orange arrows relate VHS peaks and associated quasi-bands. A white arrow highlights the interlayer coupling-induced band gap at the M point.
}
\end{figure*}

\begin{figure*}
\includegraphics[width=2\columnwidth, keepaspectratio]{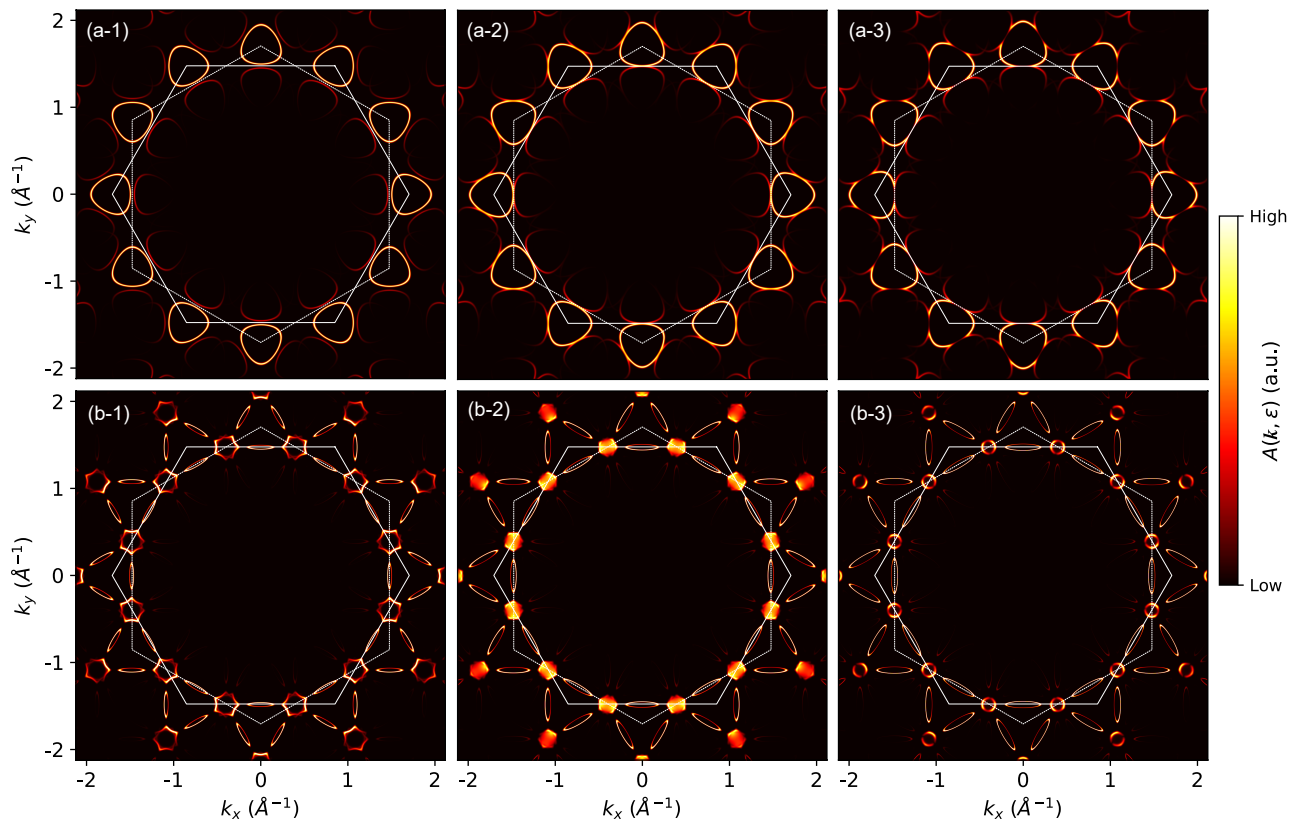}
\caption{\label{fig:30_Ecuts} Evolution of Fermi surfaces around VHS peaks labeled by (a) $\alpha$ and (b) $\beta$ in Fig.~\ref{fig:30_DOS_band}(b). For each case, the Fermi surfaces at (1) energies slightly below the VHS peak, (2) at the VHS peak, and (3) slightly above the VHS peak are presented: (a-1) $\epsilon=-1.15$ eV, (a-2) $\epsilon=-1.24$ eV, (a-3) $\epsilon=-1.28$ eV, (b-1) $\epsilon=-1.46$ eV, (b-2) $\epsilon=-1.50$ eV, and (b-3) $\epsilon=-1.53$ eV. The cutoff parameter utilized is $G_{\mathrm{cut}}=12$ \AA$^{-1}$. White dotted and solid lines represent boundaries of $\mathrm{BZ}_1$ and $\mathrm{BZ}_2$, respectively.}
\end{figure*}

\begin{figure*}
\includegraphics[width=2\columnwidth, keepaspectratio]{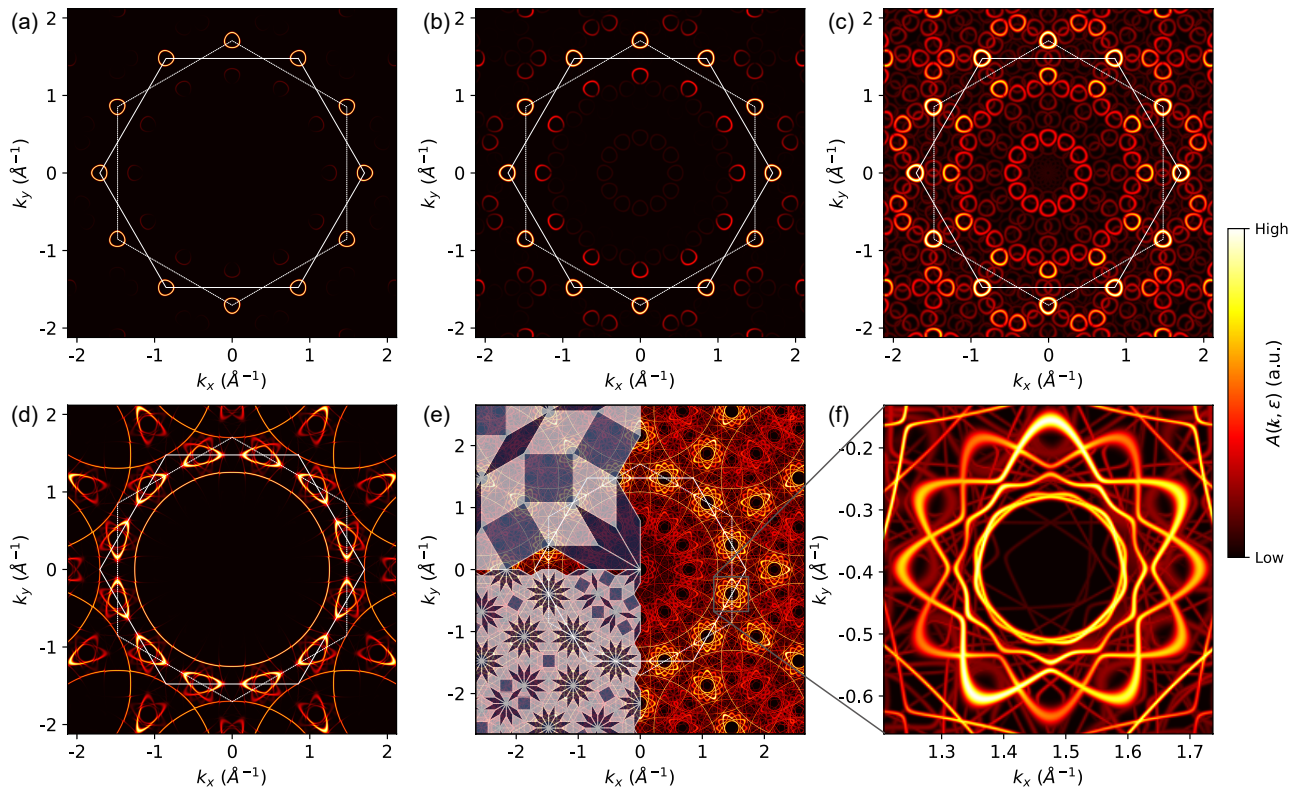}
\caption{\label{fig:30_FS} (a) The mirrored Dirac cone at $\epsilon=-0.5$ eV, (b) with $x^{1/5}$ scaling, and (c) with $x^{1/15}$ scaling. (d) The spiral Fermi surface at $\epsilon=-2.65$ eV, (e) with $x^{1/15}$ scaling, and (f) zoom-in around the Q point. The cutoff parameter utilized is $G_{\mathrm{cut}}=12$ \AA$^{-1}$. White dotted and solid lines represent boundaries of $\mathrm{BZ}_1$ and $\mathrm{BZ}_2$, respectively.}
\end{figure*}

\begin{figure*}
\includegraphics[width=2\columnwidth, keepaspectratio]{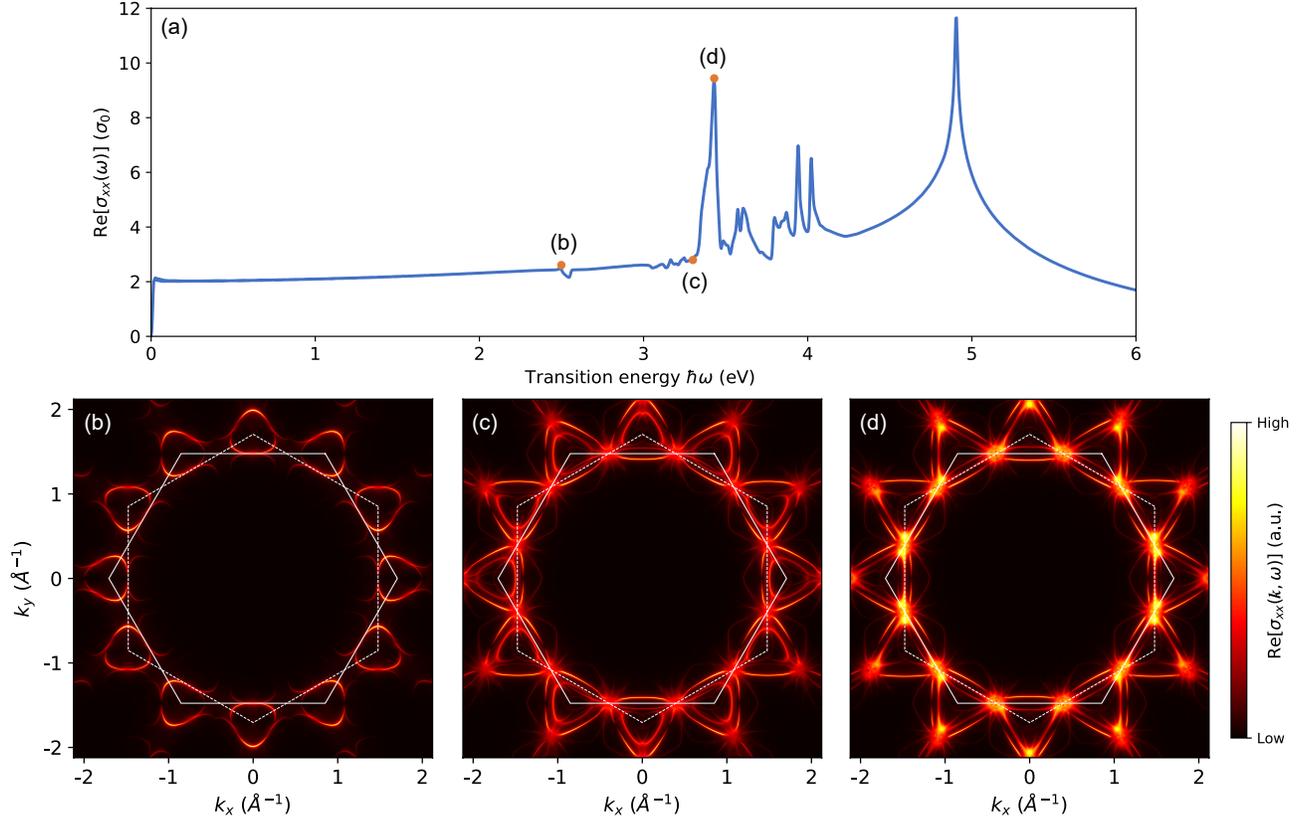}
\caption{\label{fig:30_ReS} (a) Optical conductivity $\mathrm{Re}[\sigma_{xx}(\omega)]$ for 30$^\circ$ TBG obtained from our theory with cutoffs $G_{\mathrm{cut}}=10$~\AA$^{-1}$. And profiles of $\mathrm{Re}[\sigma_{xx}(\omega)]$ at transition energies of (b) $\hbar \omega=2.49$ eV, (c) $\hbar \omega=3.30$ eV, and (d) $\hbar \omega=3.34$ eV [highlighted by orange dots in (a)]. The white dotted and solid lines represent boundaries of $\mathrm{BZ}_1$ and $\mathrm{BZ}_2$, respectively.}
\end{figure*}

\subsection{Incommensurate case}

In this section, we demonstrate our theory's application in the incommensurate case using quasicrystalline 30$^\circ$ TBG as an example.

We start by examining spectral quantities, beginning with the density of states $D(\epsilon)$. In prior research, spectral quantities in quasicrystalline 30$^\circ$ TBG are commonly computed using supercell approximants and large-scale finite quasicrystalline samples. However, computing in supercell approximants near $30^\circ$ is challenging due to the exponential increase in the number of atoms $N_{\mathrm{atom}}$ required for enhanced accuracy (see Fig.~\ref{fig:30_structure}(b) in Appendix~\ref{app:numerical_30}). Alternatively, using large-scale finite samples with an exact incommensurate quasicrystalline structure at a $30^\circ$ twist angle also remains challenging, even with two million atoms, as they are not enough to overcome the break of single-layer periodicity~\cite{yu2019dodecagonal}. Therefore, maintaining long-range 12-fold rotational symmetry and single-layer periodicity is crucial for accurately and efficiently describing the spectra of quasicrystalline 30$^\circ$ TBG.

Our theory satisfies these requirements and provides accurate results using only a small cutoff with $G_{\mathrm{cut}}=12$ \AA$^{-1}$. As shown in Fig.~\ref{fig:30_DOS_band}(a), our theory provides an accurate density of states, comparable to outcomes from large-scale calculations with ten million atoms~\cite{yu2019dodecagonal}. In addition to the central VHS peak ($\xi$) inherited from single-layer graphene, it shows dense VHS peaks around $\epsilon=-1.5$ eV ($\beta - \eta$). These peaks originate from small gaps induced by interlayer coupling, mainly located at the $\mathrm{Q}$ point. The quasi-band structure with and without interlayer coupling, as depicted in Fig.~\ref{fig:30_DOS_band}(c)(d), traces the origin of these peaks. The results clearly indicate that these VHS peaks are not a result of Dirac cone hybridization or hybridization of the isotropic bottoms of the graphene $\pi$-bands, as seen at small-twist-angle TBG and single layer graphene, respectively. Instead, they are attributed to interlayer coupling and the intervalley coupling mediated by it, which is also reported as ghost anti-crossings in a graphene/InSe bilayer system~\cite{graham2020ghost}. Additionally, the observed VHS peak ($\alpha$) and band gap at the $\mathrm{M}$ point aligns with recent ARPES measurements~\cite{science.aar8412,pnas.1720865115}, confirming our theory's faithful reproduction of experimental observations in incommensurate quasicrystalline $30^\circ$ TBG.

To gain insights into the interlayer coupling-induced VHS peak, we examine the corresponding Fermi surface evolution. As shown in Fig.~\ref{fig:30_Ecuts}(a), the density of states discontinuities marked by $\alpha$ [see Fig.~\ref{fig:30_DOS_band}(b)] is clearly attributed to merging of two electron pockets. One valley's electron pocket intersects another valley's mirrored pocket in the same layer, involving a saddle point and generating a VHS peak. This intervalley coupling, mediated by interlayer coupling, results in a band gap at the $\mathrm{M}$ point, indicated by a white arrow in Fig.~\ref{fig:30_DOS_band}(b). In addition to the intervalley coupling peak ($\alpha$), there is an interlayer coupling peak ($\beta$). Hybridization of different layers' electron pocket at the $\mathrm{Q}$ point leads to a flat quasi-band as shown in Fig.~\ref{fig:30_DOS_band}(c) and Fig.~\ref{fig:30_Ecuts}(b-2). A similar flat band emerges at the $\zeta$ VHS peak, providing a platform to explore exotic interaction effects (see Fig.~\ref{fig:30_Ecuts_more} in Appendix~\ref{app:numerical_30} and SM \footnotemark[\value{footnote}] for additional results). The $\xi$ VHS peak is associated with the hybridization of the isotropic bottoms of the graphene $\pi$-bands, leading to a Fermi surface change similar to that in single graphene [see Fig.~\ref{fig:30_Ecuts_more}(d) in Appendix~\ref{app:numerical_30}]. These modifications of Fermi surface resulting from interlayer coupling are general characteristics of twisted multilayer systems, which have been reported in recent ARPES measurements on near-30$^\circ$ TBG~\cite{hamer2022moire}, as well as in theoretical studies of small-twist-angle TBG~\cite{PhysRevResearch.4.043145} and twisted trilayer graphene~\cite{PhysRevResearch.4.L012013}.

These changes signify a topological Lifshitz transition, previously observed in TBG~\cite{kim2016charge, PhysRevLett.125.176801, PhysRevLett.128.026404}, leading to observable transport signatures~\cite{PhysRevB.94.245403}, correlation-induced gaps~\cite{wu2021chern}, de Haas–van Alphen oscillations~\cite{PhysRevB.104.085412}, and reentrant superconductivity~\cite{PhysRevB.104.174505}. Notably, these phenomena are independent of the system's commensurability, challenging previous theories that were limited to the commensurate case. Therefore, our unified theory is essential for addressing incommensurate scenarios, especially when strong interlayer coupling contradicts previous long-wavelength low-energy approximations. With recent experimental progress in extreme graphene doping~\cite{PhysRevB.100.035445, PhysRevLett.125.176403}, exploring such Lifshitz transitions and Fermi surface topology changes in TBG and twisted trilayer graphene is now experimentally feasible.

Our theory also facilitates the exploration of phenomena tightly associated with the incommensurate structure. In Fig.~\ref{fig:30_FS}(a), the mirrored Dirac cone, a distinct feature of quasicrystalline 30$^\circ$ TBG as observed in recent ARPES measurements~\cite{science.aar8412, pnas.1720865115}, is easily obtained from the spectral function calculation. Due to the incommensurate nature of the quasicrystalline 30$^\circ$ TBG, the mirrored Dirac cone and its higher-order replicas densely populate the entire reciprocal plane, as evident in our numerical results showing a power-law scaling of the spectral function [see Fig.~\ref{fig:30_FS}(b)(c)]. These patterns are anticipated to be observed with enhanced experimental accuracy, resolution, and reduced thermal noise.

With sufficiently large doping, layer-1's Fermi surface intersects layer-2's, forming a fractal pattern consistent with Stampfli's 12-fold tiling~\cite{Stampfli1986ADQ}. As shown in Fig.~\ref{fig:30_FS}(e), this reveals an intriguing duality between incommensurate quasicrystalline atomic configurations and the fractal Fermi surface~\cite{PhysRevResearch.2.033162}. Our theory also uncovers an interlayer coupling-induced spiral Fermi surface around the $\mathrm{Q}$ point, as depicted in Fig.~\ref{fig:30_FS}(f). This topologically nontrivial spiral Fermi surface, characterized by a turning number $N_t=5$, facilitates semiclassical trajectories jumping between layer-1 and layer-2, repeating 12 times before quantization. This phenomenon, proposed in previous research using an effective model~\cite{PhysRevB.100.081405} and confirmed by direct calculations in our theory, can lead to fascinating quantum oscillations.

We proceed to apply our theory to calculate wave function-related quantities for the incommensurate case, using the same approach employed for the commensurate case within the framework of our unified theoretical framework. Optical absorption $\mathrm{Re}[\sigma_{xx}(\omega)]$ and $\bm{k}$-resolved $\mathrm{Re}[\sigma_{xx}(\omega)]$ are presented in Fig.~\ref{fig:30_ReS}. As depicted in Fig.~\ref{fig:30_ReS}(a), the optical absorption in incommensurate quasicrystalline 30$^\circ$ TBG also reveals a plateau at $2\sigma_0$, stemming from the linear band regime around the Dirac cones, along with multiple absorption peaks near 4 eV. Remarkably, a minor peak appears within the plateau region at $\hbar\omega=2.49$ eV, followed by a noticeable reduction. This characteristic is absent in Fig.~\ref{fig:21.79_ReS} for the commensurate 21.79$^\circ$ TBG, as it arises from intervalley coupling mediated by interlayer interactions, as evidenced by the $\bm{k}$-resolved $\mathrm{Re}[\sigma_{xx}(\omega)]$ profile in Fig.~\ref{fig:30_ReS}(b). Thus, this optical absorption peak correlates tightly with the $\alpha$ density of states peak [see Fig.~\ref{fig:30_DOS_band}(b)] and the corresponding Lifshitz transition of the Fermi surface topology shown in Fig.~\ref{fig:30_Ecuts}(a). Other optical absorption peaks also relate to different density of states peaks in Fig.~\ref{fig:30_DOS_band}(b). For instance, the significant increase  between $\hbar \omega=3.30$ eV and $\hbar \omega=3.34$ eV is explained by the $\mathrm{Re}[\sigma_{xx}(\bm{k},\omega)]$ profiles in Fig.~\ref{fig:30_ReS}(c)(d). This rise is also attributed to bright spots around the $\mathrm{Q}$ point dominated by the interlayer coupling and is related to the $\beta$ density of states peak [see Fig.~\ref{fig:30_DOS_band}(b)] and the flat quasi-band shown in Fig.~\ref{fig:30_DOS_band}(c) and Fig.~\ref{fig:30_Ecuts}(b-2).

In summary, our theory efficiently computes wave function-related quantities for incommensurate quasicrystalline 30$^\circ$ TBG, distinguishing it from the commensurate case. Additionally, it reveals interlayer coupling-mediated physics and provides insights into Lifshitz transitions of Fermi surface topology in these twisted multilayer systems.

\section{Summary and Outlook}

We developed a unified theoretical framework for efficiently describing electronic structures covering both spectrum and wave functions in twisted multilayer systems for both commensurate and incommensurate scenarios. Our approach decomposes physical observables into contributions from individual layers and their respective $\bm{k}$ points, even in the presence of intricate interlayer coupling. we propose a local low-rank approximated Hamiltonian for accurate and numerically efficient computations, outperforming previous supercell methods. Our theory facilitates the computation of wave function-related quantities relevant to response characteristics beyond spectra, particularly valuable for incommensurate twisted multilayer systems. We validate our theory by computing spectra and optical conductivity for moir\'e TBG and demonstrate its application in incommensurate quasicrystalline TBG.

Significantly, current continuous models for twisted multilayer systems mainly focus on the low-energy range, such as around the K point of TBG, allowing the effective separation of degrees of freedom. Models like the BM model and its derivatives are successful in this context. However, challenges arise when the initially separated degrees of freedom couple again, mediated by interlayer coupling, especially in incommensurate systems. These couplings become significant near the intersections of single-layer Brillouin zones (e.g., around the Q point of TBG) with sufficiently large doping. In such cases, previous continuous models break down as degrees of freedom cannot be effectively processed separately. Given recent experimental progress in extreme graphene doping~\cite{PhysRevB.100.035445, PhysRevLett.125.176403}, a theory is urgently needed to comprehend interlayer coupling-induced physics beyond the long-wavelength low-energy case. This is essential for capturing details at length and energy scales relevant to mesoscopic twisted multilayer systems. Thus, our method is crucial for addressing this, enabling the handling of any energy and specific $\bm{k}$ point with the desired accuracy. With recent experimental advancements enabling local and continuous tuning of the twist angle~\cite{inbar2023quantum}, our theory also facilitates a comprehensive study of physics induced by twisting, covering a continuous spectrum of physical observables across the entire range of twist angles.

Besides, our unified theoretical framework remains compatible with prior theories, allowing expansion at any energy and specific $\bm{k}$ point with the desired accuracy. For example, focusing on low-energy physics around the K point in small-twist-angle TBG, our theory in the leading order is equivalent to the well-known BM model~\cite{bistritzer2011moire}. Furthermore, our approach features a simple workflow and a clear physical picture, enhancing accessibility. The generality of our theory renders details of single layers, the number of layers, stacking order, and twist angle irrelevant. Additionally, parameters for constructing the local low-rank approximated Hamiltonian can be extracted from \textit{ab initio} calculations, making it versatile for various twisted multilayer systems.

Finally, our generic theoretical framework allows a number of interesting directions for follow-up research. First, extending to interacting systems is straightforward using Hartree-Fock or dynamic mean-field theory calculations~\cite{PhysRevB.102.035136, PhysRevB.103.205413, PhysRevB.104.075150, PhysRevB.104.115167, PhysRevX.12.021064}. This enables the exploration of exotic quantum phenomena with strong electron-electron interaction, including unconventional superconductivity and strongly correlated insulating phases. Besides, an open question is how to uniformly characterize topology in our theoretical framework under the composite Bloch basis, especially for the incommensurate case. Additionally, the framework can be generalized to describe Bosonic collective excitations such as phonons~\cite{lin2018moire, PhysRevB.100.075416,quan2021phonon, PhysRevB.106.144305, PhysRevB.107.115301}, Cooper pairs~\cite{PhysRevB.99.134515, PhysRevB.99.195114, PhysRevB.106.184517, PhysRevLett.129.187001, PhysRevLett.131.016001}, and magnons~\cite{PhysRevB.102.094404, PhysRevB.102.165118, PhysRevLett.125.247201,xie2022twist, xu2022coexisting, PhysRevX.13.021016}. Thus, our findings provide a generic approach for studying the electronic structure of twisted multilayer systems and pave the way for future research.

\begin{acknowledgments}
We thank Xufeng Liu and Yanxi Wu for the valuable discussions. This work is supported by the National Key R\&D Program of China (Grant No. 2021YFA1401600) and the National Natural Science Foundation of China (Grant No. 12074006). The computational resources were supported by the high-performance computing platform of Peking University.
\end{acknowledgments}

\textit{Note added.} During the final stage of preparing the manuscript, we became aware of an independent work on arXiv~\cite{he2023energy}, which has an overlap with the incommensurate spectral part in our theory.

\appendix

\section{\label{app:convention}Conventions of Bloch basis}
It is important to acknowledge two distinct conventions in the literature when expanding the Bloch basis $\{|\bm{k} \bm{t} \alpha \rangle\}$ using a local atomic basis $\{|\bm{R} \bm{t} \alpha \rangle\}$~\cite{RevModPhys.84.1419}. The first convention is known as the ``atom gauge"~\cite{velocity_matrix_scipost}. In this convention, the Bloch basis is defined as:
\begin{equation}
|\bm{k} \bm{t} \alpha \rangle_a = \frac{1}{\sqrt{N}} \sum_{\bm{R}} e^{i \bm{k} \cdot(\bm{R}+\bm{t})}|\bm{R} \bm{t} \alpha \rangle.
\end{equation}
The second convention is referred to as the ``cell gauge"~\cite{velocity_matrix_scipost}, where the Bloch basis is given by:
\begin{equation}
| \bm{k} \bm{t} \alpha \rangle_c = \frac{1}{\sqrt{N}} \sum_{\bm{R}} e^{i \bm{k} \cdot \bm{R}} | \bm{R} \bm{t} \alpha \rangle.
\end{equation}
These conventions differ in their treatment of the phase factor $e^{i \bm{k} \cdot\bm{t}}$ related to atomic positions. The atom gauge explicitly incorporates these phase factors, while the cell gauge does not. The choice of convention depends on the specific requirements and goals of the theoretical framework in use.

In the atom gauge convention, the Hamiltonian matrix entries are given by
\begin{equation}
\langle \bm{p} \bm{t}^\prime \beta   | \mathcal{H}   | \bm{k} \bm{t}\alpha \rangle_a =  \delta_{\bm{k}, {\bm{p}}} \sum_{\bm{R}} e^{i \bm{k} \cdot ( \bm{R} +\bm{t}-\bm{t}^\prime )} \mathcal{H}_{\beta,\alpha}( \bm{R} +\bm{t}-\bm{t}^\prime ).
\end{equation}
Since the Hamiltonian is block-diagonal with each block labeled by $\bm{k}$, we can express the Hamiltonian matrix entries within each block-diagonal subspace as
\begin{equation}
\mathcal{H}^a_{\bm{t}^\prime \beta,\bm{t} \alpha}(\bm{k})=\sum_{\bm{R}} e^{i \bm{k} \cdot ( \bm{R} +\bm{t}-\bm{t}^\prime)} \mathcal{H}_{\beta,\alpha}( \bm{R} +\bm{t}-\bm{t}^\prime ).
\end{equation}
In contrast, the cell gauge convention leads to the Hamiltonian matrix entries denoted as
\begin{equation}
\langle \bm{p} \bm{t}^\prime \beta   | \mathcal{H}   | \bm{k} \bm{t}\alpha \rangle_c = \delta_{\bm{k}, {\bm{p}}} \sum_{\bm{R}} e^{i \bm{k} \cdot \bm{R}} \mathcal{H}_{\beta,\alpha}( \bm{R}+\bm{t}-\bm{t}^\prime ),
\end{equation}
with corresponding entries within each subspace represented as
\begin{equation}
\mathcal{H}^c_{\bm{t}^\prime \beta,\bm{t} \alpha}(\bm{k})=\sum_{\bm{R}} e^{i \bm{k} \cdot \bm{R}} \mathcal{H}_{\beta,\alpha}( \bm{R}+\bm{t}-\bm{t}^\prime).
\end{equation}
Thus, the connection between the Hamiltonian matrix entries in these two conventions is established by
\begin{equation}
\label{eq_appen_convention_H}
\mathcal{H}^c_{\bm{t}^\prime \beta,\bm{t} \alpha}(\bm{k})= e^{i\bm{k}\cdot(\bm{t}^\prime -\bm{t})}
\mathcal{H}^a_{\bm{t}^\prime \beta,\bm{t} \alpha}(\bm{k}).
\end{equation}

Under different conventions, the Bloch eigenstates can be expressed as follows. In the atom gauge convention, the Bloch eigenstates $|\bm{k} n\rangle$ are expanded using coefficients $c_{\bm{t}\alpha n}(\bm{k})$ as:
\begin{equation}
|\bm{k} n\rangle = \sum_{\bm{t}\alpha} c_{\bm{t}\alpha n}(\bm{k})| \bm{k} \bm{t} \alpha\rangle_a,
\end{equation}
where $n$ represents the band index, and these states possess energy eigenvalues $\epsilon_n(\bm{k})$.
Conversely, within the cell gauge convention, the Bloch eigenstates $|\bm{k} n\rangle$ are expressed using coefficients $b_{\bm{t}\alpha n}(\bm{k})$ as:
\begin{equation}
|\bm{k} n\rangle = \sum_{\bm{t}\alpha} b_{\bm{t}\alpha n}(\bm{k})|\bm{k} \bm{t} \alpha\rangle_c.
\end{equation}
The relationship between the expansion coefficients under these two conventions is given by:
\begin{equation}
\label{eq_appen_convention_c}
b_{\bm{t}\alpha n}(\bm{k}) = e^{i \bm{k} \cdot \bm{t}} c_{\bm{t}\alpha n}(\bm{k}).
\end{equation}
To establish this relation, we begin by examining the secular equations satisfied by $c_{\bm{t}\alpha n}(\bm{k})$:
\begin{equation}
\sum_{\bm{t}^\prime \beta} \mathcal{H}^a_{\bm{t}\alpha,\bm{t}^\prime \beta}(\bm{k}) c_{\bm{t}^\prime \beta n}(\bm{k}) = \epsilon_n(\bm{k}) c_{\bm{t}\alpha n}(\bm{k}).
\end{equation}
Similarly, $b_{\bm{t}\alpha n}(\bm{k})$ satisfies the following secular equations within the cell gauge convention:
\begin{equation}
\sum_{\bm{t}^\prime \beta} \mathcal{H}^c_{\bm{t}\alpha,\bm{t}^\prime\beta}(\bm{k}) b_{\bm{t}^\prime \beta n}(\bm{k}) = \epsilon_n(\bm{k}) b_{\bm{t}\alpha n}(\bm{k}).
\end{equation}
By substituting the relation
\begin{equation}
\mathcal{H}^c_{\bm{t}\alpha,\bm{t}^\prime \beta}(\bm{k}) =  e^{i \bm{k} \cdot (\bm{t}-\bm{t}^\prime )} \mathcal{H}^a_{\bm{t}\alpha,\bm{t}^\prime \beta}(\bm{k})
\end{equation}
into the cell gauge secular equation, we arrive at:
\begin{equation}
\sum_{\bm{t}^\prime \beta} e^{i \bm{k} \cdot (\bm{t}-\bm{t}^\prime )} \mathcal{H}^a_{\bm{t}\alpha,\bm{t}^\prime \beta}(\bm{k}) b_{\bm{t}^\prime \beta n}(\bm{k})= \epsilon_n(\bm{k}) b_{\bm{t}\alpha n}(\bm{k}),
\end{equation}
or equivalently:
\begin{equation}
\sum_{\bm{t}^\prime \beta} \mathcal{H}^a_{\bm{t}\alpha,\bm{t}^\prime \beta}(\bm{k}) e^{-i \bm{k} \cdot \bm{t}^\prime} b_{\bm{t}^\prime \beta n}(\bm{k})= \epsilon_n(\bm{k}) e^{-i \bm{k} \cdot \bm{t}} b_{\bm{t}\alpha n}(\bm{k}).
\end{equation}
Comparing this expression with the atom gauge secular equation, we establish the relationship of Bloch expansion coefficients under different conventions, as indicated in Eq. (\ref{eq_appen_convention_c}).

In summary, the relationship between quantities in the two conventions can be concisely expressed as Eq.~(\ref{eq_appen_convention_H}) and Eq.~(\ref{eq_appen_convention_c}). This relationship underscores that the two conventions are fundamentally connected through a unitary rotation within the internal space. One can draw an analogy between the Bloch functions, denoted as $\psi_{n \bm{k}}(\bm{r})$ and $b_{\bm{t}\alpha n}(\bm{k})$, as well as between the cell-periodic Bloch functions, represented by $u_{n \bm{k}}(\bm{r})$ and $c_{\bm{t}\alpha n}(\bm{k})$. To facilitate this comparison, we invoke the Bloch theorem~\cite{girvin_yang_2019}, which asserts: $\psi_{n \bm{k}}(\bm{r})=e^{i \bm{k} \cdot \bm{r}} u_{n \bm{k}}(\bm{r})$. For clarity, we momentarily introduce a change of notation, i.e., $c_{\bm{t}\alpha n}(\bm{k}) \rightarrow c_{\bm{k} n}(\bm{t}\alpha)$ and similarly for $b_{\bm{t}\alpha n}(\bm{k}) \rightarrow b_{\bm{k} n}(\bm{t}\alpha)$. We can now express:
\begin{eqnarray}
| \psi_{n \bm{k}}\rangle
& = &\frac{1}{\sqrt{\Omega}}\sum_{\bm{R}} \int_{\text {cell }} \mathrm{d}^d \bm{r} \psi_{n \bm{k}}(\bm{r}) e^{i \bm{k} \cdot \bm{R}}|\bm{R}+\bm{r}\rangle, \nonumber\\
& =& \frac{1}{\sqrt{N}}\sum_{\bm{R}} \sum_{\bm{t} \alpha} b_{\bm{k} n}(\bm{t}\alpha) e^{i \bm{k} \cdot \bm{R}}|\bm{R} \bm{t}\alpha\rangle.
\end{eqnarray}
Simultaneously, we have:
\begin{eqnarray}
|\psi_{n \bm{k}}\rangle
& = & \frac{1}{\sqrt{\Omega}}\sum_{\bm{R}} \int_{\text {cell }} \mathrm{d}^d \bm{r} u_{n \bm{k}}(\bm{r}) e^{i \bm{k} \cdot(\bm{R}+\bm{r})}|\bm{R}+\bm{r}\rangle, \nonumber\\
& = & \frac{1}{\sqrt{N}} \sum_{\bm{R} } \sum_{\bm{t} \alpha}  c_{\bm{k} n}(\bm{t}\alpha) e^{i \bm{k} \cdot(\bm{R}+\bm{t})}|\bm{R}\bm{t} \alpha\rangle .
\end{eqnarray}
Therefore, we establish the analogy:
\begin{eqnarray}
\psi_{n \bm{k}}(\bm{r}) & \leftrightarrow & b_{\bm{t}\alpha n}(\bm{k}),\\
u_{n \bm{k}}(\bm{r}) & \leftrightarrow & c_{\bm{t}\alpha n}(\bm{k}).
\end{eqnarray}

The cell gauge convention is commonly used due to its simplicity, neglecting the additional factors of $e^{i \bm{k} \cdot \bm{t}}$. However, the atom gauge convention offers a more natural representation for calculations involving response and Berry-phase quantities~\cite{vanderbilt_2018}. In this paper, we adopt the atom gauge unless otherwise specified.

\section{\label{app:Poisson}The Poisson summation formula}
We briefly review the Poisson summation formula, a fundamental mathematical tool applicable in various contexts, particularly in the context of relating real space lattice sums to reciprocal lattice sums~\cite{girvin_yang_2019}.
Our discussion begins with the one-dimensional (1D) case, highlighting its generalizability to higher dimensions for clarity.

Consider a 1D lattice with lattice constant $a$, corresponding to a reciprocal lattice constant of $b=2\pi / a$. Our goal is to evaluate the sum of a function $f$ over lattice sites:
\begin{equation}
F =\sum_{n=-\infty}^{\infty} f(n a),
\end{equation}
where $n \in \mathbb{Z}$. 
We establish a connection between the sum $F$ and a sum over the reciprocal lattice by introducing the function:
\begin{equation}
\label{eq_Appen_Poisson_sum}
F(x)=\sum_{n=-\infty}^{\infty} f(x+n a),
\end{equation}
which possesses the periodicity property $F(x+m a)=F(x), m\in \mathbb{Z}$. 
As a consequence, $F(x)$ can be expressed as a Fourier series:
\begin{equation}
\label{eq_Appen_Poisson_sum_F}
F(x)=\sum_{l=-\infty}^{\infty} \tilde{F}_l e^{i l b x},
\end{equation}
where $l \in \mathbb{Z}$, 
and the Fourier coefficients $\tilde{F}_l$ are defined as:
\begin{eqnarray}
\tilde{F}_l &=& \frac{1}{a} \int_0^a \mathrm{d} x e^{-i l b x} F(x), \nonumber\\
& = & \frac{1}{a} \int_0^a \mathrm{d} x e^{-i l b x} \sum_{n=-\infty}^{\infty} f(x+n a), \nonumber\\
& = & \frac{1}{a} \sum_{n=-\infty}^{\infty} \int_{n a}^{(n+1) a} \mathrm{d} x^{\prime} e^{-i l b (x^{\prime}-n a)} f(x^{\prime}).
\end{eqnarray}
Utilizing the identity $e^{i l b a}=e^{i 2 \pi l}=1$, we simplify the expression:
\begin{eqnarray}
\tilde{F}_l & = &\frac{1}{a} \sum_{n=-\infty}^{\infty} \int_{n a}^{(n+1) a} \mathrm{d} x^{\prime} e^{-i l b x^{\prime}} f(x^{\prime}), \nonumber\\
& = & \frac{1}{a} \int_{-\infty}^{\infty}  \mathrm{d} x^{\prime} e^{-i l b x^{\prime}} f(x^{\prime}), \nonumber\\
& = & \frac{1}{a} \tilde{f}(lb),
\end{eqnarray}
where $\tilde{f}$ denotes the Fourier transform of the function $f$. Now, referring back to Eq. (\ref{eq_Appen_Poisson_sum_F}), we find:
\begin{equation}
    F(0)=\sum_{l=-\infty}^{\infty} \tilde{F}_l= \frac{1}{a} \sum_{l=-\infty}^{\infty} \tilde{f}(lb) .
\end{equation}
By combining this with Eq. (\ref{eq_Appen_Poisson_sum}), we arrive at the Poisson summation formula:
\begin{equation}
\label{eq_Appen_PSF}
\sum_{n=-\infty}^{\infty} f(na)=\frac{1}{a} \sum_{l=-\infty}^{\infty} \tilde{f}(lb).
\end{equation}

The generalization of these results to higher-dimensional lattices is straightforward and can be expressed as:
\begin{equation}
\sum_{\bm{R}} f(\bm{R})=\frac{1}{\Omega} \sum_{\bm{G}} \tilde{f}(\bm{G}),
\end{equation}
where $\Omega$ represents the volume of the unit cell.

In practical applications, we often encounter functions of the form $f(\bm{x})=e^{i \bm{k} \cdot \bm{x}}$, whose Fourier coefficient is given by $\tilde{f}(\bm{x^{\prime}}) = (2\pi)^d \delta(\bm{x^{\prime}}-\bm{k})$, with $d$ denoting the spatial dimension of the system. In such cases, we obtain the following result:
\begin{equation}
\label{eq_appen_Poisson_delta}
\sum_{\bm{R}} e^{i \bm{k} \cdot \bm{R}}=\frac{(2\pi)^d}{\Omega} \sum_{\bm{G}} \delta^{(d)}(\bm{k}-\bm{G}).
\end{equation}
Utilizing the relationship between the Dirac delta function and the Kronecker delta function,
\begin{equation}
\label{eq_appen_2delta}
\frac{(2\pi)^d}{N \Omega} \delta^{(d)}(\bm{k}-\bm{G}) = \delta_{\bm{k},\bm{G}},
\end{equation}
we arrive at an alternative form of Eq. (\ref{eq_appen_Poisson_delta}):
\begin{equation}
\label{eq_appen_Poisson_delta_final}
\sum_{\bm{R}} e^{i \bm{k} \cdot \bm{R}}=N \sum_{\bm{G}} \delta_{\bm{k},\bm{G}},
\end{equation}
where, in cases where $\bm{k}$ is confined within the first BZ, only the $\bm{G}=\bm{0}$ component survives on the right-hand side, leading to the result:
\begin{equation}
\sum_{\bm{R}} e^{i \bm{k} \cdot \bm{R}}=N \delta_{\bm{k},\bm{0}}.
\end{equation}

In the derivation above, we utilized the conversion relation between the Dirac delta function and the Kronecker delta function, which arises directly from their definitions. For a given function $f(\bm{x})$, the property of these delta functions can be expressed as follows:
\begin{eqnarray}
f(\bm{x_0}) &=& \int \mathrm{d}^{d} \bm{x} \delta^{(d)}(\bm{x}-\bm{x_0} ) f(\bm{x}), \nonumber\\
&=& \sum_{\bm{x}} |\Delta \bm{x}| \delta^{(d)}(\bm{x}-\bm{x_0} ) f(\bm{x}), \\
f(\bm{x_0}) &=& \sum_{\bm{x}} \delta_{\bm{x},\bm{x_0}} f(\bm{x}), \nonumber\\
&=& \frac{1}{|\Delta \bm{x}|} \sum_{\bm{x}} |\Delta \bm{x}| \delta_{\bm{x},\bm{x_0}} f(\bm{x}), \nonumber\\
&=& \frac{1}{|\Delta \bm{x}|} \int \mathrm{d}^{d} \bm{x}  \delta_{\bm{x},\bm{x_0}} f(\bm{x}).
\end{eqnarray}
These equations lead to the relation between the Dirac delta function and the Kronecker delta function:
\begin{equation}
    |\Delta \bm{x}| \delta^{(d)}(\bm{x}-\bm{x_0})=\delta_{\bm{x},\bm{x_0}}.
\end{equation}
When $\bm{x} \rightarrow \bm{G}$, we have $|\Delta \bm{G}|=(2\pi)^d / (N \Omega)$. Thus, we arrive at the expression
\begin{equation}
(2\pi)^d/(N \Omega) \delta^{(d)}(\bm{k}-\bm{G}) = \delta_{\bm{k},\bm{G}},
\end{equation}
which corresponds to Eq.~(\ref{eq_appen_2delta}).

\section{\label{app:intralayer}Derivation of intralayer Hamiltonian matrix elements under the composite Bloch basis}
Here, we derive the intralayer Hamiltonian matrix elements in the twisted multilayer system, which is Eq.~(\ref{eq_intralayer}) in the main text. We highlight that the periodic condition allows the Poisson summation formula to decouple different single-layer Bloch states, making the $\mathcal{H}_l$ block-diagonal.

We start by expressing the single layer Bloch basis $\{|\bm{k}_{l}\bm{t}_{l}\alpha\rangle\}$ in real space atomic basis $\{|\bm{R}_{l} \bm{t}_{l} \alpha \rangle\}$, using Eq.~(\ref{eq_kbasis}) as demonstrated below:
\begin{eqnarray}
\langle\bm{k}^\prime_{l}\bm{t}^\prime_{l}\beta|\mathcal{H}_{l}|\bm{k}_{l}\bm{t}_{l}\alpha\rangle
&=& \frac{1}{N_l} \sum_{\bm{R}_l,\bm{R}^\prime_l}
e^{-i\bm{k}^\prime_l \cdot ( \bm{R}^\prime_l + \bm{t}^\prime_{l} )}
e^{i \bm{k}_l \cdot ( \bm{R}_l+\bm{t}_{l} )} \nonumber\\
& & \times \langle \bm{R}^\prime_l \bm{t}^\prime_l \beta|\mathcal{H}|\bm{R}_{l} \bm{t}_{l} \alpha \rangle.
\end{eqnarray}
As both $\bm{R}^\prime_l$ and $\bm{R}_l$ are lattice vectors of layer-$l$, we can perform the change of variables trick. More specifically, we
can replace $\bm{R}_l$ with $\bm{R}_l+\bm{R}^\prime_l$, making  $\sum_{\bm{R}_l+\bm{R}^\prime_l}$ equivalent to $\sum_{\bm{R}_l}$:
\begin{eqnarray}
\bm{R}_l &\to& \bm{R}_l+\bm{R}^\prime_l, \\
\sum_{\bm{R}_l}[\cdot] &\to&  \sum_{\bm{R}_l+\bm{R}^\prime_l}[\cdot] =  \sum_{\bm{R}_l}[\cdot].
\end{eqnarray}
This simplifies the intralayer Hamiltonian matrix elements as follows:
\begin{eqnarray}
& & \langle\bm{k}^\prime_{l}\bm{t}^\prime_{l}\beta|\mathcal{H}_{l}|\bm{k}_{l}\bm{t}_{l}\alpha\rangle \nonumber\\
&= & \frac{1}{N_l} \sum_{\bm{R}_l,\bm{R}^\prime_l}
e^{-i\bm{k}^\prime_l \cdot ( \bm{R}^\prime_l + \bm{t}^\prime_{l} )}
e^{i \bm{k}_l \cdot ( \bm{R}_l+\bm{R}^\prime_l+\bm{t}_{l} )} \nonumber\\
& & \times \langle \bm{R}^\prime_l \bm{t}^\prime_l \beta|\mathcal{H}|\bm{R}_{l}+\bm{R}^\prime_l, \bm{t}_{l} \alpha \rangle,\nonumber\\
&= & \frac{1}{N_l}  \sum_{\bm{R}^\prime_l}
e^{i(\bm{k}_l-\bm{k}^\prime_l ) \cdot \bm{R}^\prime_l}  \sum_{\bm{R}_l}
e^{i \bm{k}_l \cdot ( \bm{R}_l +\bm{t}_{l}-\bm{t}^\prime_{l} )}  \langle \bm{0} \bm{t}^\prime_l \beta|\mathcal{H}|\bm{R}_{l} \bm{t}_{l} \alpha \rangle,\nonumber\\
&= & \delta_{\bm{k}_l, \bm{k}^\prime_l} \sum_{\bm{R}_l} e^{i \bm{k}_l \cdot ( \bm{R}_l +\bm{t}_l-\bm{t}^\prime_l )} \mathcal{H}_{\beta,\alpha}(\bm{R}_{l}+\bm{t}_l-\bm{t}^\prime_l).
\end{eqnarray}

In the second step, we applied the translation invariance of the Hamiltonian $\mathcal{H}$ under local atomic basis in real space:
\begin{equation}
\label{eq_hamiltonian_element}
\langle {\bm{R}}^{\prime}_{l^{\prime}} {\bm{t}}^{\prime}_{l^\prime} \beta | \mathcal{H} | \bm{R}_{l} \bm{t}_{l}\alpha \rangle = \mathcal{H}_{\beta,\alpha}(\bm{R}_{l}-\bm{R}^\prime_{l^\prime}+\bm{t}_{l}-\bm{t}^\prime_{l^\prime}+\Delta h_{l,l^\prime} \bm{e}_z),
\end{equation}
where $\Delta h_{l,l^\prime}=h_{l}-h_{l^\prime}$.
While in the final step, we applied the Poisson summation formula, Eq.~(\ref{eq_appen_Poisson_delta_final}), to establish
\begin{equation}
\sum_{\bm{R}^\prime_l} e^{i(\bm{k}_l-\bm{k}^\prime_l ) \cdot \bm{R}^\prime_l} =N_l \delta_{\bm{k}_l-\bm{k}^\prime_l,\bm{0}}=N_l\delta_{\bm{k}_l,\bm{k}^\prime_l}.
\end{equation}
This simplification relies on the fact that $\bm{R}_l- \bm{R}^\prime_l$ is still a lattice vector of layer-$l$. As a result, the Poisson summation formula introduces a delta function, effectively decoupling the $|\bm{k}^\prime_{l} \rangle$ and $| \bm{k}_{l} \rangle$ states with different single layer Bloch wavevectors. Thus, $\mathcal{H}_l$ still possesses a block-diagonal structure.

\section{\label{app:interlayer}Derivation of interlayer Hamiltonian matrix elements under the composite Bloch basis}
Here, we derive the intralayer Hamiltonian matrix elements in the twisted multilayer system, which is Eq.~(\ref{eq_interlayer}) in the main text. We also give a remark on the original derivation in Ref.~\cite{Koshino_2015} about the Dirac delta function and the Kronecker delta function as discussed in Appendix~\ref{app:Poisson}.

We first introduce the Fourier transformation of the interlayer hopping integral $\mathcal{H}_{\beta,\alpha}(\bm{R}_{l}-\bm{R}^\prime_{l^\prime}+\bm{t}_{l}-\bm{t}^\prime_{l^\prime}+\Delta h_{l,l^\prime} \bm{e}_z)$. Consider $\mathcal{H}_{\beta,\alpha}(\bm{r}+\Delta h_{l,l^\prime} \bm{e}_z)$ with $\bm{r} \in \mathbb{R}^2$ defined as:
\begin{equation}
\label{eq_H_fourier}
\mathcal{H}_{\beta,\alpha}(\bm{r}+\Delta h_{l,l^\prime} \bm{e}_z) =  \frac{S}{\sqrt{N_l {N}_{l^\prime}}} \int \frac{\mathrm{d}^2 \bm{q}}{(2\pi)^2}  e^{-i \bm{q} \cdot \bm{r}} \mathcal{V}^{l^\prime,l}_{\beta,\alpha}(\bm{q}),
\end{equation}
where $S=N_l \Omega_l={N}_{l^\prime} {\Omega}_{l^\prime}$ is the total area of each layer with $\Omega_l$ and ${\Omega}_{l^\prime}$ denotes the area of the unit cell of layer-$l$ and layer-$l^\prime$, respectively. The Fourier components $\mathcal{V}^{l^\prime,l}_{\beta,\alpha}(\bm{q})$ are defined through the inverse Fourier transformation:
\begin{equation}
\label{eq_H_fourier_inv}
\mathcal{V}^{l^\prime,l}_{\beta,\alpha}(\bm{q})
= \frac{1}{\sqrt{\Omega_l \Omega_{l^\prime}}}
\int \mathrm{d}^2 \bm{r}
e^{i \bm{q} \cdot \bm{r}}
\mathcal{H}_{\beta,\alpha}( \bm{r} + \Delta h_{l,l^\prime} \bm{e}_z ).
\end{equation}

Using this expansion, we can simplify Eq.~(\ref{eq_interlayer_direct}) as follows:
\begin{widetext}
\begin{eqnarray}
\langle\bm{k}^\prime_{l^\prime}\bm{t}^\prime_{l^\prime}\beta|\mathcal{V}_{l,l^\prime}|\bm{k}_{l}\bm{t}_{l}\alpha\rangle
&=& \frac{S}{N_l {N}_{l^\prime}} \sum_{\bm{R}_l,\bm{R}^\prime_{l^\prime}} e^{-i\bm{k}^\prime_{l^\prime} \cdot ( \bm{R}^\prime_{l^\prime} + \bm{t}^\prime_{l^\prime} )} e^{i\bm{k}_l \cdot (\bm{R}_l+\bm{t}_l)}  \int \frac{\mathrm{d}^2 \bm{q}}{(2\pi)^2}  e^{-i \bm{q} \cdot (\bm{R}_{l}-\bm{R}^\prime_{l^\prime}+\bm{t}_{l}-\bm{t}^\prime_{l^\prime})} \mathcal{V}^{l^\prime,l}_{\beta,\alpha}(\bm{q}), \nonumber \\
&=&
\frac{S}{N_l N_{l^\prime}}
\int\frac{\mathrm{d}^2 \bm{q}}{(2\pi)^2}
\mathcal{V}^{l^\prime,l}_{\beta,\alpha}(\bm{q})
e^{i(\bm{k}_l-\bm{q}) \cdot \bm{t}_{l} -i (\bm{k}^\prime_{l^\prime}-\bm{q}) \cdot {\bm{t}}^\prime_{l^\prime}} \times\sum_{\bm{R}_l} e^{i(\bm{k}_l-\bm{q}) \cdot \bm{R}_l}
\sum_{\bm{R}^\prime_{l^\prime}} e^{i(-\bm{k}^\prime_{l^\prime} +\bm{q}) \cdot{\bm{R}_{l^\prime}}}.
\end{eqnarray}
Employing the Poisson summation formula, Eq.~(\ref{eq_appen_Poisson_delta}) and Eq.~(\ref{eq_appen_Poisson_delta_final}), we have
\begin{eqnarray}
& &\sum_{\bm{R}_l} e^{i(\bm{k}_l - \bm{q}) \cdot \bm{R}_l}
=  \frac{(2\pi)^2}{\Omega_l} \sum_{\bm{G}_l} \delta^{(2)}(\bm{k}_l -\bm{q}-\bm{G}_l), \\
& &\sum_{\bm{R}^\prime_{l^\prime}} e^{i(-\bm{k}^\prime_{l^\prime} +\bm{q}) \cdot \bm{R}^\prime_{l^\prime}}
= N_{l^\prime} \sum_{\bm{G}^\prime_{l^\prime}} \delta_{-\bm{k}^\prime_{l^\prime}+\bm{q},\bm{G}^\prime_{l^\prime}}.
\end{eqnarray}
This simplifies the expression for the interlayer coupling to:
\begin{eqnarray}
\langle\bm{k}^\prime_{l^\prime}\bm{t}^\prime_{l^\prime}\beta|\mathcal{V}_{l,l^\prime}|\bm{k}_{l}\bm{t}_{l}\alpha\rangle
&=& \frac{S}{N_l N_{l^\prime}} \frac{(2\pi)^2}{\Omega_l} N_{l^\prime}
\sum_{\bm{G}_l,\bm{G}^\prime_{l^\prime}} \int \frac{\mathrm{d}^2 \bm{q}}{(2\pi)^2}
\mathcal{V}^{l^\prime,l}_{\beta,\alpha}(\bm{q})
e^{i(\bm{k}_l-\bm{q}) \cdot \bm{t}_{l} -i (\bm{k}^\prime_{l^\prime}-\bm{q}) \cdot {\bm{t}}^\prime_{l^\prime}}
\delta^{(2)}(\bm{k}_l -\bm{q}-\bm{G}_l)
\delta_{-\bm{k}^\prime_{l^\prime}+\bm{q},\bm{G}^\prime_{l^\prime}}, \nonumber \\
&=& \sum_{\bm{G}_l,\bm{G}^\prime_{l^\prime}}
\mathcal{V}^{l^\prime,l}_{\beta,\alpha}(\bm{k}_l - \bm{G}_l)
e^{i\bm{G}_l \cdot \bm{t}_{l} -i (\bm{k}^\prime_{l^\prime}-\bm{k}_l+\bm{G}_l) \cdot \bm{t}^\prime_{l^\prime}}
\delta_{-\bm{k}^\prime_{l^\prime}+\bm{k}_l-\bm{G}_l,\bm{G}^\prime_{l^\prime}}.
\end{eqnarray}
\end{widetext}

Subsequently, by changing the variable from $\bm{G}_l$ to $-\bm{G}_l$, equating $\sum_{-\bm{G}_l}$ to $\sum_{\bm{G}_l}$, the resulting Kronecker delta function can be rewritten as
\begin{eqnarray}
\delta_{-\bm{k}^\prime_{l^\prime}+\bm{k}_l-\bm{G}_l,\bm{G}^\prime_{l^\prime}} \to
\delta_{-\bm{k}^\prime_{l^\prime}+\bm{k}_l+\bm{G}_l,\bm{G}^\prime_{l^\prime}}=\delta_{\bm{k}_l+\bm{G}_l,\bm{k}^\prime_{l^\prime}+\bm{G}^\prime_{l^\prime}},
\end{eqnarray}
Similarly, the phase factor subject to the delta function constraint can be rewritten as follows:
\begin{eqnarray}
e^{-i (\bm{k}^\prime_{l^\prime}-\bm{k}_l+\bm{G}_l) \cdot \bm{t}^\prime_{l^\prime}} \to
e^{-i (\bm{k}^\prime_{l^\prime}-\bm{k}_l-\bm{G}_l) \cdot \bm{t}^\prime_{l^\prime}} =e^{i \bm{G}^\prime_{l^\prime} \cdot \bm{t}^\prime_{l^\prime}}.
\end{eqnarray}
Then, we arrive at the final expression for the interlayer coupling matrix elements in a general twisted multilayer system:
\begin{eqnarray}
\langle\bm{k}^\prime_{l^\prime}\bm{t}^\prime_{l^\prime}\beta|\mathcal{V}_{l,l^\prime}|\bm{k}_{l}\bm{t}_{l}\alpha\rangle
&=& \sum_{\bm{G}_l,\bm{G}^\prime_{l^\prime}}
\mathcal{V}^{l^\prime,l}_{\beta,\alpha}(\bm{k}_l + \bm{G}_l)
e^{-i\bm{G}_l \cdot \bm{t}_{l} +i \bm{G}^\prime_{l^\prime} \cdot \bm{t}^\prime_{l^\prime}} \nonumber \\
& & \times \delta_{\bm{k}_l+\bm{G}_l,\bm{k}^\prime_{l^\prime}+\bm{G}^\prime_{l^\prime}}.
\end{eqnarray}

Note that the normalization coefficients utilized in the Fourier transform exhibit slight differences from the expressions found in the original reference~\cite{Koshino_2015}. The author of Ref.~\cite{Koshino_2015} seems to have incorrectly applied the Kronecker delta function to perform the $\bm{q}$ integral (see Eq.~(9) in Ref.~\cite{Koshino_2015}). However, the correct procedure involves first converting the Kronecker delta function to the Dirac delta function and subsequently performing the integral over $\bm{q}$. This conversion introduces an additional factor, which is accommodated by redefining the normalization factor in the Fourier transform, as outlined in Eq.~(\ref{eq_H_fourier}).

\section{\label{app:Koshino}Findings and remaining ambiguities in Koshino's scheme}
Here, we provide a brief review of Koshino's scheme for computing the density of states $D(\epsilon)$ for incommensurate TBG~\cite{Koshino_2015}. We highlight the key findings and remaining ambiguities that motivate our work.

In this scheme, it is assumed that one can directly divide $D(\epsilon)$ into two distinct components, each representing contributions from specific layers, and further into single $\bm{k}$ point contributions. Specifically, this intuitive decomposition of $D(\epsilon)$ can be expressed as:
\begin{eqnarray}
\label{eq_Koshino_D}
D(\epsilon) &=& D_1(\epsilon)+D_2(\epsilon), \\
\label{eq_Koshino_D1}
D_1(\epsilon) &=& \sum_{\bm{k}} A_1(\bm{k},\epsilon), \\
\label{eq_Koshino_A}
A_1(\bm{k},\epsilon) &=& \sum_{n} g_{n\bm{k}}^{(1)} \delta(\epsilon-\epsilon_{n \bm{k}}).
\end{eqnarray}
Here, $D_1(\epsilon)$ represents the density of states contributed by layer-1, with summation over the first BZ of layer-1. $A_1(\bm{k},\epsilon)$ is the spectral function of layer-1 in $\bm{k}$, where $g_{n\bm{k}}^{(1)}$ signifies the total wave amplitudes on layer-1 within the eigenstate $|n\bm{k}\rangle$, associated with eigenvalue $\epsilon_{n \bm{k}}$. The essential ingredients, $|n\bm{k}\rangle$ and $\epsilon_{n \bm{k}}$, are obtained by diagonalizing an approximated Hamiltonian matrix constructed according to the method outlined in Ref.~\cite{Koshino_2015}:
\begin{itemize}
\item Starting with a specific state $|\bm{k}\rangle$ from layer-1, one consider all $|\tilde{\bm{k}}\rangle$ states in layer-2 that are directly coupled to $|\bm{k}\rangle$.
\item Neglecting exponentially small matrix elements, the basis of this matrix only contains a single state $|\bm{k}\rangle$ from layer-1 and a selected set of states $|\tilde{\bm{k}}\rangle$ from layer-2.
\item Diagonalizing the resulting approximated matrix provides energy eigenvalues $\epsilon_{n\bm{k}}$ and eigenstates $|n\bm{k}\rangle$ labeled by $n$.
\end{itemize}
With these eigenvalues and eigenstates, one can calculate the density of states $D(\epsilon)$ following Eq.~(\ref{eq_Koshino_D})-(\ref{eq_Koshino_A}), where the spectral function is defined through an ensemble-like average without any further explanation.

Koshino's scheme is highly successful. It provides valuable insights into the effect of interlayer coupling on the spectrum of bilayer systems. The central idea of this scheme is the direct partitioning of $D(\epsilon)$ into $A_1(\bm{k},\epsilon)$ and $A_2(\tilde{\bm{k}},\epsilon)$, obtained from an approximate truncated Hamiltonian. At the first time, it describes within the first order approximation of the quasi-band structure and the density of states of an incommensurate honeycomb lattice bilayer with a large rotation angle, which cannot be treated as a long-range moir\'e superlattice. However, this scheme relies heavily on physical intuition and involves several unproven assumptions, leading to remaining ambiguities.

One fundamental ambiguity is the legality of dividing the density of states for the entire bilayer into contributions from individual layers and further into contributions from each $\bm{k}$ and $\tilde{\bm{k}}$ points. Such decomposition is not expected on the first sight, as tight coupling exists between different layers so that $\mathcal{H} \neq \mathcal{H}_1 \oplus \mathcal{H}_2$. Furthermore, the appearance of the total wave amplitudes $g_{n\bm{k}}$ in Eq.~(\ref{eq_Koshino_A}) requires thorough examination. Can this term be derived without resorting to ensemble-like assumptions? Is this specific form of the weight factor unique?

Another ambiguity concerns the construction of the approximated Hamiltonian matrix. It seems like one can decompose the Hilbert space into non-orthogonal subspaces with substantial overlaps leading to redundancy. How can this decomposition be equivalent to the original Hilbert space?

Finally, one may ask whether this scheme exclusively applies to the density of states and, if so, what distinguishes it. Conversely, if this scheme is generally applicable, how can it be extended to calculate other physical observables?

These inquiries naturally arise in the pursuit of a comprehensive description of the electronic structure in multilayer systems. Our theory provides satisfactory responses to these queries, and clarifies all ambiguities with rigorous formalism, as elaborated in the main text.

\section{\label{app:velocity}Derivation of velocity operator matrix elements under the composite Bloch basis}
In this appendix, we provide a detailed derivation of the matrix elements of the velocity operator $\bm{v}$ under the composite Bloch basis $\coprod_{l}\{|\bm{k}_{l} \bm{t}_{l}\alpha\rangle\}$ for twisted multilayer systems.

We begin by deriving the intralayer velocity matrix elements, as expressed in Eq.~(\ref{eq_app_v_k}), starting from the definition of the velocity operator $\bm{v}=(i/\hbar)[\mathcal{H},\bm{r}]$ as follows:
\begin{widetext}
\begin{eqnarray}
\label{eq_intralayer_v0}
\langle\bm{k}^\prime_{l}\bm{t}^\prime_{l}\beta|\bm{v}|\bm{k}_{l}\bm{t}_{l}\alpha\rangle
&= & \frac{i}{N_l \hbar} \sum_{\bm{R}_l,\bm{R}^\prime_l}
e^{-i\bm{k}^\prime_l \cdot ( \bm{R}^\prime_l + \bm{t}^\prime_{l} )}
e^{i \bm{k}_l \cdot ( \bm{R}_l+\bm{t}_{l} )}
\langle \bm{R}^\prime_l \bm{t}^\prime_l \beta|\mathcal{H}\bm{r}- \bm{r}\mathcal{H}|\bm{R}_{l}\bm{t}_{l} \alpha \rangle, \nonumber \\
&=& \frac{i}{N_l \hbar} \sum_{\bm{R}_l,\bm{R}^\prime_l}
e^{-i\bm{k}^\prime_l \cdot ( \bm{R}^\prime_l + \bm{t}^\prime_{l} )}
e^{i \bm{k}_l \cdot ( \bm{R}_l+\bm{t}_{l} )}
\langle \bm{R}^\prime_l \bm{t}^\prime_l \beta|\mathcal{H}|\bm{R}_{l}\bm{t}_{l}\alpha \rangle (\bm{R}_{l}+\bm{t}_{l}-\bm{R}^\prime_l- \bm{t}^\prime_l ).
\end{eqnarray}
By applying the same variable transformation technique as used in Eq.~(\ref{eq_intralayer}), where we replace $\bm{R}_l$ with $\bm{R}_l+\bm{R}^\prime_l$ rendering $\sum_{\bm{R}_l+\bm{R}^\prime_l}$ equivalent to $\sum_{\bm{R}_l}$ and utilize the Poisson summation formula as $\sum_{\bm{R}^\prime_l} e^{i(\bm{k}_l-\bm{k}^\prime_l ) \cdot \bm{R}^\prime_l} =N_l\delta_{\bm{k}_l,\bm{k}^\prime_l}$, we have
\begin{eqnarray}
\langle\bm{k}^\prime_{l}\bm{t}^\prime_{l}\beta|\bm{v}|\bm{k}_{l}\bm{t}_{l}\alpha\rangle
&= & \frac{i}{N_l \hbar} \sum_{\bm{R}_l,\bm{R}^\prime_l} e^{-i\bm{k}^\prime_l \cdot ( \bm{R}^\prime_l + \bm{t}^\prime_{l} )} e^{i \bm{k}_l \cdot ( \bm{R}_l+\bm{R}^\prime_l+\bm{t}_{l} )} \langle \bm{R}^\prime_l \bm{t}^\prime_l \beta|\mathcal{H}|\bm{R}_{l}+\bm{R}^\prime_l, \bm{t}_{l} \alpha \rangle(\bm{R}_{l}+\bm{R}^\prime_l+\bm{t}_{l}-\bm{R}^\prime_l- \bm{t}^\prime_l ),\nonumber\\
&= & \frac{i}{N_l \hbar}  \sum_{\bm{R}^\prime_l} e^{i(\bm{k}_l-\bm{k}^\prime_l ) \cdot \bm{R}^\prime_l}  \sum_{\bm{R}_l} e^{i \bm{k}_l \cdot ( \bm{R}_l +\bm{t}_{l}-\bm{t}^\prime_{l} )} \langle \bm{0} \bm{t}^\prime_l \beta|\mathcal{H}|\bm{R}_{l} \bm{t}_{l} \alpha \rangle(\bm{R}_{l}+\bm{t}_{l}- \bm{t}^\prime_l ),\nonumber\\
&= &\frac{i}{\hbar} \delta_{\bm{k}_l, \bm{k}^\prime_l} \sum_{\bm{R}_l} e^{i \bm{k}_l \cdot ( \bm{R}_l +\bm{t}_l-\bm{t}^\prime_l )} \mathcal{H}_{\beta,\alpha}(\bm{R}_{l}+\bm{t}_l-\bm{t}^\prime_l)(\bm{R}_{l}+\bm{t}_{l}- \bm{t}^\prime_l ),\nonumber\\
&= &\frac{1}{\hbar} \nabla_{\bm{k}_l} \left(\sum_{\bm{R}_l} e^{i \bm{k}_l \cdot ( \bm{R}_l +\bm{t}_l-\bm{t}^\prime_l )} \mathcal{H}_{\beta,\alpha}(\bm{R}_{l}+\bm{t}_l-\bm{t}^\prime_l)\right) \delta_{\bm{k}_l, \bm{k}^\prime_l}.
\end{eqnarray}
This is Eq.~(\ref{eq_app_v_k}) after substituting Eq.~(\ref{eq_intralayer}) into it.

Now, we derive the interlayer velocity matrix elements Eq.~(\ref{eq_app_v_kp}). By definition, we have
\begin{eqnarray}
\label{eq_interlayer_v0}
\langle\bm{k}^\prime_{l^\prime}\bm{t}^\prime_{l^\prime}\beta|\bm{v}|\bm{k}_{l}\bm{t}_{l}\alpha\rangle
&=& \frac{i}{\sqrt{N_l {N}_{l^\prime}}\hbar}
\sum_{\bm{R}_l,\bm{R}^\prime_{l^\prime}}
e^{-i\bm{k}^\prime_{l^\prime} \cdot ( \bm{R}^\prime_{l^\prime} + \bm{t}^\prime_{l^\prime} )} e^{i\bm{k}_l \cdot (\bm{R}_l+\bm{t}_l)}
\langle \bm{R}^\prime_{l^\prime} \bm{t}^\prime_{l^\prime}\beta|\mathcal{H}\bm{r}- \bm{r}\mathcal{H}|\bm{R}_{l}\bm{t}_{l} \alpha \rangle
\nonumber\\
&=& \frac{i}{\sqrt{N_l {N}_{l^\prime}}\hbar}
\sum_{\bm{R}_l,\bm{R}^\prime_{l^\prime}}
e^{-i\bm{k}^\prime_{l^\prime} \cdot ( \bm{R}^\prime_{l^\prime} + \bm{t}^\prime_{l^\prime} )} e^{i\bm{k}_l \cdot (\bm{R}_l+\bm{t}_l)}
\mathcal{H}_{\beta,\alpha}(\bm{R}_{l}-\bm{R}^\prime_{l^\prime}+\bm{t}_{l}-\bm{t}^\prime_{l^\prime}+\Delta h_{l,l^\prime} \bm{e}_z) \nonumber\\
&\times&(\bm{R}_{l}-\bm{R}^\prime_{l^\prime}+\bm{t}_{l}-\bm{t}^\prime_{l^\prime}+\Delta h_{l,l^\prime} \bm{e}_z).
\end{eqnarray}
As we focus solely on the in-plane velocity components ($\mu,\nu=x,y$), we can omit the term involving $\Delta h_{l,l^\prime} \bm{e}_z$ and establish the Fourier transform of the dipole of interlayer hopping integral, following a similar approach to Eq.~(\ref{eq_H_fourier}):
\begin{equation}
\label{eq_rH_fourier}
\bm{r} \mathcal{H}_{\beta,\alpha}(\bm{r}+\Delta h_{l,l^\prime} \bm{e}_z) =  \frac{S}{\sqrt{N_l {N}_{l^\prime}}} \int \frac{\mathrm{d}^2 \bm{q}}{(2\pi)^2}  e^{-i \bm{q} \cdot \bm{r}} \bm{\mathcal{D}}^{l^\prime,l}_{\beta,\alpha}(\bm{q}),
\end{equation}
 where the dipole Fourier components $\bm{\mathcal{D}}^{l^\prime,l}_{\beta,\alpha}(\bm{q})$ are defined through the inverse transformation as follows:
\begin{equation}
\label{eq_rH_fourier_inv}
\bm{\mathcal{D}}^{l^\prime,l}_{\beta,\alpha}(\bm{q})
= \frac{1}{\sqrt{\Omega_l \Omega_{l^\prime}}}
\int \mathrm{d}^2 \bm{r}
e^{i \bm{q} \cdot \bm{r}}
\bm{r} \mathcal{H}_{\beta,\alpha}( \bm{r} + \Delta h_{l,l^\prime} \bm{e}_z ).
\end{equation}
Similar to the derivation of Eq.~(\ref{eq_interlayer}), the matrix element of the interlayer velocity operator is given by:
\begin{eqnarray}
\label{eq_interlayer_v1}
\langle\bm{k}^\prime_{l^\prime}\bm{t}^\prime_{l^\prime}\beta|\bm{v}|\bm{k}_{l}\bm{t}_{l}\alpha\rangle
&=& \frac{i}{\sqrt{N_l {N}_{l^\prime}}\hbar}
\sum_{\bm{R}_l,\bm{R}^\prime_{l^\prime}}
e^{-i\bm{k}^\prime_{l^\prime} \cdot ( \bm{R}^\prime_{l^\prime} + \bm{t}^\prime_{l^\prime} )} e^{i\bm{k}_l \cdot (\bm{R}_l+\bm{t}_l)}
\frac{S}{\sqrt{N_l {N}_{l^\prime}}} \int \frac{\mathrm{d}^2 \bm{q}}{(2\pi)^2}
e^{-i \bm{q} \cdot (\bm{R}_{l}-\bm{R}^\prime_{l^\prime}+\bm{t}_{l}-\bm{t}^\prime_{l^\prime})} \bm{\mathcal{D}}^{l^\prime,l}_{\beta,\alpha}(\bm{q}), \nonumber \\
&=&
\frac{iS}{N_l N_{l^\prime}\hbar}
\int\frac{\mathrm{d}^2 \bm{q}}{(2\pi)^2}
\bm{\mathcal{D}}^{l^\prime,l}_{\beta,\alpha}(\bm{q})
e^{i(\bm{k}_l-\bm{q}) \cdot \bm{t}_{l} -i (\bm{k}^\prime_{l^\prime}-\bm{q}) \cdot {\bm{t}}^\prime_{l^\prime}} \times\sum_{\bm{R}_l} e^{i(\bm{k}_l-\bm{q}) \cdot \bm{R}_l}
\sum_{\bm{R}^\prime_{l^\prime}} e^{i(-\bm{k}^\prime_{l^\prime} +\bm{q}) \cdot{\bm{R}_{l^\prime}}}, \nonumber \\
&=& \frac{iS}{N_l N_{l^\prime}\hbar} \frac{(2\pi)^2}{\Omega_l} N_{l^\prime}
\sum_{\bm{G}_l,\bm{G}^\prime_{l^\prime}} \int \frac{\mathrm{d}^2 \bm{q}}{(2\pi)^2}
\bm{\mathcal{D}}^{l^\prime,l}_{\beta,\alpha}(\bm{q})
e^{i(\bm{k}_l-\bm{q}) \cdot \bm{t}_{l} -i (\bm{k}^\prime_{l^\prime}-\bm{q}) \cdot {\bm{t}}^\prime_{l^\prime}}
\delta^{(2)}(\bm{k}_l -\bm{q}-\bm{G}_l)
\delta_{-\bm{k}^\prime_{l^\prime}+\bm{q},\bm{G}^\prime_{l^\prime}}, \nonumber \\
&=& \frac{i}{\hbar} \sum_{\bm{G}_l,\bm{G}^\prime_{l^\prime}}
\bm{\mathcal{D}}^{l^\prime,l}_{\beta,\alpha}(\bm{k}_l - \bm{G}_l)
e^{i\bm{G}_l \cdot \bm{t}_{l} -i (\bm{k}^\prime_{l^\prime}-\bm{k}_l+\bm{G}_l) \cdot \bm{t}^\prime_{l^\prime}}
\delta_{-\bm{k}^\prime_{l^\prime}+\bm{k}_l-\bm{G}_l,\bm{G}^\prime_{l^\prime}}, \nonumber \\
&= & \frac{i}{\hbar} \sum_{\bm{G}_l,\bm{G}^\prime_{l^\prime}}
\bm{\mathcal{D}}^{l^\prime,l}_{\beta,\alpha}(\bm{k}_l + \bm{G}_l)
e^{-i\bm{G}_l \cdot \bm{t}_{l} +i \bm{G}^\prime_{l^\prime} \cdot \bm{t}^\prime_{l^\prime}}
\delta_{\bm{k}_l+\bm{G}_l,\bm{k}^\prime_{l^\prime}+\bm{G}^\prime_{l^\prime}}.
\end{eqnarray}
\end{widetext}
Note that Eq.~(\ref{eq_rH_fourier_inv}) can be alternatively expressed as:
\begin{equation}
\label{eq_rH_fourier_inv_partial}
i \bm{\mathcal{D}}^{l^\prime,l}_{\beta,\alpha}(\bm{q})
= \nabla_{\bm{q}}
\left[\frac{1}{\sqrt{\Omega_l \Omega_{l^\prime}}}
\int \mathrm{d}^2 \bm{r}
e^{i \bm{q} \cdot \bm{r}}
\mathcal{H}_{\beta,\alpha}( \bm{r} + \Delta h_{l,l^\prime} \bm{e}_z )\right].
\end{equation}
Comparing it with Eq.~(\ref{eq_H_fourier_inv}), we immediately obtain:
\begin{equation}
i\bm{\mathcal{D}}^{l^\prime,l}_{\beta,\alpha}(\bm{q}) = \nabla_{\bm{q}} \left[\mathcal{V}^{l^\prime,l}_{\beta,\alpha}(\bm{q})\right].
\end{equation}
Together with Eq.~(\ref{eq_interlayer_v1}), this leads to Eq.~(\ref{eq_app_v_kp}), concluding the derivation.

\section{\label{app:TBG_matrix}Construction of local low-rank approximated Hamiltonian in incommensurate TBG}

In this appendix, we demonstrate the construction of local low-rank approximated Hamiltonian in incommensurate TBG. We omit considerations of internal cell degrees of freedom here for clarity, but their inclusion is straightforward. To simplify notation, we use $\bm{k}$ for wave vectors in layers-1 and $\bm{p}$ for layers-2. Therefore, the composite Bloch basis for layers-1 and layers-2 is represented as $\{ |\bm{k}\rangle, |\bm{p}\rangle |\bm{k} \in \mathrm{BZ}_1,\bm{p} \in \mathrm{BZ}_2 \}$. Within this basis, the Hamiltonian is represented as a $2N \times 2N$ matrix, denoted as $\mathcal{H} \in \mathbb{C}^{2N \times 2N}$. Here, $N$ is the rank of $\{ |\bm{k}\rangle\}$ (or $\{ |\bm{p}\rangle\}$), equivalent to the number of $\bm{k}$ points in the $\mathrm{BZ}_1$ with $N \to \infty$. Regardless, $\mathcal{H}$ can be expressed concisely as follows,
\begin{equation}
\mathcal{H} =
\begin{bmatrix}
    \mathcal{H}_1 & \mathcal{V}_{12}  \\
    \mathcal{V}_{21} & \mathcal{H}_2
\end{bmatrix}.
\end{equation}
Here, $\mathcal{H}_1,\mathcal{H}_2\in\mathbb{C}^{N\times N}$, are diagonal matrices representing the Hamiltonian for each layer. We also introduce the interlayer coupling matrices denoted as $\mathcal{V}_{12}={\mathcal{V}_{21}}^\dagger \in \mathbb{C}^{N\times N}$. Notably, the entries of these coupling matrices are generally non-zero, capturing the intricate interlayer coupling. Alternatively, the complete Hamiltonian $\mathcal{H}$ can be expanded as follows:
\begin{widetext}
\begin{equation}
\mathcal{H}=
\begin{bmatrix}
\mathcal{H}^1_{\bm{k}_1} & 0 & \dots & 0  & \mathcal{V}_{\bm{k}_1,\bm{p}_1} & \mathcal{V}_{\bm{k}_1,\bm{p}_2} & \dots & \mathcal{V}_{\bm{k}_1,\bm{p}_N} \\
0 & \mathcal{H}^1_{\bm{k}_2} & \dots & 0  & \mathcal{V}_{\bm{k}_2,\bm{p}_1} & \mathcal{V}_{\bm{k}_2,\bm{p}_2} & \dots & \mathcal{V}_{\bm{k}_2,\bm{p}_N} \\
\vdots & \vdots & \ddots & \vdots  & \vdots & \vdots & \ddots & \vdots \\
0 & 0 & \dots &  \mathcal{H}^1_{\bm{k}_N}  & \mathcal{V}_{\bm{k}_N,\bm{p}_1} & \mathcal{V}_{\bm{k}_N,\bm{p}_2} & \dots & \mathcal{V}_{\bm{k}_N,\bm{p}_N} \\
\mathcal{V}_{\bm{p}_1,\bm{k}_1} & \mathcal{V}_{\bm{p}_1,\bm{k}_2} & \dots & \mathcal{V}_{\bm{p}_1,\bm{k}_N} & \mathcal{H}^2_{\bm{p}_1} & 0 & \dots & 0 \\
\mathcal{V}_{\bm{p}_2,\bm{k}_1} & \mathcal{V}_{\bm{p}_2,\bm{k}_2} & \dots & \mathcal{V}_{\bm{p}_2,\bm{k}_N} & 0 & \mathcal{H}^2_{\bm{p}_2} & \dots & 0 \\
\vdots & \vdots & \ddots & \vdots  & \vdots & \vdots & \ddots & \vdots   \\
\mathcal{V}_{\bm{p}_N,\bm{k}_1} & \mathcal{V}_{\bm{p}_N,\bm{k}_2} & \dots & \mathcal{V}_{\bm{p}_N,\bm{k}_N} & 0 & 0 & \dots & \mathcal{H}^2_{\bm{p}_N}
\end{bmatrix}.
\end{equation}
\end{widetext}

Given a cutoff $G_{\mathrm{cut}}$, we obtain $N_{\mathrm{cut}}=M$ dominant interlayer elements of $\mathcal{V}_{\bm{k}_i,\bm{p}_j}$ ($j = i_1, i_2, \dots, i_M$) based on a cutoff of the coupling strength. This selection results in $M+1$ relevant columns and rows for $\bm{k}_i$. Without loss of generality, let's consider $\bm{k}_i=\bm{k}_N$, and rearrange the indices of the chosen $\bm{p}_j$ that are coupled to $\bm{k}_N$ from $j \in \{ N_1, N_2, \dots, N_M\}$ to $j\in \{ 1, 2, \dots, M\}$. This rearrangement results in the following column matrix $\mathcal{C}$
and the row submatrix $\mathcal{R}$:
\begin{widetext}
\setcounter{MaxMatrixCols}{20}
\begin{equation}
\label{eq_full_R}
\mathcal{C}^{\dagger}=\mathcal{R} =
\begin{bmatrix}
0&\dots &0&\mathcal{H}^1_{\bm{k}_N}&\mathcal{V}_{\bm{k}_N,\bm{p}_1}&\mathcal{V}_{\bm{k}_N,\bm{p}_2}&\dots&\mathcal{V}_{\bm{k}_N,\bm{p}_M}&\mathcal{V}_{\bm{k}_N,\bm{p}_{M+1}} &\dots & \mathcal{V}_{\bm{k}_N,\bm{p}_N} \\
\mathcal{V}_{\bm{p}_1,\bm{k}_1}&\dots &\mathcal{V}_{\bm{p}_1,\bm{k}_{N-1}}&\mathcal{V}_{\bm{p}_1,\bm{k}_N}&\mathcal{H}^2_{\bm{p}_1}       &0    &\dots &0&0&\dots & 0 \\
\mathcal{V}_{\bm{p}_2,\bm{k}_1}&\dots &\mathcal{V}_{\bm{p}_2,\bm{k}_{N-1}}&\mathcal{V}_{\bm{p}_2,\bm{k}_N}&0&\mathcal{H}^2_{\bm{p}_2}&\dots &0  &0&\dots & 0 \\
\vdots&\ddots&\vdots&\vdots&\vdots&\vdots&\ddots&\vdots&\vdots&\ddots& \vdots\\
\mathcal{V}_{\bm{p}_M,\bm{k}_1}&\dots &\mathcal{V}_{\bm{p}_M,\bm{k}_{N-1}}&\mathcal{V}_{\bm{p}_M,\bm{k}_N}&0&0&\dots &\mathcal{H}^2_{\bm{p}_M}  &0&\dots & 0
\end{bmatrix}.
\end{equation}
\end{widetext}
The remaining challenge is to construct the approximated Hamiltonian $\widetilde{H}(\bm{k}_N)$ using the column submatrix $\mathcal{C}\in \mathbb{C}^{2N\times(M+1)}$ and the row submatrix $\mathcal{R}\in \mathbb{C}^{(M+1)\times 2N}$.

As discussed in the main text, one direct approach of constructing $\widetilde{\mathcal{H}}(\bm{k}_N)$ is to take their intersection, denoted as $\widetilde{\mathcal{H}}_0(\bm{k}_N) \in \mathbb{C}^{(M+1)\times(M+1)}$, as proposed in Ref.~\cite{Koshino_2015} as the first-order approximation:
\begin{equation}
\label{eq_simple_Hk}
\widetilde{\mathcal{H}}_0(\bm{k}_N) =
\begin{bmatrix}
\mathcal{H}^1_{\bm{k}_N} & \mathcal{V}_{\bm{k}_N,\bm{p}_1} & \mathcal{V}_{\bm{k}_N,\bm{p}_2} & \dots & \mathcal{V}_{\bm{k}_N,\bm{p}_M} \\
\mathcal{V}_{\bm{p}_1,\bm{k}_N} & \mathcal{H}^2_{\bm{p}_1} & 0 & \dots & 0 \\
\mathcal{V}_{\bm{p}_2,\bm{k}_N} & 0 & \mathcal{H}^2_{\bm{p}_2} & \dots & 0 \\
\vdots  & \vdots & \vdots & \ddots & \vdots \\
\mathcal{V}_{\bm{p}_M,\bm{k}_M} & 0 & 0 & \dots & \mathcal{H}^2_{\bm{p}_M}
\end{bmatrix}.
\end{equation}
However, such method of obtaining $\widetilde{\mathcal{H}}_0(\bm{k}_N)$ by intersecting $\mathcal{C}$ and $\mathcal{R}$ may be considered overly simplistic. The combined column and row submatrices contain $(N+1)(M+1)$ non-zero entries, comprising $M+1$ intralayer entries and $N(M+1)$ independent interlayer entries. These interlayer entries can be divided into two categories: $M^2$ dominant relevant entries and $(N-M)(M+1)$ irrelevant ones. In contrast, Eq.(\ref{eq_simple_Hk}) utilizes only $M+1$ intralayer entries and $M$ interlayer entries, thereby neglecting $M(M-1)$ dominant relevant intralayer entries, which can lead to significant errors. To address this, we must account for these entries when constructing $\widetilde{\mathcal{H}}(\bm{k}_N)$.

In this work, we propose a method to construct the local approximated Hamiltonian $\widetilde{\mathcal{H}}(\bm{k}_N) \in \mathbb{C}^{2M\times2M}$ using $\bm{k}_N$ and its $M$ relevant counterparts $\bm{p}_j \in \{\bm{p}\}_{\bm{k}_N}$. Furthermore, we include the remaining $M-1$ \textit{most} dominant $\bm{k}_{i_j}=\bm{p}_j$ ($j=2,\dots,M$) coupled to $\bm{p}_j$. This consideration is based on the fact that the leading order coupling is given by $\bm{G}_1=\bm{G}^\prime_2=\bm{0}$, indicating the most dominate $|\bm{p}\rangle$ coupled to $|\bm{k}\rangle$ is located at $\bm{p}=\bm{k}$. Thus, the primary $\bm{k}$ associated with $\bm{p}_1$ is simply $\bm{k}_N$ itself. Consequently, the resulting local approximated Hamiltonian is
\begin{widetext}
\begin{equation}
\widetilde{\mathcal{H}}(\bm{k}_N) =
\begin{bmatrix}
\mathcal{H}^1_{\bm{k}_{i_M}} & \dots & 0 & 0 & \mathcal{V}_{\bm{k}_{i_M},\bm{p}_1} & \mathcal{V}_{\bm{k}_{i_M},\bm{p}_2} & \dots & \mathcal{V}_{\bm{k}_{i_M},\bm{p}_M} \\
\vdots & \ddots & \vdots & \vdots  & \vdots & \vdots & \ddots & \vdots   \\
0 & \dots & \mathcal{H}^1_{\bm{k}_{i_2}} & 0 & \mathcal{V}_{\bm{k}_{i_2},\bm{p}_1} & \mathcal{V}_{\bm{k}_{i_2},\bm{p}_2} & \dots & \mathcal{V}_{\bm{k}_{i_2},\bm{p}_M} \\
0 & \dots &0 & \mathcal{H}^1_{\bm{k}_N} & \mathcal{V}_{\bm{k}_N,\bm{p}_1} & \mathcal{V}_{\bm{k}_N,\bm{p}_2} & \dots & \mathcal{V}_{\bm{k}_N,\bm{p}_M} \\
\mathcal{V}_{\bm{p}_1,\bm{k}_{i_M}} & \dots & \mathcal{V}_{\bm{p}_1,\bm{k}_{i_2}} & \mathcal{V}_{\bm{p}_1,\bm{k}_N} & \mathcal{H}^2_{\bm{p}_1} & 0 & \dots & 0 \\
\mathcal{V}_{\bm{p}_2,\bm{k}_{i_M}} & \dots & \mathcal{V}_{\bm{p}_2,\bm{k}_{i_2}} & \mathcal{V}_{\bm{p}_2,\bm{k}_N} & 0 & \mathcal{H}^2_{\bm{p}_2} & \dots & 0 \\
 \vdots & \ddots & \vdots & \vdots  & \vdots & \vdots & \ddots & \vdots   \\
\mathcal{V}_{\bm{p}_1,\bm{k}_{i_M}} & \dots & \mathcal{V}_{\bm{p}_2,\bm{k}_{i_2}} & \mathcal{V}_{\bm{p}_M,\bm{k}_M} & 0 & 0 & \dots & \mathcal{H}^2_{\bm{p}_M}
\end{bmatrix}.
\end{equation}
\end{widetext}
Typically, the set $\{\bm{k}_N,\bm{k}_{i_2},\dots,\bm{k}_{i_M}\}$ is just the $M$ dominant $\bm{k}$ for $\bm{p}_1 = \bm{k}_N$, i.e., $\{\bm{k}\}_{\bm{p}_1 = \bm{k}_N}$.

Our method involves further selection of the most relevant interlayer coupling entries from the remaining entries in each column or row, serving as corrections to $\mathcal{H}_0(\bm{k}_i)$. Importantly, this approach only marginally increases computational complexity, dealing with a $2M\times2M$ matrix instead of a $(M+1)\times(M+1)$ one. Nevertheless, this modest computational cost yields a significant improvement in results, as demonstrated in the following simple example, where the necessity of introducing $\widetilde{\mathcal{H}}(\bm{k}_i)$ beyond $\widetilde{\mathcal{H}}_0(\bm{k}_i)$ becomes evident.

Consider a Hermitian matrix $\mathcal{H}$ in basis $\{|1\rangle,|2\rangle,|3\rangle,|4\rangle,|5\rangle,|6\rangle,|7\rangle,|8\rangle \}$:
\begin{equation}
\mathcal{H}=
\begin{bmatrix}
10 & 0 & 0 & 0 & 0.2 & 0.4 & 2 & 4 \\
0 & 22 & 0 & 0 & 0.4 & 0.2 & 4 & 2 \\
0 & 0 & 16 & 0 & 2 & 4 & 0.2 & 0.4 \\
0 & 0 & 0 & 28 & 4 & 2 & 0.4 & 0.2 \\
0.2 & 0.4 & 2 & 4 & 19 & 0 & 0 & 0\\
0.4 & 0.2 & 4 & 2 & 0 & 25 & 0 & 0\\
2 & 5 & 0.2 & 0.4 & 0 & 0 & 13 & 0\\
5 & 2 & 0.4 & 0.2 & 0 & 0 & 0 & 31\\
\end{bmatrix}.
\end{equation}
Suppose we are interested in the spectral function for state $|4\rangle$, donates as $A(4,\epsilon)$. This function's exact form is defined by
\begin{equation}
A(4,\epsilon)=\sum_{n=1}^8 | \langle4|\psi_n\rangle |^2 \delta(\epsilon-\epsilon_n),
\end{equation}
with $\{|\psi_n\rangle\}$ and $\{\epsilon_n\}$ representing the eigenvectors and energy eigenvalues of $\mathcal{H}$.
Upon analyzing the coupling entries, we identify $4$ and $2$ as dominant relevant coupling entries and consider $0.4$ and $0.2$ as irrelevant.
Based on the division of coupling entries, we select relevant columns and rows, resulting in column submatrix and row submatrix
\begin{equation}
\mathcal{C}^{\dagger}=
\mathcal{R}=
\begin{bmatrix}
0 & 0 & 0 & 28 & 4 & 2 & 0.4 & 0.2 \\
0.2 & 0.4 & 2 & 4 & 19 & 0 & 0 & 0\\
0.4 & 0.2 & 4 & 2 & 0 & 25 & 0 & 0
\end{bmatrix}.
\end{equation}
Then $\widetilde{\mathcal{H}}_0(4)$ is the intersection of $\mathcal{C}$ and $\mathcal{R}$:
\begin{equation}
\widetilde{\mathcal{H}}_0(4)=
\begin{bmatrix}
28 & 4 & 2 \\
4 & 19 & 0 \\
2 & 0 & 25
\end{bmatrix},
\end{equation}
Our local approximated matrix around $|4\rangle$ is given by
\begin{equation}
\widetilde{\mathcal{H}}(4)=
\begin{bmatrix}
16 & 0 & 2 & 4 \\
0  & 28 & 4 & 2 \\
2 & 4 & 19 & 0 \\
4 & 2 & 0 & 25
\end{bmatrix}.
\end{equation}

\begin{figure*}
\includegraphics[width=1.7\columnwidth, keepaspectratio]{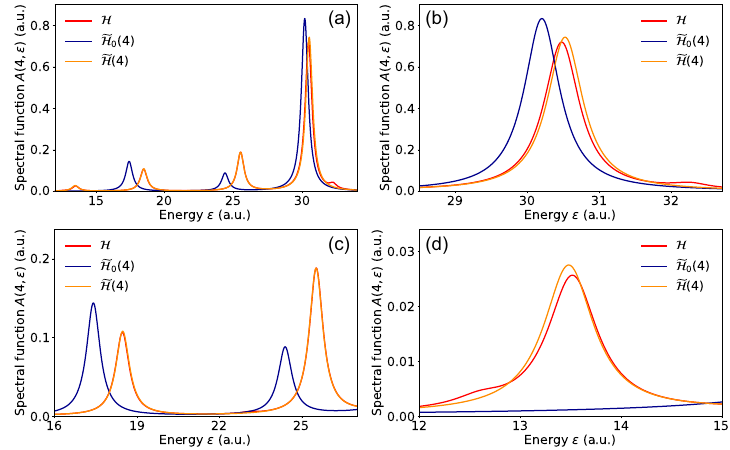}
\caption{\label{fig:8x8example} Spectral function $A(4,\epsilon)$ of matrix $\mathcal{H}$ (red line) along with its approximations $\widetilde{\mathcal{H}}_0(4)$ (blue line) and $\widetilde{\mathcal{H}}(4)$ (orange line). (a) Overview of $A(4,\epsilon)$ reveals a primary peak, two minor peaks, and two negligible peaks. Focusing on the primary peak (b) and the two minor peaks (c), and the left negligible peak (d) provides insights into the accuracy of the approximations.}
\end{figure*}

Our calculations yield $A(4,\epsilon)$ based on both $\widetilde{\mathcal{H}}(4)$ and $\widetilde{\mathcal{H}}_0(4)$, providing the approximated spectral function with and without high-order correction. The results are depicted in Fig.~\ref{fig:8x8example}, where we employ a Lorentzian function approximation of the Dirac delta function:
\begin{equation}
\delta(\epsilon-\epsilon_n) \to L(\epsilon)= \frac{1}{\pi} \frac{\Gamma}{(\epsilon-\epsilon_n)^2+\Gamma^2},
\end{equation}
with the broaden parameter $\Gamma=0.3$ for better illustration.

As demonstrated in Fig.~\ref{fig:8x8example}, we calculate the spectral function $A(4,\epsilon)$ calculated for matrix $\mathcal{H}$ (red line) along with its approximations $\widetilde{\mathcal{H}}_0(4)$ (blue line) and $\widetilde{\mathcal{H}}(4)$ (orange line). From the overview, $A(4,\epsilon)$ shows a primary peak, two minor peaks, and two negligible peaks, as shown in Fig.~\ref{fig:8x8example}(a).
At the primary peak from $|4\rangle$, both $\widetilde{\mathcal{H}}_0(4)$ and $\widetilde{\mathcal{H}}(4)$ offers a good approximation to $\mathcal{H}$, while $\widetilde{\mathcal{H}}(4)$ is better (see Fig.~\ref{fig:8x8example}(b)).
However, in the vicinity of the two minor peaks associated with states $|5\rangle$ and $|6\rangle$, $\widetilde{\mathcal{H}}_0(4)$ introduces significant errors due to the omission of relevant couplings between $|6\rangle$ and $|3\rangle$, also $|5\rangle$ and $|3\rangle$. When these couplings are reintroduced via correction, $\widetilde{\mathcal{H}}(4)$ becomes a notably better approximation, as shown in Fig.~\ref{fig:8x8example}(c). Additionally, at the left negligible peak, $\widetilde{\mathcal{H}}_0(4)$ provides nothing due to its oversimplification, while $\widetilde{\mathcal{H}}(4)$ faithfully captures the primary details, as shown in Fig.~\ref{fig:8x8example}(d). It's worth noting that $\widetilde{\mathcal{H}}(4)$ still can't describe the right negligible peak near $\epsilon=32$ (see Fig.~\ref{fig:8x8example}(b)), but it offers dimensionality reduction as a trade-off. Nevertheless, $\widetilde{\mathcal{H}}(4)$ serves as a sufficiently accurate local approximation that effectively characterizes the essential features of $|4\rangle$.

\section{\label{app:Ritz}Rayleigh-Ritz methods for eigenvalue problems}
In numerical linear algebra, the Rayleigh-Ritz method~\cite{trefethen2022numerical} approximates solutions to eigenvalue problems like $\mathcal{H} \bm{\psi}=\epsilon \bm{\psi}$, where $\mathcal{H}  \in \mathbb{C}^{N \times N}$. It provides approximations to eigenvalues and eigenvectors using a lower-dimensional projected matrix $\widetilde{\mathcal{H} } \in \mathbb{C}^{m \times m}$ where $m<N$.
This matrix $\widetilde{\mathcal{H}}$ is generated from a given transform matrix $\mathcal{S}\in \mathbb{C}^{N \times m}$ with orthonormal columns. In general, the Rayleigh-Ritz method involves these steps:
\begin{itemize}
\item Given a transform matrix $\mathcal{S}$, compute the projected matrix
\begin{equation}
\widetilde{\mathcal{H}}=\mathcal{S}^{\dagger} \mathcal{H}\mathcal{S}.
\end{equation} 
\item Solve the reduced eigenvalue problem $\widetilde{\mathcal{H}} \bm{\phi}_i=\mu_i \bm{\phi}_i$, obtaining eigenvalues $\mu_i$ and eigenvectors $\bm{\phi}_i$, where $i=1,2,\dots,m$.
\item Use the solutions to derive Ritz vectors and the associated Ritz values, denoted as
\begin{eqnarray}
\tilde{\bm{\psi}}_i=\mathcal{S} \bm{\phi}_i,\quad \text{and}\quad \tilde{\epsilon}_i=\mu_i.
\end{eqnarray}
The Ritz pairs $(\tilde{\epsilon}_i, \tilde{\bm{\psi}}_i)$ provide valuable estimates for the eigenvalues and eigenvectors of the original matrix $\mathcal{H}$.
\end{itemize}

Note that for a given Hamiltonian $\mathcal{H}$, all the physical considerations are encapsulated in the choice of $\mathcal{S}$, connecting the original Hilbert space with the subspace of interest.
In the context of the Rayleigh-Ritz method, this subspace is characterized by an orthonormal basis formed by the columns of the matrix $\mathcal{S}\in \mathbb{C}^{N \times m}$.
If this subspace includes $l \leq m$ vectors closely resembling the eigenvectors of the matrix $\mathcal{H}$, the Rayleigh-Ritz procedure can identify $l$ Ritz vectors that provide highly accurate approximations of these eigenvectors.
The accuracy of each Ritz pair can be quantified by $\|\mathcal{H}\tilde{\bm{\psi}}_i-\tilde{\epsilon}_i \tilde{\bm{\psi}}_i\|$, which measures the quality of approximation associated for each specific Ritz pair.

In the simplest scenario where $m=1$, $\mathcal{S}$ reduces to a unit column-vector $\bm{s}$. Consequently, $\widetilde{\mathcal{H}}$ collapses into a scalar quantity, which is precisely the Rayleigh quotient~\cite{SymmetricEigenvalue}:
\begin{equation}
\rho(\bm{s})=\bm{s}^\dagger\mathcal{H}\bm{s} / \bm{s}^\dagger \bm{s}.
\end{equation}
An insightful link to the Rayleigh quotient is established through the relationship $\mu_i=\rho(\bm{s}_i)$, that holds true for each Ritz pair $(\tilde{\epsilon}_i, \tilde{\bm{\psi}}_i)$. This connection enables us to deduce several properties of Ritz values $\mu_i$ by leveraging the well-established theory associated with the Rayleigh quotient~\cite{SymmetricEigenvalue}. For instance, when $\mathcal{H}$ is a Hermitian matrix, its Rayleigh quotient, and consequently, each Ritz value, assumes a real value and resides within the closed interval bounded by the smallest and largest eigenvalues of $\mathcal{H}$. Therefore, the spectrum of $\mathcal{H}$ can be effectively described by the set of Ritz values.
\begin{figure*}
\includegraphics[width=1.6\columnwidth, keepaspectratio]{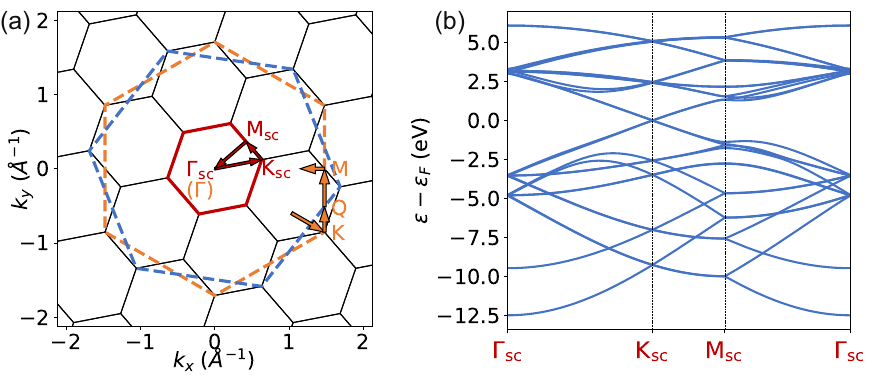}
\caption{\label{fig:21.79_BZ} (a) The moir\'e BZ of 21.79$^\circ$ commensurate TBG is outlined in red solid lines, with black solid lines for extending it. The blue and orange dotted lines represent boundaries of $\mathrm{BZ}_1$ and $\mathrm{BZ}_2$, respectively. (b) Band structure of 21.79$^\circ$ TBG calculated within the moir\'e supercell. The $\bm{k}$-path in the moir\'e BZ is highlighted by red arrows in (b). The parameters used are $V^0_{pp\pi}=-2.7$ eV, $V^0_{pp\sigma}=0.48$ eV, and $d_0=0.184 a$.}
\end{figure*}

\section{\label{app:numerical_TBG}Details of calculations for commensurate TBG}
Here, we present the details of calculations for commensurate TBG. The atomic structure of commensurate TBG is obtained by twisting an AA stacking bilayer graphene with a commensurate angle $\theta$, resulting in 2D primitive real lattice vectors $\bm{a}^{(1,2)}_{1}= a(\sqrt{3}/2,\mp 1/2)$ for layer-1 and $\bm{a}^{(1,2)}_{2}=\mathcal{R}(\theta) \bm{a}^{(1,2)}_{1}$ for layer-2 with $a=2.46$~\AA. Note that all possible commensurate angles with $0<\theta<\pi /3$ in TBG can be determined by two co-prime positive integers~\cite{PhysRevB.86.155449}:
\begin{equation}
\label{eq_app_theta_mr}
\theta(m,r) = \arccos \left(\frac{3m^2+3mr+r^2/2}{3m^2+3mr+r^2}\right).
\end{equation}
For the specific case of $m=r=1$, we obtain $\theta(1,1) \approx 21.79^\circ$, which leads to the formation of a moir\'e superlattice, depicted in Fig.~\ref{fig:GmG_both}(a). The corresponding moir\'e BZ is also depicted in Fig.~\ref{fig:21.79_BZ}(a).

With the atomic structure, we construct the tight-binding model utilizing a single $p_z$ orbital, where the hopping integral is given by Slater-Koster parameterization~\cite{PhysRev.94.1498}:
\begin{equation}
\mathcal{H}_{p_z,p_z}(\bm{d})= (1-n^2)V_{pp\pi}+n^2 V_{pp\sigma},
\end{equation}
where $\bm{d}$ is the distance vector, $n=d_z/|\bm{d}|$ is the direction cosine along $z$ axis, and $V_{pp\pi}$, $V_{pp\sigma}$ are Slater-Koster parameters with exponential concerning distance:
\begin{eqnarray}
V_{pp\pi}&=&V_{pp\pi}^0 e^{-(|\bm{d}|-a_0)/d_0}\Theta(d_{\mathrm{max}}-|\bm{d}|), \\
V_{pp\sigma}&=&V_{pp\sigma}^0 e^{-(|\bm{d}|-h_0)/d_0}\Theta(d_{\mathrm{max}}-|\bm{d}|).
\end{eqnarray}
Here, $a_0=a/\sqrt{3}$ is the bond length of the nearest-neighboring atoms, $h_0=3.35$~\AA~is the layer distance, $d_0$ is the decay length of the hopping integral chosen to be $0.184a$, and $d_{\mathrm{max}}$ is the neighboring cutoff chosen to be 5 \AA.

\begin{figure}
\includegraphics[width=1\columnwidth, keepaspectratio]{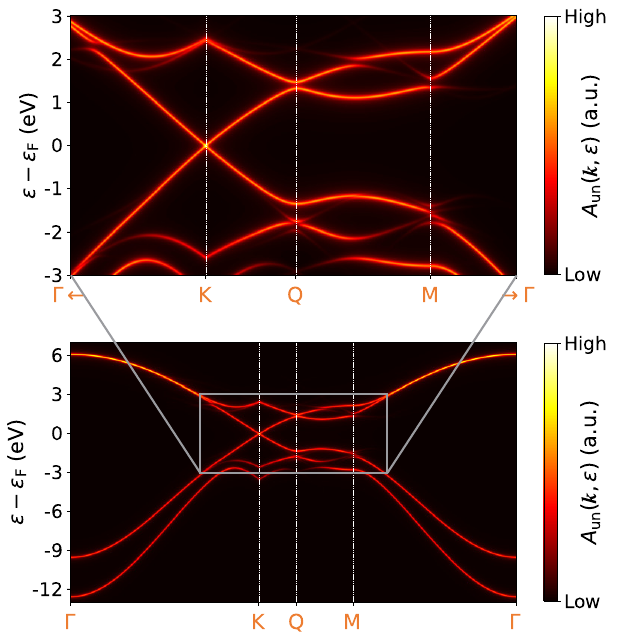}
\caption{\label{fig:21.79_unfold} Unfolding quasi-band structure of 21.79$^\circ$ TBG calculated within the moir\'e supercell. The $\bm{k}$-path in the $\mathrm{BZ}_1$ is highlighted by orange arrows in Fig.~\ref{fig:21.79_BZ}(a). The parameters used are same as Fig.~\ref{fig:21.79_BZ}(b).}
\end{figure}

The band structure of the $21.79^\circ$ TBG is shown in Fig.~(\ref{fig:21.79_BZ})(b). This band structure is calculated in a large supercell, resulting in a smaller BZ and folded dense energy bands. This complicates direct comparisons with ARPES measurements. To address this, we need to perform band unfolding. However, traditional unfolding techniques simply assume a single primitive-cell BZ, leading to unphysical ghost bands~\cite{PhysRevMaterials.2.010801}. To overcome this problem, the projection methods have been generalized to the two-periodicity case in previous research~\cite{PhysRevMaterials.2.010801, PhysRevB.95.085420}. In this work, we propose an unfolding spectra formula based on basis transformation for multi-periodicity twisted multilayer systems. The unfolding spectral function $A_{\mathrm{un}}(\bm{k},\epsilon)$ is defined as follows:
\begin{widetext}
\begin{equation}
\label{eq_unfold_A}
 A_{\mathrm{un}}(\bm{k},\epsilon) =  \sum_{l}\frac{N_{\mathrm{sc}}}{N_l}
\sum_{n \bm{t}_{l}\alpha}
\sum_{ \bm{R}^{\mathrm{sc}}_{l},\bm{R}^{\prime\mathrm{sc}}_{l}}
\delta(\epsilon-\epsilon_{\bm{k}_{\mathrm{sc}}(\bm{k}) n})
e^{i \bm{k} \cdot( \bm{R}^{\mathrm{sc}}_{l}-\bm{R}^{\prime\mathrm{sc}}_{l})}
b_{ n \bm{R}^{\mathrm{sc}}_{l}\bm{t}_{l} \alpha}^{*}(\bm{k}_{\mathrm{sc}}(\bm{k}))
b_{ n \bm{R}^{\prime\mathrm{sc}}_{l}\bm{t}_{l} \alpha}(\bm{k}_{\mathrm{sc}}(\bm{k})).
\end{equation}
\end{widetext}
For detailed derivation and notations, please refer to the following Appendix~\ref{app:unfold}. We calculate the unfolding spectra using Eq.(\ref{eq_unfold_A}) and present the results in Fig.~(\ref{fig:21.79_unfold}). This result serves as a benchmark of the quasi-band structure to validate our theory with supercell calculations, as discussed in the main text.

\begin{figure*}
\includegraphics[width=2\columnwidth, keepaspectratio]{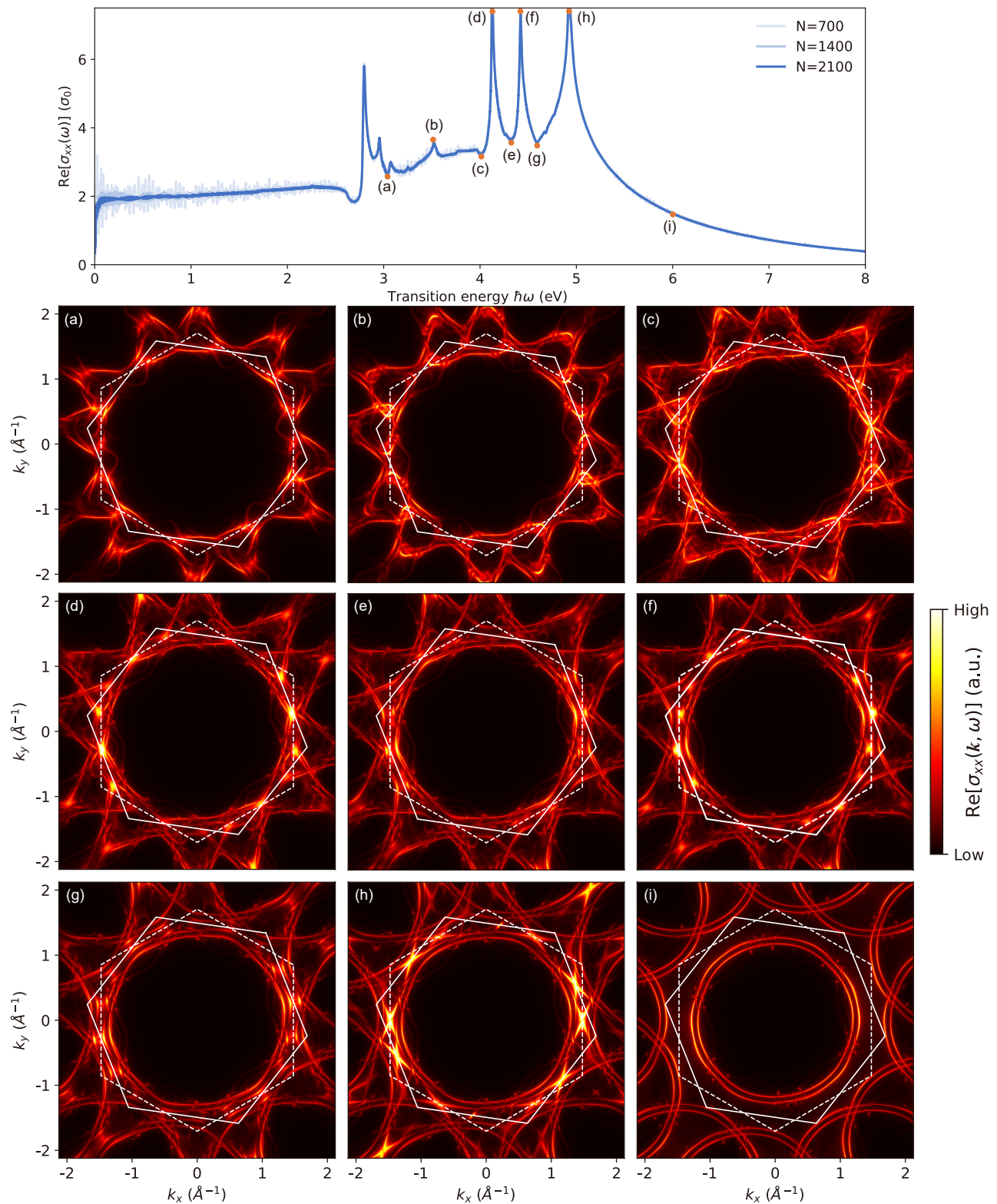}
\caption{\label{fig:more_ReS_cuts} Optical conductivity $\mathrm{Re}[\sigma_{xx}(\omega)]$ and its profiles at various transition energies (highlighted by orange dots) for 21.79$^\circ$ TBG obtained from our theory with cutoffs $G_{\mathrm{cut}}=12$~\AA$^{-1}$: (a) $\hbar \omega=3.04$ eV, (b) $\hbar \omega=3.52$ eV, (c) $\hbar \omega=4.01$ eV, (d) $\hbar \omega=4.13$ eV, (e) $\hbar \omega=4.33$ eV, (f) $\hbar \omega=4.42$ eV, (g) $\hbar \omega=4.59$ eV, (h) $\hbar \omega=4.92$ eV, and (i) $\hbar \omega=6.00$ eV. The white solid and dotted lines represent boundaries of $\mathrm{BZ}_1$ and $\mathrm{BZ}_2$, respectively.}
\end{figure*}

\section{\label{app:unfold}Band unfolding from the perspective of basis transformation}
Here, we explore band unfolding through basis transformation. Our objective is to compute unfolding spectra using supercell eigenstates and energy eigenvalues. We'll start by defining the spectral function and then explore its unfolding version.

The definition of spectral function naturally arises from the calculation of density of states, denoted as $D(\epsilon)$. Consider the density of states operator $\mathcal{D}=\delta(\epsilon - \mathcal{H})$, we obtain the density of states as follows:
\begin{equation}
\label{eq_app_Dn}
D(\epsilon) = \sum_{n} \delta(\epsilon-\epsilon_{n}),
\end{equation}
where $n$ labels the eigenstates $|n\rangle$ with energy eigenvalues $\epsilon_{n}$, and we can express it alternatively as a trace of $\mathcal{D}$:
\begin{equation}
\label{eq_app_De}
D(\epsilon) = \mathrm{Tr} (\delta(\epsilon- \mathcal{H})).
\end{equation}
When the trace operation is related to the eigenstates basis $\{|n\rangle \}$, Eq.~(\ref{eq_app_De}) reduces to Eq.~(\ref{eq_app_Dn}) as:
\begin{eqnarray}
D(\epsilon) = \sum_{n} \langle n| \delta(\epsilon-\mathcal{H})|n\rangle = \sum_{n} \delta(\epsilon-\epsilon_{n}).
\end{eqnarray}
Additionally, when the eigenstates basis is labeled by $\bm{k} n$ instead of $n$, we perform the trace operation using the eigenstates basis $\{ |\bm{k}n\rangle \}$, leading to
\begin{equation}
D(\epsilon) = \sum_{\bm{k}} \sum_{n}
\delta(\epsilon-\epsilon_{\bm{k} n}),
\end{equation}
where the spectral function $A(\bm{k},\epsilon)$ naturally emerges as the weight of $\bm{k}$-resolved density of states at $\bm{k}$:
\begin{equation}
A(\bm{k},\epsilon) = \sum_{n} \delta(\epsilon-\epsilon_{\bm{k} n}).
\end{equation}
However, the trace operation is invariant under orthonormal basis transformations. Therefore, the eigenstate basis is not a necessity, and we can perform the trace operation using any orthonormal basis, consistently yielding the same results. With this in mind, we proceed to derive the unfolding spectral function.

To introduce the unfolding spectral function, we use information from the supercell eigenstates basis $\{ |\bm{k}_{\mathrm{sc}} n\rangle \}$ and corresponding energy eigenvalues $\{ \epsilon_{\bm{k}_\mathrm{sc} n} \}$. This yields the folding spectral function, or the supercell spectral function, denoted as $A_{\mathrm{sc}}(\bm{k}_{\mathrm{sc}},\epsilon)$:
\begin{equation}
\label{eq_app_A_sc}
A_{\mathrm{sc}}(\bm{k}_{\mathrm{sc}},\epsilon) = \sum_{n} \delta(\epsilon-\epsilon_{\bm{k}_{\mathrm{sc}} n}).
\end{equation}
We can alternatively use the composite Bloch basis $\coprod_{l}\{|\bm{k}_{l} \bm{t}_{l}\alpha\rangle\}$ of the twisted multilayer system to perform the trace operation, resulting in:
\begin{equation}
\label{eq_app_D_l}
D(\epsilon) = \sum_{l}\sum_{\bm{k}_l}\sum_{\bm{t}_{l}\alpha} \langle \bm{k}_{l} \bm{t}_{l}\alpha| \delta(\epsilon-\mathcal{H})|\bm{k}_{l} \bm{t}_{l}\alpha\rangle.
\end{equation}
Similarly, we can define the unfolding spectral function $A_l(\bm{k}_l,\epsilon)$ for layer-$l$ to represent the weight of $D(\epsilon)$ at $\bm{k}_l$ for that layer:
\begin{equation}
\label{eq_app_A_l}
A_l(\bm{k}_l,\epsilon) = \sum_{\bm{t}_{l}\alpha} \langle \bm{k}_{l} \bm{t}_{l}\alpha| \delta(\epsilon-\mathcal{H})|\bm{k}_{l} \bm{t}_{l}\alpha\rangle.
\end{equation}
Therefore, the summation of all layers at the same $\bm{k}$ point yields the final unfolding spectral function $A_{\mathrm{un}}(\bm{k},\epsilon) $ consistent with ARPES results:
\begin{equation}
\label{eq_app_A}
A_{\mathrm{un}}(\bm{k},\epsilon) = \sum_{l} A_l(\bm{k}_l=\bm{k},\epsilon).
\end{equation}

Next, we demonstrate the calculation of Eq.~(\ref{eq_app_A_l}) using supercell eigenstates, adopting the cell gauge for simplicity. Thus, the reciprocal basis of the supercell is denoted as $\{ |\bm{k}_{\mathrm{sc}}  \bm{R}^{\mathrm{sc}}_l  \bm{t}_l \alpha \rangle\}$ with:
\begin{equation}
|\bm{k}_{\mathrm{sc}}  \bm{R}^{\mathrm{sc}}_l  \bm{t}_l \alpha \rangle
= \frac{1}{\sqrt{N_{\mathrm{sc}}}} \sum_{\bm{R}_{\mathrm{sc}}} e^{i\bm{k}_{\mathrm{sc}} \cdot \bm{R}_{\mathrm{sc}}} |\bm{R}_{\mathrm{sc}}  \bm{R}^{\mathrm{sc}}_l \bm{t}_l \alpha \rangle,
\end{equation}
where $\bm{R}_{\mathrm{sc}}$ represents the lattice vectors of the supercell, and $\bm{R}^{\mathrm{sc}}_l$ donates the single-layer lattice vectors of layer-$l$ within the supercell.  Let's consider the eigenstates of the supercell, which are denoted as $\{ |\bm{k}_{\mathrm{sc}} n\rangle \}$:
\begin{equation}
|\bm{k}_{\mathrm{sc}} n\rangle
= \sum_{ l \bm{R}^{\mathrm{sc}}_l \bm{t}_l \alpha } b_{ n \bm{R}_l \bm{t}_l \alpha}(\bm{k}_{\mathrm{sc}}) |\bm{k}_{\mathrm{sc}}  \bm{R}^{\mathrm{sc}}_l  \bm{t}_l \alpha \rangle.
\end{equation}
The first step involves inserting identity operator $\sum_{\bm{k}_{\mathrm{sc}} n} |\bm{k}_{\mathrm{sc}} n\rangle \langle \bm{k}_{\mathrm{sc}} n| = \mathbbm{1}$ into Eq.~(\ref{eq_app_A_l}):
\begin{eqnarray}
\label{eq_app_Al}
A_l(\bm{k}_l,\epsilon)
&=& \sum_{\bm{t}_{l}\alpha} \langle \bm{k}_{l} \bm{t}_{l}\alpha| \delta(\epsilon-\mathcal{H}) \sum_{\bm{k}_{\mathrm{sc}} n} |\bm{k}_{\mathrm{sc}} n\rangle \langle \bm{k}_{\mathrm{sc}} n|\bm{k}_{l} \bm{t}_{l}\alpha\rangle , \nonumber \\
&=& \sum_{\bm{t}_{l}\alpha} \sum_{\bm{k}_{\mathrm{sc}} n}
|\langle \bm{k}_{l} \bm{t}_{l}\alpha  |\bm{k}_{\mathrm{sc}} n \rangle|^2
\delta(\epsilon-\epsilon_{\bm{k}_{\mathrm{sc}} n}).
\end{eqnarray}
Thus, we only need to calculate the overlap between the supercell eigenstates and the single-layer Bloch basis $\langle \bm{k}_{l} \bm{t}_{l}\alpha  |\bm{k}_{\mathrm{sc}} n \rangle$, which can be obtained by directly expanding:
\begin{equation}
\label{eq_app_overlap_kn}
\langle \bm{k}_{l} \bm{t}_{l}\alpha  |\bm{k}_{\mathrm{sc}} n \rangle
= \sum_{ l^\prime \bm{R}^{\prime\mathrm{sc}}_{l^\prime} \bm{t}^\prime_{l^\prime} \beta}
b_{n \bm{R}^\prime_{l^\prime}\bm{t}^\prime_{l^\prime} \beta}(\bm{k}_{\mathrm{sc}})
\langle\bm{k}_{l}\bm{t}_{l}\alpha|\bm{k}_{\mathrm{sc}}\bm{R}^{\prime\mathrm{sc}}_{l^\prime}\bm{t}^\prime_{l^\prime}\beta\rangle.
\end{equation}
\begin{widetext}
Therefore, the question becomes calculating $\langle\bm{k}_{l}\bm{t}_{l}\alpha|\bm{k}_{\mathrm{sc}}\bm{R}^{\prime\mathrm{sc}}_{l^\prime}\bm{t}^\prime_{l^\prime}\beta\rangle$, which is the Bloch basis overlap between the supercell and single-layer cell:
\begin{eqnarray}
\label{eq_app_overlap_kk}
\langle\bm{k}_{l}\bm{t}_{l}\alpha|\bm{k}_{\mathrm{sc}}\bm{R}^{\prime\mathrm{sc}}_{l^\prime}\bm{t}^\prime_{l^\prime} \beta\rangle
&=& \frac{1}{\sqrt{N_l}} \sum_{\bm{R}_l} e^{-i \bm{k}_l \cdot \bm{R}_l} \langle \bm{R}_l \bm{t}_l \alpha |
\frac{1}{\sqrt{N_{\mathrm{sc}}}} \sum_{\bm{R}_{\mathrm{sc}}}
e^{i\bm{k}_{\mathrm{sc}} \cdot \bm{R}_{\mathrm{sc}}} |\bm{R}_{\mathrm{sc}} \bm{R}^{\prime\mathrm{sc}}_{l^\prime}\bm{t}^\prime_{l^\prime}\beta\rangle, \nonumber\\
&=& \frac{1}{\sqrt{N_l N_{\mathrm{sc}}}}  \sum_{\bm{R}_l} \sum_{\bm{R}_{\mathrm{sc}}}
e^{-i \bm{k}_l \cdot \bm{R}_l} e^{i\bm{k}_{\mathrm{sc}} \cdot \bm{R}_{\mathrm{sc}}}
\langle\bm{R}_l\bm{t}_l \alpha |\bm{R}_{\mathrm{sc}} \bm{R}^{\prime\mathrm{sc}}_{l^\prime}\bm{t}^\prime_{l^\prime}\beta\rangle.
\end{eqnarray}
So, we just need the real space overlap of atomic basis, denoted as  $\langle \bm{R}_l\bm{t}_l \alpha |\bm{R}_{\mathrm{sc}} \bm{R}^{\prime\mathrm{sc}}_{l^\prime}\bm{t}^\prime_{l^\prime}\beta\rangle$, to obtain all the necessary quantities leading to:
\begin{equation}
A_l(\bm{k}_l,\epsilon)
= \sum_{\bm{t}_{l}\alpha} \sum_{\bm{k}_{\mathrm{sc}} n}
\left|\sum_{ l^\prime \bm{R}^{\prime\mathrm{sc}}_{l^\prime} \bm{t}^\prime_{l^\prime} \beta}
b_{n \bm{R}^\prime_{l^\prime}\bm{t}^\prime_{l^\prime} \beta}(\bm{k}_{\mathrm{sc}}) \frac{1}{\sqrt{N_l N_{\mathrm{sc}}}}  \sum_{\bm{R}_l} \sum_{\bm{R}_{\mathrm{sc}}}
e^{-i \bm{k}_l \cdot \bm{R}_l} e^{i\bm{k}_{\mathrm{sc}} \cdot \bm{R}_{\mathrm{sc}}}
\langle\bm{R}_l\bm{t}_l \alpha |\bm{R}_{\mathrm{sc}} \bm{R}^{\prime\mathrm{sc}}_{l^\prime}\bm{t}^\prime_{l^\prime}\beta\rangle\right|^2
\delta(\epsilon-\epsilon_{\bm{k}_{\mathrm{sc}} n}).
\end{equation}

Specifically, in the tight-binding model, we work with the orthonormal real-space atomic basis that meets the orthonormality conditions:
\begin{equation}
\langle \bm{R}_l \bm{t}_l \alpha |\bm{R}_{\mathrm{sc}} \bm{R}^{\prime\mathrm{sc}}_{l^\prime}\bm{t}^\prime_{l^\prime}\beta\rangle =
\delta_{l,l^\prime}
\delta_{\bm{R}_l,\bm{R}_{\mathrm{sc}}+\bm{R}^{\prime\mathrm{sc}}_{l^\prime}}
\delta_{\bm{t}_l,\bm{t}^\prime_{l^\prime}}
\delta_{\alpha,\beta}.
\end{equation}
Thus, we can simplify Eq.~(\ref{eq_app_overlap_kk}) as
\begin{eqnarray}
\langle\bm{k}_{l}\bm{t}_{l}\alpha|\bm{k}_{\mathrm{sc}}\bm{R}^{\prime\mathrm{sc}}_l\bm{t}^\prime_l\beta\rangle
&=& \frac{1}{\sqrt{N_l N_{\mathrm{sc}}}}  \sum_{\bm{R}_l} \sum_{\bm{R}_{\mathrm{sc}}}
e^{-i \bm{k}_l \cdot \bm{R}_l} e^{i\bm{k}_{\mathrm{sc}} \cdot \bm{R}_{\mathrm{sc}}}
\delta_{l,l^\prime}
\delta_{\bm{R}_l,\bm{R}_{\mathrm{sc}}+\bm{R}^{\prime\mathrm{sc}}_{l^\prime}}
\delta_{\bm{t}_l,\bm{t}^\prime_{l^\prime}}
\delta_{\alpha,\beta}, \nonumber\\
&=& \frac{1}{\sqrt{N_l N_{\mathrm{sc}}}} \sum_{\bm{R}_{\mathrm{sc}}}
e^{-i \bm{k}_l \cdot (\bm{R}_{\mathrm{sc}}+\bm{R}^{\prime\mathrm{sc}}_{l^\prime})} e^{i\bm{k}_{\mathrm{sc}} \cdot \bm{R}_{\mathrm{sc}}}
\delta_{l,l^\prime}
\delta_{\bm{t}_l,\bm{t}^\prime_{l^\prime}}
\delta_{\alpha,\beta},\nonumber\\
&=& \sqrt{\frac{N_{\mathrm{sc}}}{N_l }} \sum_{\bm{G}_{\mathrm{sc}}}
e^{-i \bm{k}_l \cdot \bm{R}^{\prime\mathrm{sc}}_{l^\prime}}
\delta_{\bm{k}_{\mathrm{sc}}-\bm{k}_l,\bm{G}_{\mathrm{sc}}}
\delta_{l,l^\prime}
\delta_{\bm{t}_l,\bm{t}^\prime_{l^\prime}}
\delta_{\alpha,\beta},
\end{eqnarray}
where we use the Poisson summation formula, Eq.~(\ref{eq_appen_Poisson_delta_final}), as $\sum_{\bm{R}_{\mathrm{sc}}} e^{i(\bm{k}_{\mathrm{sc}}-\bm{k}_l)\cdot \bm{R}_{\mathrm{sc}}}=N_{\mathrm{sc}} \sum_{\bm{G}_{\mathrm{sc}}} \delta_{\bm{k}_{\mathrm{sc}}-\bm{k}_l,\bm{G}_{\mathrm{sc}}}$. This result will further simplify Eq.~(\ref{eq_app_overlap_kn}) as
\begin{eqnarray}
\langle \bm{k}_{l} \bm{t}_{l}\alpha  |\bm{k}_{\mathrm{sc}} n \rangle
& = & \sum_{ l^\prime \bm{R}^{\prime\mathrm{sc}}_{l^\prime} \bm{t}^\prime_{l^\prime} \beta }
b_{n \bm{R}^{\prime\mathrm{sc}}_{l^\prime} \bm{t}^\prime_{l^\prime} \beta}(\bm{k}_{\mathrm{sc}})
\sqrt{\frac{N_{\mathrm{sc}}}{N_l }} \sum_{\bm{G}_{\mathrm{sc}}}
e^{-i \bm{k}_l \cdot \bm{R}^{\prime\mathrm{sc}}_{l^\prime}}
\delta_{\bm{k}_{\mathrm{sc}}-\bm{k}_l,\bm{G}_{\mathrm{sc}}}
\delta_{l,l^\prime}
\delta_{\bm{t}_l,\bm{t}^\prime_{l^\prime}}
\delta_{\alpha,\beta},\nonumber\\
& = &  \sqrt{\frac{N_{\mathrm{sc}}}{N_l }}
\sum_{ \bm{R}^{\prime\mathrm{sc}}_{l}}
\sum_{\bm{G}_{\mathrm{sc}}}
b_{ n \bm{R}^\prime_{l}\bm{t}_{l} \alpha}(\bm{k}_{\mathrm{sc}})
e^{-i \bm{k}_l \cdot \bm{R}^{\prime\mathrm{sc}}_{l}}
\delta_{\bm{k}_{\mathrm{sc}}-\bm{k}_l,\bm{G}_{\mathrm{sc}}}.
\end{eqnarray}
Therefore, we arrive at the final expression for unfolding spectra of layer-$l$ by substituting the above result into Eq.~(\ref{eq_app_Al}), which reads
\begin{eqnarray}
A_l(\bm{k}_l,\epsilon)
&=& \sum_{\bm{t}_{l}\alpha} \sum_{\bm{k}_{\mathrm{sc}} n}
\delta(\epsilon-\epsilon_{\bm{k}_{\mathrm{sc}} n})
\frac{N_{\mathrm{sc}}}{N_l }
\sum_{ \bm{R}^{\mathrm{sc}}_{l},\bm{R}^{\prime\mathrm{sc}}_{l}}
\sum_{\bm{G}_{\mathrm{sc}},\bm{G}^\prime_{\mathrm{sc}}}
b_{ n \bm{R}^{\mathrm{sc}}_{l}\bm{t}_{l} \alpha}^{*}(\bm{k}_{\mathrm{sc}})
b_{ n \bm{R}^{\prime\mathrm{sc}}_{l}\bm{t}_{l} \alpha}(\bm{k}_{\mathrm{sc}})
e^{i \bm{k}_l \cdot( \bm{R}^{\mathrm{sc}}_{l}-\bm{R}^{\prime\mathrm{sc}}_{l})}
\delta_{\bm{k}_{\mathrm{sc}}-\bm{k}_l,\bm{G}^\prime_{\mathrm{sc}}}
\delta_{\bm{k}_{\mathrm{sc}}-\bm{k}_l,\bm{G}_{\mathrm{sc}}},\nonumber\\
&=& \frac{N_{\mathrm{sc}}}{N_l}
\sum_{n \bm{t}_{l}\alpha}
\sum_{ \bm{R}^{\mathrm{sc}}_{l},\bm{R}^{\prime\mathrm{sc}}_{l}}
\sum_{\bm{G}_{\mathrm{sc}}}
\delta(\epsilon-\epsilon_{\bm{k}_l+\bm{G}_{\mathrm{sc}}, n})
e^{i \bm{k}_l \cdot( \bm{R}^{\mathrm{sc}}_{l}-\bm{R}^{\prime\mathrm{sc}}_{l})}
b_{ n \bm{R}^{\mathrm{sc}}_{l}\bm{t}_{l} \alpha}^{*}(\bm{k}_l+\bm{G}_{\mathrm{sc}})
b_{ n \bm{R}^{\prime\mathrm{sc}}_{l}\bm{t}_{l} \alpha}(\bm{k}_l+\bm{G}_{\mathrm{sc}}).
\end{eqnarray}
Note that the summation over $\bm{G}_{\mathrm{sc}}$ is constrained by the condition that $\bm{k}_l+\bm{G}_{\mathrm{sc}} \in \mathrm{BZ}_{\mathrm{sc}}$. Thus, for a given $\bm{k}_l$, there exists a unique $\bm{k}_{\mathrm{sc}} \in \mathrm{BZ}_{\mathrm{sc}}$ that will be unfolded into $\bm{k}_l$. We can denote this unique $\bm{k}_{\mathrm{sc}}$ that unfolds to $\bm{k}_l$ as $\bm{k}_{\mathrm{sc}}(\bm{k}_l)$, which simplifies the above equation as follows:
\begin{equation}
A_l(\bm{k}_l,\epsilon) =
\frac{N_{\mathrm{sc}}}{N_l}
\sum_{n \bm{t}_{l}\alpha}
\sum_{ \bm{R}^{\mathrm{sc}}_{l},\bm{R}^{\prime\mathrm{sc}}_{l}}
\delta(\epsilon-\epsilon_{\bm{k}_{\mathrm{sc}}(\bm{k}_l) n})
e^{i \bm{k}_l \cdot( \bm{R}^{\mathrm{sc}}_{l}-\bm{R}^{\prime\mathrm{sc}}_{l})}
b_{ n \bm{R}^{\mathrm{sc}}_{l}\bm{t}_{l} \alpha}^{*}(\bm{k}_{\mathrm{sc}}(\bm{k}_l))
b_{ n \bm{R}^{\prime\mathrm{sc}}_{l}\bm{t}_{l} \alpha}(\bm{k}_{\mathrm{sc}}(\bm{k}_l)).
\end{equation}
Finally, when considering contributions from all layers, the final spectral function $A_{\mathrm{un}}(\bm{k},\epsilon)$ is given directly by Eq.~(\ref{eq_app_A}):
\begin{equation}
 A_{\mathrm{un}}(\bm{k},\epsilon) =  \sum_{l}\frac{N_{\mathrm{sc}}}{N_l}
\sum_{n \bm{t}_{l}\alpha}
\sum_{ \bm{R}^{\mathrm{sc}}_{l},\bm{R}^{\prime\mathrm{sc}}_{l}}
\delta(\epsilon-\epsilon_{\bm{k}_{\mathrm{sc}}(\bm{k}) n})
e^{i \bm{k} \cdot( \bm{R}^{\mathrm{sc}}_{l}-\bm{R}^{\prime\mathrm{sc}}_{l})}
b_{ n \bm{R}^{\mathrm{sc}}_{l}\bm{t}_{l} \alpha}^{*}(\bm{k}_{\mathrm{sc}}(\bm{k}))
b_{ n \bm{R}^{\prime\mathrm{sc}}_{l}\bm{t}_{l} \alpha}(\bm{k}_{\mathrm{sc}}(\bm{k})).
\end{equation}

Note that for $l=2$, our unfolding spectra for bilayer systems through basis transformation is equivalent to the result in Ref.~\cite{PhysRevMaterials.2.010801} using the projection operator decomposition approach, demonstrating the generality of our result.
\end{widetext}

\section{\label{app:numerical_30}Details of calculations for incommensurate quasicrystalline 30$^\circ$ TBG}
\begin{figure*}[ht]
\includegraphics[width=2\columnwidth, keepaspectratio]{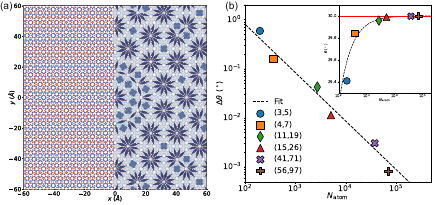}
\caption{\label{fig:30_structure} (a) Comparison of the first type of Stampfli's 12-fold tiling with 30$^\circ$ TBG. (b) Accuracy ($\Delta \theta = |\theta(m,r) - 30^\circ|$) as a function of the number of atoms ($N_{\mathrm{atom}}$) in supercell approximations for 30$^\circ$ TBG, with an additional insert presenting the data differently.}
\end{figure*}

\begin{figure*}
\includegraphics[width=2.0\columnwidth, keepaspectratio]{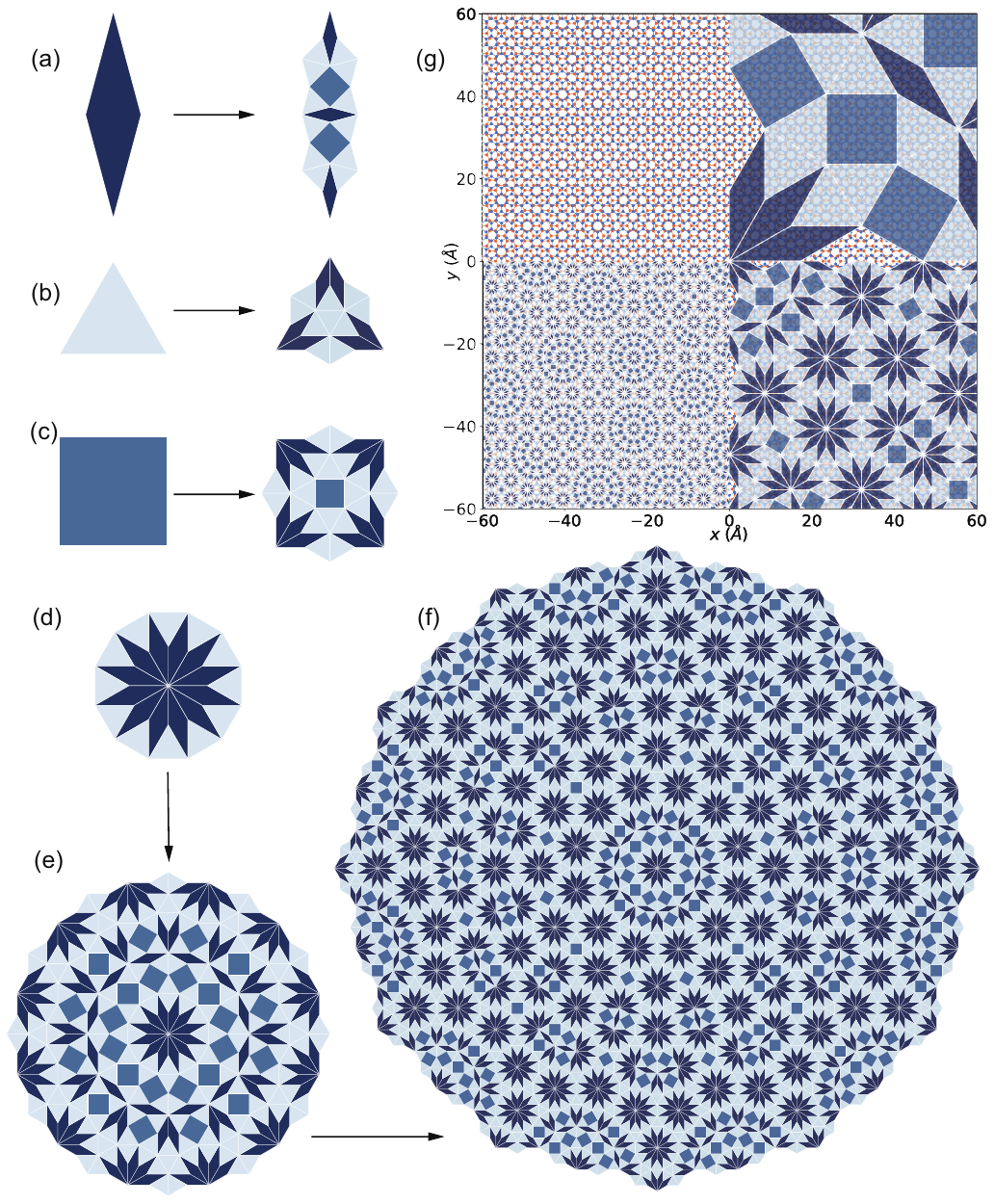}
\caption{\label{fig:stampfli_1} Substitution rules for color-coded (a) rhombus, (b) triangle, and (c) square tiles in the first type of Stampfli's 12-fold tiling. Color-coded (d) Initial configuration, (e) First inflation, and (f) Second inflation of the first type of Stampfli's 12-fold tiling. (g) Comparison between the first type of Stampfli's 12-fold tiling and 30$^\circ$ TBG at different inflation stages.}
\end{figure*}

\begin{figure*}
\includegraphics[width=2\columnwidth, keepaspectratio]{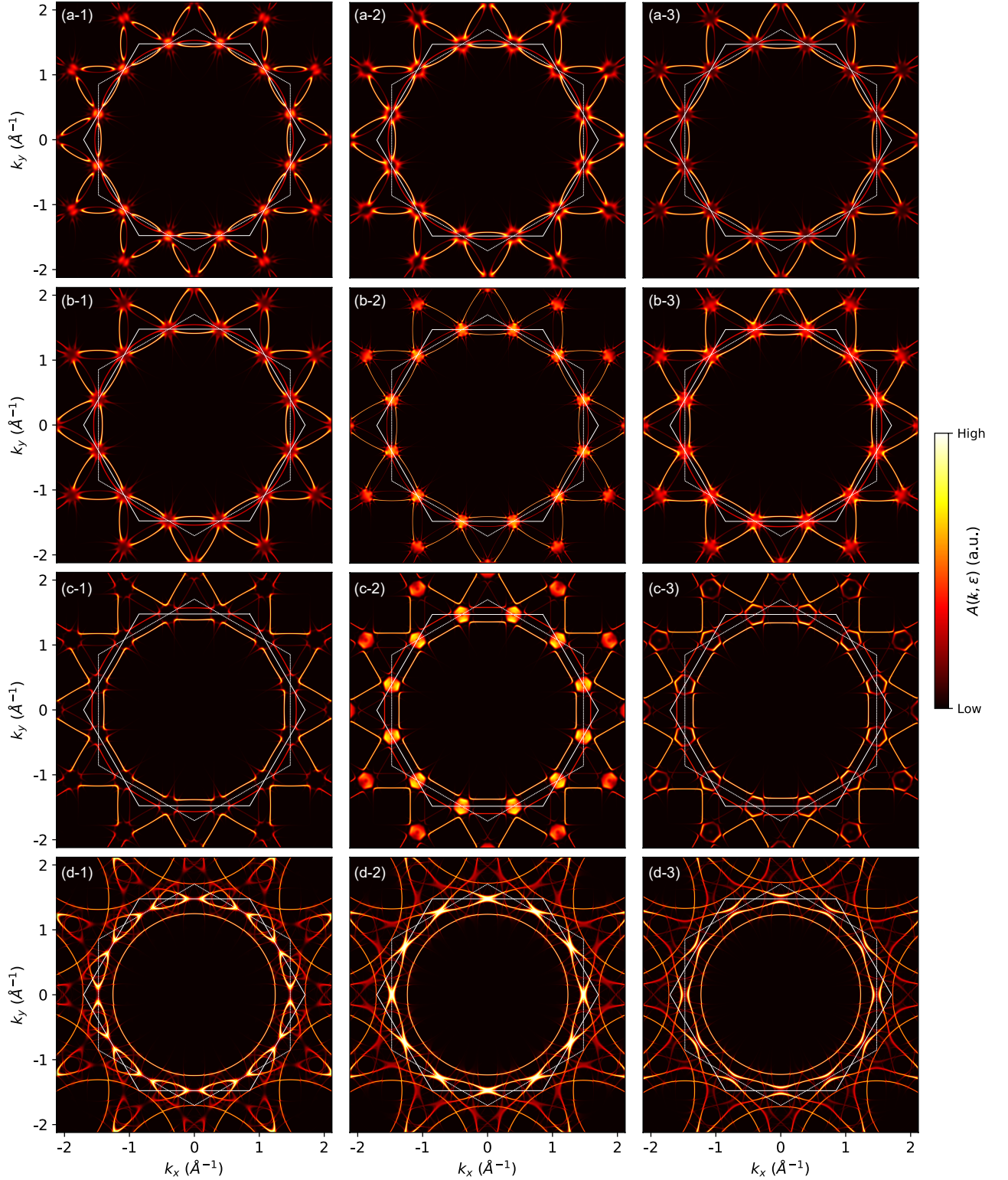}
\caption{\label{fig:30_Ecuts_more} Evolution of Fermi surfaces around VHS peaks labeled by (a) $\gamma$, (b) $\zeta$, (c) $\eta$, and (d) $\xi$ in Fig.~\ref{fig:30_DOS_band}(b). For each case, the Fermi surfaces at (1) energies slightly below the VHS peak, (2) at the VHS peak, and (3) slightly above the VHS peak are presented: (a-1) $\epsilon=-1.60$ eV, (a-2) $\epsilon=-1.64$ eV, (a-3) $\epsilon=-1.69$ eV, (b-1) $\epsilon=-1.71$ eV, (b-2) $\epsilon=-1.75$ eV, (b-3) $\epsilon=-1.78$ eV, (c-1) $\epsilon=-1.86$ eV, (c-2) $\epsilon=-1.96$ eV, (c-3) $\epsilon=-2.05$ eV, (d-1) $\epsilon=-2.69$ eV, (d-2) $\epsilon=-2.74$ eV, and (d-3) $\epsilon=-2.79$ eV. The cutoff parameter utilized is $G_{\mathrm{cut}}=12$~\AA$^{-1}$. White dotted and solid lines represent boundaries of $\mathrm{BZ}_1$ and $\mathrm{BZ}_2$, respectively.}
\end{figure*}

Here, we present the details of calculations for incommensurate quasicrystalline 30$^\circ$ TBG. The atomic structure of 30$^\circ$ TBG is shown in Fig.~\ref{fig:GmG_both}(d), where the quasicrystalline nature is evident as shown in Fig.~\ref{fig:30_structure}(a) and Fig.~\ref{fig:stampfli_1}(g) by comparing with the first type of Stampfli's 12-fold tiling~\cite{Stampfli1986ADQ}.

Typically, there are three types of Stampfli's 12-fold tiling~\cite{Stampfli1986ADQ,Frettlöh_Harriss_Gähler}, each with an inflation ratio of $2+\sqrt{3}$. The first type, illustrated in Fig.~\ref{fig:stampfli_1}, employs three types of tiles (Rhombus, Triangle, and Square), color-coded for distinction. Corresponding substitution rules are displayed in Fig.~\ref{fig:stampfli_1}(a)-(c). A comparison between the first type of Stampfli's 12-fold tiling and 30$^\circ$ TBG at different inflation stages is presented in Fig.~\ref{fig:stampfli_1}(g), highlighting the self-similarity.

Using supercell approximants near $30^\circ$ to approximate the electronic structure of quasicrystalline 30$^\circ$ TBG faces challenges due to the exponential growth in the number of atoms $N_{\mathrm{atom}}$ needed for improved accuracy. In Table~\ref{tab:30_N_atoms} and Fig.~\ref{fig:30_structure}(b), we present co-prime integers $(m,r)$ characterizing the resulting supercell, along with $N_{\mathrm{atom}}$ and accuracy $\Delta \theta = |\theta(m,r) - 30^\circ|$ based on Eq.(\ref{eq_app_theta_mr}). The exponential rise in $N_{\mathrm{atom}}$ presents difficulties for large-scale supercell calculations, a drawback inherent in pursuing a fully periodic approximation. In contrast, our theory avoids this drawback by utilizing single-layer periodicity, delivering accurate results with only a small cutoff, as demonstrated in Fig.~\ref{fig:30_DOS_band}.

\begin{table}[b]
\caption{\label{tab:30_N_atoms}%
Minimal supercell approximants for quasicrystalline $30^\circ$ TBG.}
\begin{ruledtabular}
\begin{tabular}{ccc}
$(m,r)$ & $\Delta \theta~(^\circ)$ & $N_{\mathrm{atom}}$ \\
\hline
(3,5)& 5.907$\times 10^{-1}$ & 194 \\
(4,7)& 1.583$\times 10^{-1}$ & 362 \\
(11,19)& 4.241$\times 10^{-2}$ & 2702 \\
(15,26)& 1.136$\times 10^{-2}$ & 5042 \\
(41,71)& 3.045$\times 10^{-3}$ & 37634 \\
(56,97)& 8.159$\times 10^{-4}$ & 70226 \\
(153,265)& 2.186$\times 10^{-4}$ & 524174 \\
\end{tabular}
\end{ruledtabular}
\end{table}

%

\end{document}